\documentclass[preprint,flushrt]{aastex}
\usepackage{natbib}
\usepackage{lineno}
\newcommand\pflux{\mbox{${\rm \, ph \,\, cm^{-2} \, s^{-1}}$}}
\def\gsim{\mathrel{\hbox{\rlap{\hbox{\lower6pt\hbox{$\sim$}}}\hbox{$>$}}}}
\slugcomment{Submitted to The Astrophysical Journal}

\shorttitle{Fermi LAT detected blazars}
\shortauthors{Abdo et al.}

\title{The Second Catalog of Active Galactic Nuclei Detected by the  {\it Fermi} Large Area Telescope}
\author{
M.~Ackermann\altaffilmark{1}, 
M.~Ajello\altaffilmark{1}, 
A.~Allafort\altaffilmark{1}, 
E.~Antolini\altaffilmark{2,3}, 
W.~B.~Atwood\altaffilmark{4}, 
M.~Axelsson\altaffilmark{5,6,7}, 
L.~Baldini\altaffilmark{8}, 
J.~Ballet\altaffilmark{9}, 
G.~Barbiellini\altaffilmark{10,11}, 
D.~Bastieri\altaffilmark{12,13}
% ...
}
\altaffiltext{1}{W. W. Hansen Experimental Physics Laboratory, Kavli Institute for Particle Astrophysics and Cosmology, Department of Physics and SLAC National Accelerator Laboratory, Stanford University, Stanford, CA 94305, USA}
\altaffiltext{2}{Istituto Nazionale di Fisica Nucleare, Sezione di Perugia, I-06123 Perugia, Italy}
\altaffiltext{3}{Dipartimento di Fisica, Universit\`a degli Studi di Perugia, I-06123 Perugia, Italy}
\altaffiltext{4}{Santa Cruz Institute for Particle Physics, Department of Physics and Department of Astronomy and Astrophysics, University of California at Santa Cruz, Santa Cruz, CA 95064, USA}
\altaffiltext{5}{Department of Astronomy, Stockholm University, SE-106 91 Stockholm, Sweden}
\altaffiltext{6}{The Oskar Klein Centre for Cosmoparticle Physics, AlbaNova, SE-106 91 Stockholm, Sweden}
\altaffiltext{7}{Department of Physics, Royal Institute of Technology (KTH), AlbaNova, SE-106 91 Stockholm, Sweden}
\altaffiltext{8}{Istituto Nazionale di Fisica Nucleare, Sezione di Pisa, I-56127 Pisa, Italy}
\altaffiltext{9}{Laboratoire AIM, CEA-IRFU/CNRS/Universit\'e Paris Diderot, Service d'Astrophysique, CEA Saclay, 91191 Gif sur Yvette, France}
\altaffiltext{10}{Istituto Nazionale di Fisica Nucleare, Sezione di Trieste, I-34127 Trieste, Italy}
\altaffiltext{11}{Dipartimento di Fisica, Universit\`a di Trieste, I-34127 Trieste, Italy}
\altaffiltext{12}{Istituto Nazionale di Fisica Nucleare, Sezione di Padova, I-35131 Padova, Italy}
\altaffiltext{13}{Dipartimento di Fisica ``G. Galilei", Universit\`a di Padova, I-35131 Padova, Italy}
% ...

\begin{abstract}
The second catalog of  active galactic nuclei (AGNs) detected by the  {\it Fermi} Large Area Telescope (LAT) in two years of scientific operation is presented.
The Second LAT AGN Catalog (2LAC) includes 1017 \mbox{$\gamma$-ray} sources located at high Galactic latitudes ($|b|>10\arcdeg$) that are detected with a
test statistic ($TS$) greater than 25 and associated statistically with AGNs. However some of these are affected by analysis issues and some are associated with multiple AGNs. Consequently we define a clean sample which includes 886 AGNs, comprising 395 BL~Lacertae objects (BL~Lacs), 310
flat-spectrum radio quasars (FSRQs), 157 candidate blazars of unknown type (i.e., with broad-band blazar characteristics but with no optical spectral measurement yet), eight misaligned AGNs, four narrow-line Seyfert 1 (NLS1s), 10 AGNs of other types and two starburst galaxies.
Where possible, the blazars have been further classified based on their spectral energy distributions (SEDs) as archival radio, optical, and \mbox{X-ray} data permit. While almost all FSRQs have a synchrotron-peak frequency $<10^{14}$ Hz, about half of the BL~Lacs  have a synchrotron-peak frequency $>10^{15}$ Hz.  The 2LAC represents a significant improvement relative to the First LAT AGN Catalog  (1LAC), with 52\% more associated sources. The full characterization of the newly detected sources will require more broad-band data. Various properties, such as \mbox{$\gamma$-ray} fluxes and photon power law spectral indices, redshifts, \mbox{$\gamma$-ray} luminosities, variability, and
archival radio luminosities---and their correlations are presented and discussed for the different blazar classes. The general trends observed in 1LAC are confirmed. 

\end{abstract}


\begin{document}


\section{\label{sec:intro}Introduction}

This paper presents a catalog of active galaxy nuclei (AGNs) associated through formal probabilities with high-energy $\gamma$-ray sources detected in the first two years of the {\it Fermi} Gamma-ray Space Telescope mission by the Large Area Telescope (LAT).  This catalog is based on the larger second LAT  catalog, 2FGL \citep{2FGL} and is a follow-up of the first LAT AGN catalog, 1LAC \citep{1LAC}.  The second LAT AGN catalog, 2LAC includes a number of analysis refinements and additional association methods which have substantially increased the number of associations over 1LAC. 
 
The high sensitivity and nearly uniform sky coverage of the LAT make it a powerful tool for investigating the properties of large populations.  The first list of bright AGNs detected by the LAT, the LAT Bright AGN Sample \citep[LBAS;][]{LBAS} included AGNs at high Galactic latitude ($|b|>10\arcdeg$) detected with high significance (Test Statistic\footnote{The Test Statistic is defined as $TS$ = 2(log $\mathcal{L}$(source)- log $\mathcal{L}$(nosource)), where  $\mathcal{L}$ represents the likelihood of the data given the model with or without a source present at a given position on the sky,}, $TS>100$) during the first three months of scientific operation.  This list comprised 58 flat-spectrum radio quasars (FSRQs), 42 BL~Lacs, two radio galaxies, and four AGNs of unknown type.           
The next evolution, 1LAC, based on the first 11 months of data included 671 sources detected with $TS > 25$ at high Galactic latitudes ($|b| > 10\arcdeg$). The 1LAC Clean Sample (sources with single associations and not affected by analysis issues) comprised 599 sources:  248 FSRQs, 275 BL~Lacs, 26 other AGNs and 50 blazars of unknown type.
The main findings of 1LAC, summarized below, were consistent with those found with the LBAS.
\begin{enumerate}
\item Only a small number of non-blazar AGNs detected;
\item redshift distributions peaking at $z \approx 1$ for 1LAC FSRQs and at low
redshift for 1LAC BL~Lacs with known redshifts (only 60\% of the total);
\item similar numbers of BL~Lacs and FSRQs; 
\item high-synchrotron-peaked sources representing the largest subclass among BL~Lacs;
\item little evidence for different variability properties for FSRQs and
BL~Lacs using monthly light curves;  a more detailed analysis based on weekly light curves \citep{Abdo_var} showed that bright FSRQs exhibit larger fractional variability than do BL~Lacs. 
\item  the detected high-synchrotron-peaked sources have harder spectra and lower $\gamma$-ray luminosity than lower synchrotron-peaked sources.
\end{enumerate}

The 1LAC catalog has proven to be an invaluable resource opening the way to numerous studies on the blazar sequence and the BL~Lac-FSRQ dichotomy issue \citep{Ghi11,Ghi10t,Bjo10,Che11,Tra10}, blazar evolution \citep{Ino10}, the comparison of properties of $\gamma$-ray loud and  $\gamma$-ray quiet blazars \citep{Mahony2010,Linford2011,Kar10, Cha11}, the contribution of AGNs to the extragalactic diffuse $\gamma$-ray background \citep{Abdo_EDB,Sin11,Ven11}, the correlation between AGNs and the sources of ultra high-energy cosmic rays  \citep{Jia10,Der10,Nem10,Kim10}, the timing correlations between the activity in the  $\gamma$-ray bands and other bands \citep{Pus10,Ric10}, and the attenuation of $\gamma$-rays by Extragalactic Background Light \citep{Abdo_EBL,Rau10}. The release of the 1LAC also triggered TeV observations leading to discoveries of new TeV-emitting blazars \citep[e.g., ][]{RBS0413:ATel2272}.  
 
Here we report on the AGNs associated with LAT sources detected after 24 months of scientific operation. The second LAT AGN catalog comprises a total of 1017 sources detected with $TS>25$ at high Galactic latitudes ($|b| > 10\arcdeg$). Due to some analysis issues, some sources were flagged in the 2FGL catalog and 26 sources have two possible associations, so we define a Clean Sample, which includes 886 sources. An additional 104 sources at $|b| < 10\arcdeg$ are also presented here.

In Section 2, the observations by the LAT and the analysis employed to produce
the two-year catalog are described.  In Section 3, we explain the methods for associating
\mbox{$\gamma$-ray} sources with AGN counterparts and present the results of these methods.  Section 4 describes the different schemes for
classifying 2LAC AGNs. Section 5  provides a brief census of the 2LAC sample.
Section 6 summarizes some of the properties of the 2LAC, including the
\mbox{$\gamma$-ray} flux distribution, the \mbox{$\gamma$-ray} photon spectral
index distribution, the \mbox{$\gamma$-ray} variability properties, the
redshift distribution, and the \mbox{$\gamma$-ray} luminosity distribution.  In
Section 7, we discuss some radio, optical and  TeV properties of the 2LAC AGNs.  We discuss the implications of the 2LAC results in Section 8
and conclude in Section 9.

In the following, we use a $\Lambda$CDM cosmology with values within $1\sigma$ of
the {\it Wilkinson Microwave Anisotropy Probe} ({\it WMAP}) results
\citep{WMAP11}; in particular, we use $h = 0.70$, $\Omega_m = 0.27$, and
$\Omega_\Lambda = 0.73$, where the Hubble constant
$H_0=100h$~km~s$^{-1}$~Mpc$^{-1}$.  We also define the radio spectral indices
such that $S(\nu) \propto \nu^{-\alpha}$.

\section{\label{sec:obs}Observations with the Large Area Telescope --- Analysis Procedures}

The  2LAC sources are a subset of those in the 2FGL catalog, so we only briefly summarize the  analysis here and we refer the reader to the paper describing the 2FGL catalog \citep{2FGL} for details. The data were collected over the first 24 months of the mission from 2008 August 4 to 2010 August 1, with an overall data-taking efficiency of 74\%. Time intervals during which the rocking angle of the LAT was greater than 52$^{\circ}$ were excluded (leading to a reduction in exposure of less than 2\%).  A cut on the zenith-angle of $\gamma$-rays of 100$^{\circ}$ was applied. The Pass 7\_V6 Source event class \citep{2FGL} was used, with photon energies between 100 MeV and 100 GeV.  In the study of  the highest-energy photons detected for each source,  presented in \S \ref{sec:hep}, photons belonging to the purest (i.e., with the lowest instrumental background) class  (Pass 7\_V6 Ultraclean) were used, without any high-energy cut. 

The source detection procedure considered seed sources taken from 1FGL and the results of three point-source detection methods, described in \cite{1FGL} were used: {\it mr\_filter} \citep{sp98},  {\it PGWave} \citep{ciprini07} and the minimal spanning tree method \citep{Cam08}. With these seeds, an all-sky likelihood analysis produced an ``optimized'' model, where parameters characterizing the diffuse components\footnote{The Galactic diffuse model and isotropic background model (including the $\gamma$-ray diffuse and residual instrumental backgrounds) are described in \cite{2FGL}. Alternative Galactic diffuse models were tested as well.} in addition to sources were fitted. The analysis of the residual TS map provided new seeds that were included in the model for a new all-sky likelihood analysis. This iterative procedure yielded 3499 seeds that were then passed on to the maximum likelihood analysis for source characterization. 

The analysis was performed with the binned likelihood method implemented in the  {\it pyLikelihood} library  of the Science Tools\footnote{http://fermi.gsfc.nasa.gov/ssc/data/analysis/documentation/Cicerone/} (v9r23p0).   Different spectral fits were carried out with a single power-law function (\mbox{$dN/dE=N_0\:(E/E_{0})^{-\Gamma}$}) and  a LogParabola function \\ (\mbox{$dN/dE=N_0\:(E/E_{0})^{-\alpha-\beta \log(E/E_0)}$}), where $E_0$ is an arbitrary reference energy adjusted on a source-by-source basis to minimize the correlation between $N_0$ and the other fitted parameters over the whole energy range (0.1 to 100 GeV). Whenever the difference in log(likelihood) between these two fits was greater than 8 \citep[i.e., $TS_{curve}$, defined as twice this difference, see ][ was greater than 16]{2FGL} the LogParabola results were retained. The photon spectral index ($\Gamma$) presented in this paper was obtained from the single power-law fit for all sources. A threshold of $TS=25$ was applied to all sources, corresponding to a significance of approximately 4 $\sigma$. At the end of this procedure, 1873 sources survived the cut on TS. 
  Power-law fits were also performed in five different energy bands (0.1-0.3, 0.3-1, 1-3, 3-10 and
10-100~GeV), from which the energy flux was derived.  A variability index \citep[{\it TS$_{VAR}$}, see ][]{2FGL} was constructed from a likelihood test based on the monthly light curves,  with the null (alternative) hypothesis corresponding to the source being steady (variable). A source is identified as being variable at the 99\% level if the variability index is equal or greater than 41.6.   

Some of the 2FGL sources were flagged as suspicious when certain issues arose during their analysis \citep[see][ for a full list of these flags]{2FGL}.  The issues that most strongly affected the 2LAC list are: i) sources moving beyond their 95\% error ellipse when changing the model of Galactic diffuse emission, ii) sources with  $TS >$ 35 going down to $TS <$ 25 when changing the diffuse model, iii) sources located closer than $\theta_{ref}$ \citep[defined in Table 2 of ][]{2FGL} to a brighter neighbor, iv) source $Spectral\_Fit\_Quality >$ 16.3 ($\chi^{2}$ between spectral model and flux in five energy bands). Therefore, we applied a selection on sources to build a clean sample of AGNs.

\begin{figure}
\centering
\resizebox{16cm}{!}{\rotatebox[]{0}{\includegraphics{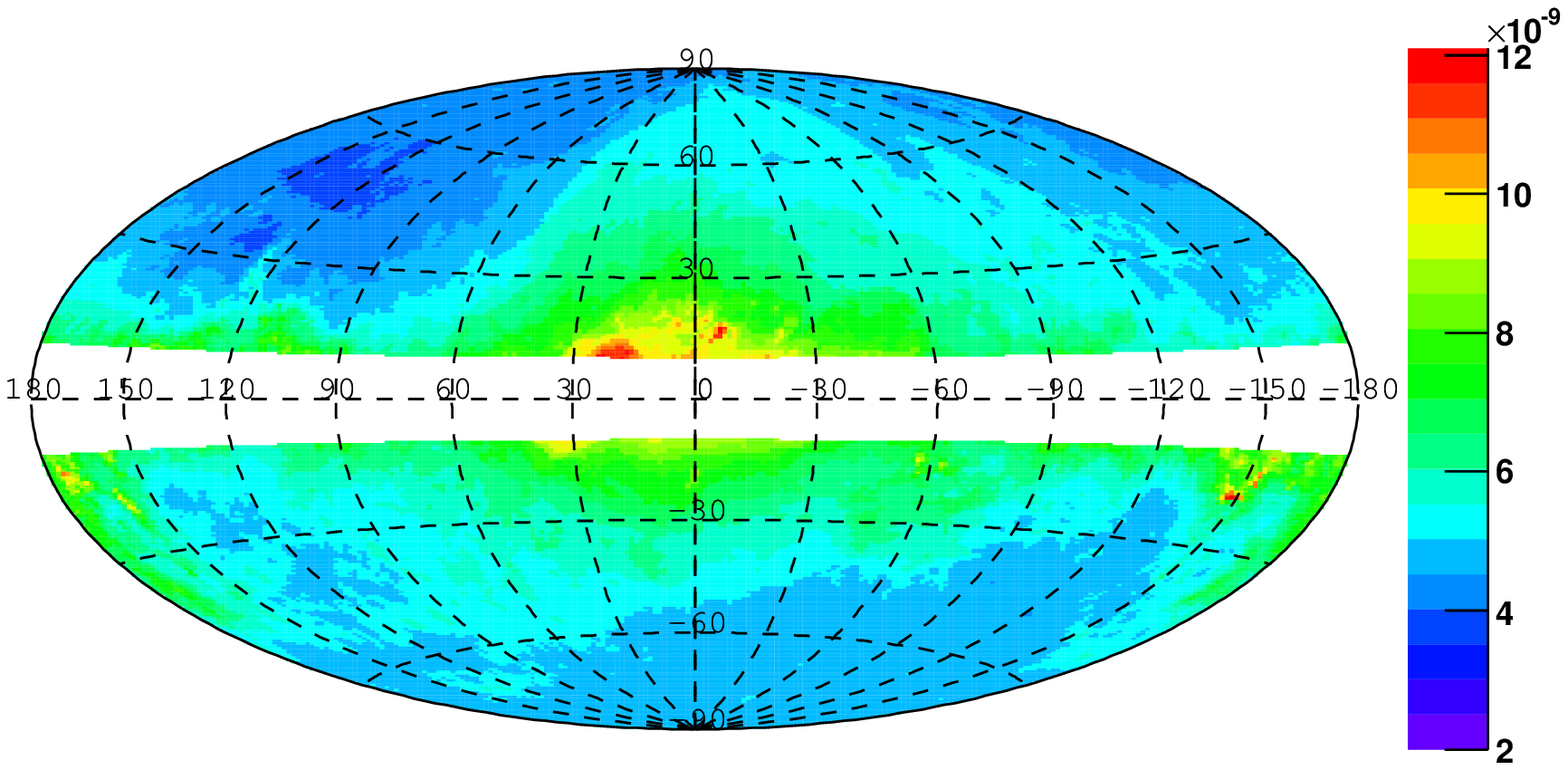}}}
\caption{Point-source flux limit in units of \pflux{} for $E > 100$~MeV and photon spectral index
$\Gamma = 2.2$ as a function of sky location (in Galactic coordinates).}
\label{fig:sens}
\end{figure}

\begin{figure}
\centering
\resizebox{16cm}{!}{\rotatebox[]{0}{\includegraphics{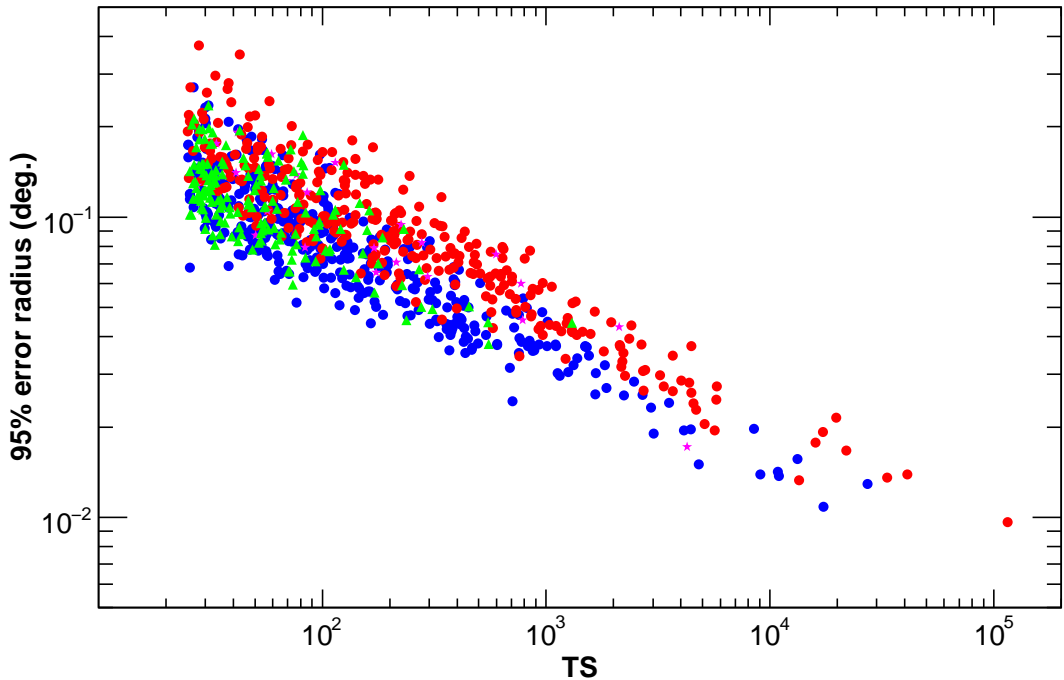}}}
\caption{ 95\% containment radius versus TS. Red: FSRQs, blue: BL~Lacs, green: unknown type, magenta: non-blazar AGNs.}
\label{fig:r95_TS}
\end{figure}

Thanks to its large field of view and sky survey mode, the LAT sensitivity is relatively uniform at large Galactic latitudes, although the switch from a rocking angle of 35$\arcdeg$ to 50$\arcdeg$ in September 2009 reduced this uniformity \citep{2FGL}.  A  map of the flux limit, calculated for the two-year period covered by this catalog, a  $TS=25$ and a photon index of 2.2, is shown in Galactic coordinates in Figure \ref{fig:sens}. 
The 95\% error radius \citep[defined as the geometric mean  of the semimajor and semiminor axes of the ellipse fitted to the $TS$ map, see][]{2FGL} is plotted as a function of $TS$ in Figure \ref{fig:r95_TS}. It ranges from about $0\fdg01$ for 3C~454.3, the brightest LAT blazar, to $0\fdg2$ on average for sources just above the detection threshold (similar to 1LAC).    

\section{Source Association}

The LAT localization accuracy is not precise enough to permit the
determination of a lower-energy counterpart based only on positional
coincidence.  We assert a firm counterpart identification only if the
variability detected by the LAT corresponds with variability at other
wavelengths.  In practice, such identifications have been  made only for 28 2FGL AGNs
(see Table~5 in \citealt{2FGL}).  For the rest, we use statistical approaches for finding associations between LAT sources and AGNs.  

In 1FGL, several sources were flagged as {\it affiliated} AGNs (and thus not included in 1LAC) as the methods providing associations were not able to give a quantitative association probability. 
Moreover some LAT-detected blazars turn out to be fainter in radio than the flux limit of catalogs of flat-spectrum radio sources. In order to improve over the results of 1LAC by including these faint radio sources,  the association procedure for building the 2LAC list makes use of three different methods: the Bayesian Method (used in 1FGL/1LAC) and two additional methods, namely the Likelihood Ratio Method and the $\log N - \log S$ Method. These procedures are described respectively in \S \ref{sec:gtsrcid}, \ref{sec:like}, \ref{sec:steve}. For a counterpart to be considered as associated, its association probability must be $>$ 0.8 for at least one method.

The two additional methods improve the association results through the use of physical properties of the candidate counterparts, such as the surface density and the spectral shape in the radio energy band, in addition to the positional coincidence with the $\gamma$-ray source. Considering potential counterparts with lower radio flux  enables more high-synchrotron peaked BL~Lacs to be selected but the number of FSRQs is also increased. This is achieved through the use of surveys and serendipitous findings, as the  available catalogs (used by the Bayesian Method) are not deep enough.

\subsection{\label{sec:gtsrcid}The Bayesian Association Method}

The Bayesian method \citep{deRuiter77,ss92}, implemented by the {\it gtsrcid}
tool in the LAT {\it ScienceTools}, is similar to that used by \citet{mhr01} to
associate EGRET sources with flat-spectrum radio sources.  A more
complete description is given in the Appendix of \cite{1FGL} and in \cite{2FGL}, but we provide a basic
summary here.  The method uses Bayes' theorem to calculate the posterior
probability that a source from a catalog of candidate counterparts is truly an
emitter of \mbox{$\gamma$-rays} detected by the LAT.  The significance of a
spatial coincidence between a candidate counterpart from a catalog $C$ and a
LAT-detected \mbox{$\gamma$-ray} source is evaluated by examining the local
density of counterparts from $C$ in the vicinity of the LAT source.  We can
then estimate the likelihood that such a coincidence is due to random chance
and establish whether the association is likely to be real.  To each catalog
$C$, we assign a prior probability, assumed for simplicity to be the same for
all sources in $C$, for detection by the LAT.  The prior probability for each
catalog can be tuned to give the desired number of false positive associations
for a given threshold on the posterior probability, above which the
associations are considered reliable (see \S~\ref{sec:cat}). 
The posterior probability threshold for high-confidence associations was set to 80\%.

Candidate counterparts were drawn from a number of source catalogs.  
With respect to 1FGL, all catalogs for which more comprehensive compilations became available have been updated.
The catalogs used are the 13th edition of the Veron catalog \citep{AGNcatalog},
version 20 of BZCAT \citep{bzcat},
the 2010 December 5 version of the VLBA Calibrator Source List\footnote{
The VLBA Calibrator Source List  can be downloaded from http://www.vlba.nrao.edu/astro/calib/vlbaCalib.txt.},
and
the most recent version of the TeVCat catalog\footnote{http://tevcat.uchicago.edu}.
We also added new counterpart catalogs,
the Australia Telescope 20-GHz Survey (AT20G) \citep{AT20G_CAT,Mas11} and
the {\it Planck} Early Release Catalogs \citep{Planckcatalog}.

\subsection{\label{sec:like}The Likelihood-Ratio (LR) Association Method}

The Likelihood Ratio method has been introduced to make use of uniform surveys in the radio and in X-ray  bands in order to search for possible counterparts among the faint radio and X-ray sources. The main differences with the Bayesian method are that i) the LR makes use of counterpart densities through the $\log N - \log S$  and therefore the source flux, ii) the LR assumes, in this paper, that the counterpart density is constant over the survey region. An improved version of the LR should take into consideration the local density, which is mandatory in the case of optical counterparts but not for radio and X-ray because of their lower surface densities. We assigned $\gamma$-ray associations and estimate their reliability using a likelihood ratio  analysis which has frequently been used to assess identification probabilities for radio, infrared and optical sources \citep[e.g., ][]{deRuiter77,Pre83,ss92,Lon98,Mas01}. 

We made use of a number of relatively uniform radio surveys. Almost all radio AGN candidates of possible interest are detected either in the NRAO VLA Sky Survey  \citep[NVSS; ][]{NVSScatalog} or the Sydney University Molonglo Sky Survey \citep[SUMSS; ][]{SUMSScatalog}. We added the 4.85 GHz Parkes-MIT-NRAO (PMN) Surveys  \citep{PMNcat,Gri95,Wri94,Wri96}, with a typical flux limit of about 40 mJy which varies as a function of declination, as well as 
 the recently released AT20G source catalog \citep{AT20G_CAT,Mas11}, which contains entries for 5890 sources observed at declination $\delta <$0. In this way we are able to look for counterparts with radio flux down to 5 mJy. To look for additional possible counterparts we cross-correlated the LAT sources with the most sensitive all-sky X-ray survey, the ROSAT All Sky Survey Bright and Faint Source Catalogs \citep{RASSbright,RASS_FAINT_CAT}.
A source is considered as a likely counterpart of the $\gamma$-ray source if its reliability (see Eq. \ref{rel}) is $>$0.8 in at least one survey.

The method, which computes the probability that a suggested association is the `true' counterpart, is outlined as follows. For each candidate counterpart $i$ in the search area neighboring a 2FGL  $\gamma$-ray source
$j$, we calculate the normalized distance between $\gamma$-ray
and radio/X-ray positions:
\begin{equation}
r_{ij}=\frac{\Delta}{(\sigma_{a}^{2} + \sigma_{b}^{2})^{1/2}}
\label{rij}
\end{equation}
where $\Delta$ is the angular distance between the $\gamma$-ray source and its prospective counterpart and $\sigma_{a}$ and $\sigma_{b}$ represent the errors on $\gamma$-ray and counterpart positions respectively.

Given $r_{ij}$, we must now distinguish between two mutually exclusive possibilities: i) the candidate is a confusing background object that happens to lie at distance $r_{ij}$ from the $\gamma$-ray source ii) the candidate is the `true' counterpart that appears at distance $r_{ij}$ owing solely to the $\gamma$-ray and radio/X-ray positional uncertainties. We assume that the $\gamma$-ray and radio/X-ray positions would coincide if these uncertainties were negligibly small \citep{Mas01}.

To distinguish between these cases, we compute the likelihood ratio $LR_{ij}$,
defined as:
\begin{equation}
LR_{ij}=\frac{e^{-r_{ij}^{2}/2}}{N(>S_{i})~A}
\label{LR}
\end{equation}
where $N(>S_{i})$ is the surface density of objects brighter than candidate $i$ (i.e., the $\log N - \log S$) and $A$ is the solid angle spanned by the 95\% confidence LAT error ellipse.
The likelihood ratio $LR_{ij}$ is therefore simply the ratio of the probability of an association (the Rayleigh distribution: $r\exp{(-r^{2}/2)}$), to that of a chance association at $r$. 
$LR_{ij}$ therefore represents a `relative weight' for each match $ij$, and our aim is to find an optimum cutoff value $LR_{c}$ above which a source is considered to be a reliable candidate. 

The value of LR$_c$ can be evaluated using simulations as described in \cite{Lon98}. We generate a truly random background population with respect to the $\gamma$-ray sources by randomly displacing $\gamma$-ray sources within an annulus with inner and outer radii of $2^{\circ}$ and $10^{\circ}$ respectively around their true positions. In addition to extragalactic sources,  2FGL contains a population of Galactic $\gamma$-ray emitters that follows a rather narrow latitude distribution.  We
limit the  source displacement in Galactic latitude to  $ b~\pm~b_{max}$, where
\begin{equation}
b_{max}=r_{max}(1-sech^{2}\frac{b}{b_0})
\end{equation}
$r_{max}=10^{\circ}$, $b$ is the Galactic latitude of the $\gamma$-ray source, and $b_{0}=5^{\circ}$ is the angular scale height above the Galactic plane for which the latitude displacement is reduced. We further require that $b_{max}>0\fdg2$ to allow for a non-zero latitude displacement of sources in the Galactic plane, and require any source to be shifted by at least $r_{min}=2^{\circ}$ away from its original location. The results derived here do not critically depend on the exact values of $r_{max}$, $b_{max}$ and $b_0$ chosen for the simulations.

We generated 100 realizations of this fake $\gamma$-ray sky and for each of the 100 fake $\gamma$-ray catalogs, we calculated the respective LR value for all counterparts. Then we compared the number of associations for ({\it true}) $\gamma$-ray source positions with the number of associations found for ({\it random}) $\gamma$-ray source positions, which enabled us to determine a critical value LR$_c$ for reliable association. From these distributions, we computed the reliability as a function of LR.
\begin{equation}
R(LR_{ij})=1-\frac{N_{random}(LR_{ij})}{N_{random}(LR_{ij})+N_{true}(LR_{ij})}
\label{rel}
\end{equation}
where $N_{true}$ and $N_{random}$ are the number of associations with $\gamma$-ray sources in the {\it true} sky and those in the simulated ({\it random}) one respectively. The reliability computed in this way also represents an approximate measure of the association probability for a candidate with given LR.

\begin{figure}
\centering
\resizebox{14cm}{!}{\rotatebox[]{-90}{\includegraphics{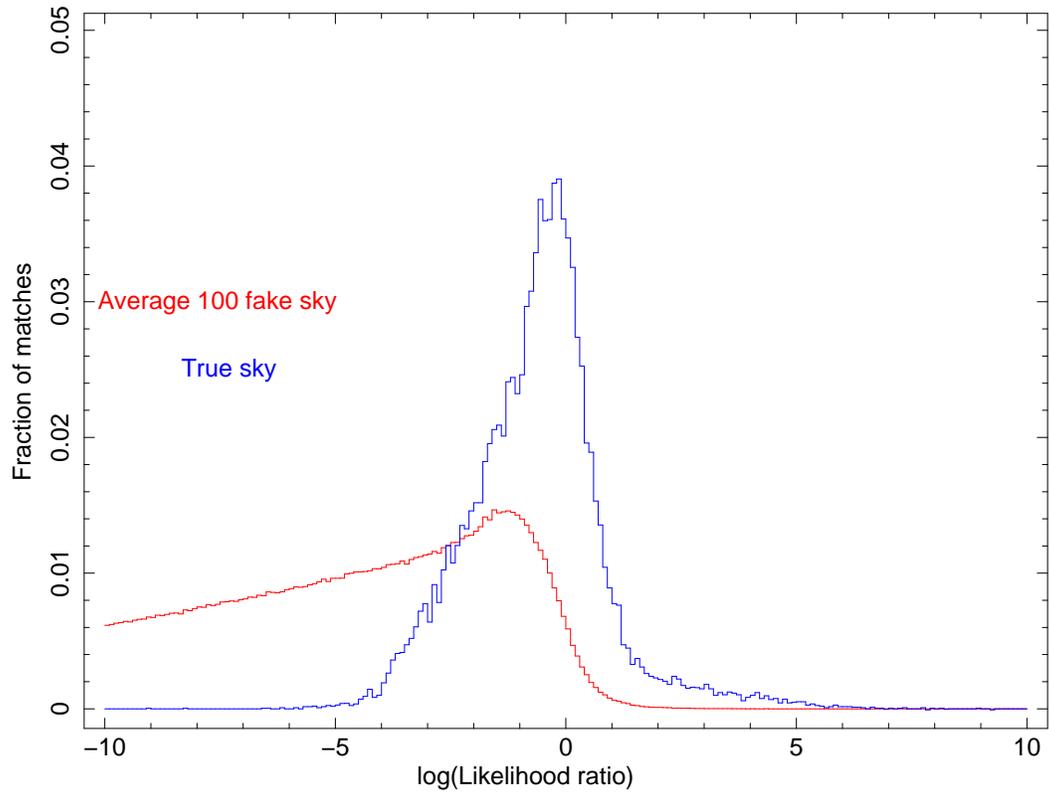}}}
\caption{Distribution of likelihood ratio (LR) for radio - $\gamma$-ray matches at {\it true} $\gamma$-ray positions (blue histogram), and at {\it fake} $\gamma$-ray positions (red histogram), for the NVSS survey.}
\label{fig:nvss_LR}
\end{figure}

Figure \ref{fig:nvss_LR} shows the two distributions of {\it true} (blue) and {\it fake} (red) LR values for the NVSS survey, which we report as an example. In order to obtain $R$ as a function of $LR$  we parametrize the reliability curve with the following function:
\begin{equation}
f(LR)=1-a ~ exp(-b ~ LR)
\end{equation}
The $a$ and $b$ parameters are given in Table \ref{tbl-ab} for the different surveys. We use this function to calculate the reliability for each value of LR and select high-confidence counterparts. The values of log (LR$_c$) above which the reliability is  greater than $80\%$ are  listed in Table \ref{tbl-ab} as well for the different surveys.

After having calculated the reliability of the association with the use of the LR based on the $\log N - \log S$ cited above, we look for typical blazar characteristics of a source taking into consideration its optical class and radio spectrum slope. The 2LAC being a list of AGN candidate counterparts for 2FGL sources, we include only AGN-type sources.
We therefore looked at their optical spectra through an extensive program of optical follow-up (M.~S. Shaw et al., 2011, in preparation and S. Piranomonte et al., 2011, in preparation) and the BZCAT list. Moreover we evaluated their spectral slopes in the radio through a cross-correlation with  catalogs of flat-spectrum radio sources.

\subsection{\label{sec:steve}$\log N - \log S$ Method}

The $\log N - \log S$ association method is a modified version of the Bayesian
method for blazars.  The Bayesian method assesses the probability of
association between a $\gamma$-ray source and a candidate counterpart using the
local density of such candidates; this local density is estimated simply by
counting candidates in a nearby region of the sky.  The $\log N - \log S$
method differs in one small but important way: the density of ``competing''
candidates is estimated by using a model of the radio $\log N - \log S$
distribution of the candidate population.  Specifically, the density $\rho$
that goes into the Bayesian calculation for a candidate $k$ with radio flux
density $S_k$ and radio spectral index $\alpha_k$ is
$\rho(S > S_k, \, \alpha < \alpha_k)$, the density of sources that are at
least as bright and have spectra at least as flat as source $k$.  \citep[This
attrition-based approach---considering only those sources that are as ``good''
as or ``better'' than the candidate in question---was used in practically the
same way by ][]{mat97,mhr01}.  The
$\log N - \log S$ method has the distinct advantage of being extensible to radio
data not found in any formal catalog.  In particular, the method can be applied
to new radio observations that explicitly target unassociated LAT sources with
no loss of statistical validity.

In order to exploit the size and uniformity of the CRATES catalog and its
proven utility as a source of radio/$\gamma$-ray blazar associations, we sought
a model of the 8.4~GHz $\log N - \log S$ distribution of the flat-spectrum
radio population.  For $S \ga 85$~mJy, CRATES itself provides sufficient
coverage of this population that the $\log N - \log S$ distribution can be
directly examined and modeled.  Below this flux density, however, the CRATES
coverage declines rapidly.  By definition, CRATES only includes sources with
4.85~GHz flux densities of at least 65~mJy, so the faint population is
explicitly disfavored.  In addition, because of this 4.85~GHz flux density
limit, CRATES sources that are faint at 8.4~GHz are far more likely to be
steep-spectrum objects.

Because the LAT selects $\gamma$-ray sources with radio counterparts fainter than those in radio catalogs of flat-spectrum radio sources such as  CRATES,  we required another source of 8.4~GHz data to study the faint end of the
$\log N - \log S$ distribution.  For this purpose, we looked to the Cosmic
Lens All-Sky Survey \citep[CLASS;][]{class1,class2}.  While CLASS did target sources down to a fainter limit than
CRATES, we were able to push to even lower flux densities by considering
serendipitous CLASS detections (i.e., sources that were not explicitly targeted
by CLASS but which were detected in CLASS pointings).  We assembled this sample
by taking CLASS detections that were at least 60$\arcsec$ away from any CLASS
pointing position in order to ensure that we were not using any component of
the ``real'' CLASS target (e.g., a jet).  We also considered only those sources
with $S > 10$~mJy at 8.4~GHz to avoid sidelobes or other mapping errors.

\begin{figure}
\centering
\resizebox{14cm}{!}{\rotatebox[]{0}{\includegraphics{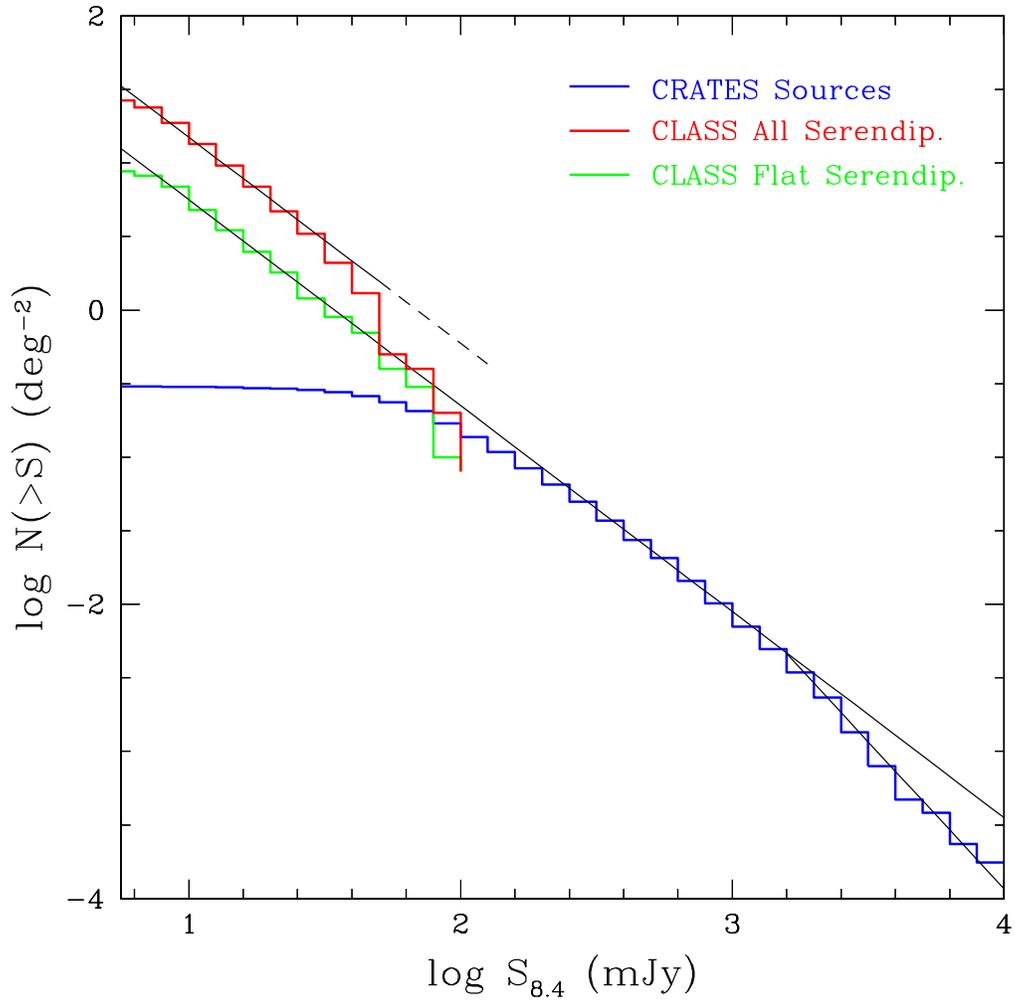}}}
\caption{$\log N - \log S$ for CRATES and serendipitous CLASS sources. The lines correspond to the parametrization mentioned in the text.  }
\label{fig:logNlogS_steve}
\end{figure}

Because the serendipitous sources were not intentionally targeted and appear in
the CLASS data purely by a coincidence of their locations on the sky, they
represent a statistically unbiased sample of the 8.4~GHz population, unaffected
by any selection criterion other than their ability to be detected cleanly by
the VLA.  In order to model just the flat-spectrum members of this population,
we computed spectral indices using 1.4~GHz data from NVSS and imposed a spectral
index cut of $\alpha < 0.5$ (the same cut as for CRATES).  In the end, we had
a sample of $\sim$ 300 flat-spectrum sources with flux densities ranging from
10~mJy to $\sim$110~mJy.  However, while the shape of the $\log N - \log S$
distribution for this sample could be studied, the sky area of this ``survey''
was not well defined, so the $\log N - \log S$ was not properly normalized.
Fortunately, the flux density range of the CRATES coverage overlapped
sufficiently with that of the serendipitous sample to allow us to scale the
latter until it agreed with the former in the overlap region.  We then had
a full characterization of the 8.4~GHz $\log N - \log S$ distribution of the
flat-spectrum population from 10~mJy to $\sim$10~Jy (see Figure \ref{fig:logNlogS_steve}).  The integral form of the
distribution is well modeled piecewise by

\begin{eqnarray}
\log N(>S) = 4.07 - 2.0 \log S \: \mathrm{for} \: \log S > 3.2 \\
\log N(>S) = 2.15 - 1.4 \log S \: \mathrm{for} \: \log S < 3.2
\end{eqnarray}
where $N(>S)$ is the number of sources per square degree with flux density
greater than $S$ at 8.4~GHz, expressed here in mJy .

With an understanding of the flux density distribution in hand, we turned to
the second component of the attrition, the spectral indices.  In particular, we
sought to characterize how the spectral index distribution varied with
increasing flux density.  We sorted the radio data into logarithmic bins in
flux density centered on $10$~mJy, $10^{1.5}$~mJy, and so on up to
$10^4$~mJy, and we examined the spectral index distribution for each bin.  In
every case, the spectral index distribution was very well approximated by a
Gaussian, and as it turned out, the widths of these Gaussians were very nearly
the same, never deviating from the mean value of 0.29 by more than 0.01.  Since
these deviations are statistically insignificant, we adopt 0.29 as the fiducial
standard deviation of the $\alpha$ distribution for all flux densities.  The
centers of the Gaussians increased with increasing flux density; we
approximated the flux density dependence of the mean $\alpha$ as

\begin{equation}
\mu_\alpha(S) = 0.527 - 0.187 \log S
\end{equation}

Thus, for a candidate counterpart $k$ with flux density $S_k$ and spectral
index $\alpha_k$, the fraction $F_\alpha$ of competing counterparts that have
spectra at least as flat as $k$ is the area to the left of $\alpha_k$ under a
Gaussian with $\sigma_\alpha = 0.29$ centered on $\alpha = \mu_\alpha(S)$.  The
sought-after density of competing counterparts,
$\rho(S > S_k, \, \alpha < \alpha_k)$, is then given simply by

\begin{equation}
\rho(S > S_k, \, \alpha < \alpha_k) = F_\alpha \times N(>S)
\end{equation}

Once the attrition-based value is used for $\rho$, the rest of the Bayesian
method is unchanged.  The prior probability can be calibrated in exactly the
same way; for this approach, we find that a value of 0.0475 gives the proper
number of false positives.

\subsection{Association Results}

\begin{figure}
\centering
\resizebox{14cm}{!}{\rotatebox[]{0}{\includegraphics{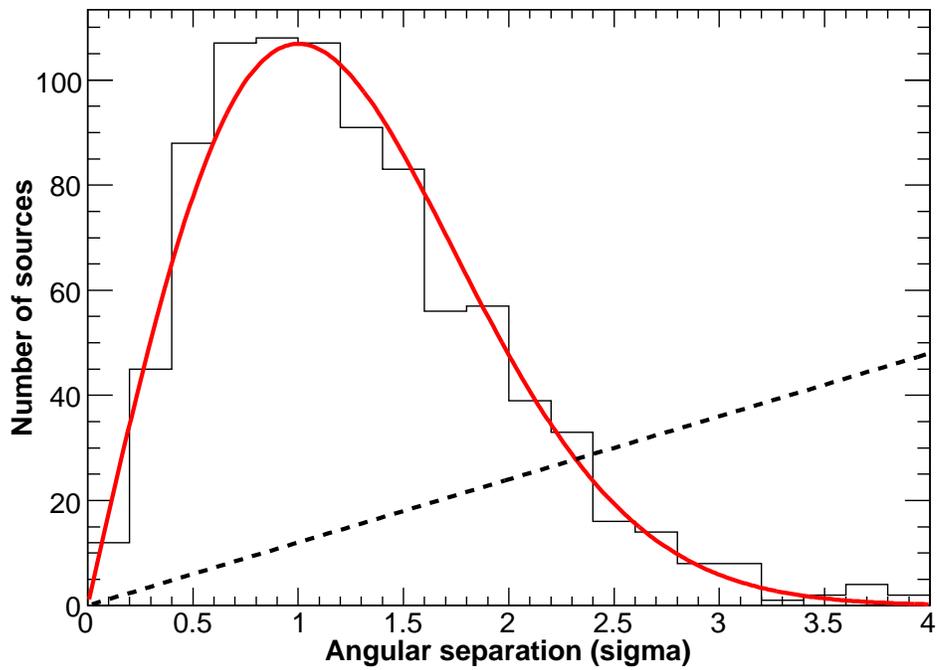}}}
\caption{Distribution of angular separation between 2LAC sources and assigned counterparts. The red curve corresponds to the expected distribution for real associations, the dashed line to that expected for spurious associations.}
\label{fig:separation}
\end{figure}

Using three different methods has increased the fraction of formally associated counterparts with respect to the 1LAC work. In total we found that 1095 2FGL sources have been associated with at least one counterpart source at other wavelengths (corresponding to a total of 1120 counterparts). Only 26 2FGL sources have been associated with more than one counterpart. A total of 1017 counterparts are at high Galactic latitude ($|b|>10\arcdeg$), comprising the full 2LAC sample. Of these 1017 sources, 704 sources (69\%) are associated by all three methods.  We found that 886 2LAC sources have a single counterpart and  are free of the analysis issues mentioned in \S \ref{sec:obs} (103 sources were discarded on these grounds), defining the Clean Sample. We note that 640 sources of the Clean Sample (72\%) are associated by all three methods.  
Table \ref{tbl-assoc} compares the performance of the different methods in terms of total number of associations, number of false associations $N_{false}$, calculated  as $N_{false}=\sum_i (1-P_i)$ and number of sources solely associated via a given method, $N_S$, for the full and Clean samples. The largest probability from the three methods has been used to evaluate the overall value of $N_{false}$. The contamination is found to be less than 2\% in both 2LAC and the Clean Sample. The distribution of separation distance between 2LAC sources and their assigned counterparts is shown in Figure \ref{fig:separation}.  

The probabilities given by the three methods are listed in Tables \ref{tab:clean} and \ref{tab:lowlat} for the high- and low-latitude sources respectively.
Where possible, counterpart names have been chosen to adhere to the NASA/IPAC Extragalactic Database\footnote{http://ned.ipac.caltech.edu/} nomenclature.
 In these tables,  a redshift z=0 means that the redshift could not be evaluated even though an optical spectrum was available, e.g., for BL~Lacs without redshifts, while no mentioned redshift  means that no optical spectrum was available.

\section{\label{sec:classif}Source Classification }

The ingredients of the classification procedure are optical spectrum or other blazar characteristics (radio loudness, flat radio spectrum, broad band emission, variability, and polarization). We made use of different surveys, including the VLBA Calibrator Survey \citep[VCS;][]{Bea02,Fom03,Pet05,Pet06,Pet08,Kov07}. PMN-CA \citep{Wri97} is a simultaneous 4.8 GHz and 8.64 GHz survey of PMN sources in the region $-87\arcdeg <\delta< -38.5\arcdeg$ observed with the Australia Telescope Compact Array. CRATES-Gaps is an extension of the CRATES sample to areas of the sky not covered by CRATES due to a lack of PMN coverage from which to draw targets.  It consists of an initial 4.85 GHz finding survey performed with the Effelsberg 100-m telescope and follow-up at 8.4 GHz with the VLA \citep{Hea09}. FRBA, standing for Finding and Rejecting Blazar Associations, is a VLA survey at 8.4 GHz that explicitly targeted otherwise unidentified 1FGL sources.

\begin{itemize}
\item
To classify a source optically  we made use of, in decreasing order of precedence: optical spectra from our intensive follow-up programs, the BZCAT list (i.e., FSRQs and BL~Lacs in this list), spectra available in the literature. The latter information was used only if we found a published spectrum.
\item
If an optical spectrum was not available, we looked for the evidence of typical blazar characteristics, such as radio loudness, a flat radio spectrum at least between 1.4 GHz and 5 GHz, broad band emission (i.e., detection of the candidate counterpart at a frequency outside the radio band). We did not take into account the optical polarization. In this context we made use of, in decreasing order of precedence: BZCAT (i.e., the BZU objects in this list), detection from high frequency surveys and catalogs (AT20G, VCS, CRATES, FRBA, PMN-CA, CRATES-Gaps, CLASS lists), radio and X-ray coincidence association with probability $\ga$ 0.8.
\end{itemize}

The classes are the following:

\begin{itemize}
\item FSRQ, BL Lac, radio galaxy, steep-spectrum radio quasar (SSRQ), Seyfert, NLS1, starburst galaxy -- for sources with well-established classes in literature and/or through an optical spectrum with a good evaluation of emission lines.
\item AGU -- for sources without a good optical spectrum or without an optical spectrum at all: 
\subitem $a$) BZU objects in the BZCAT list; 
\subitem $b$) sources in AT20G, VCS, CRATES, FRBA, PMN-CA, CRATES-Gaps, or CLASS lists, selected by the $\log N - \log S$ method (see \S \ref{sec:steve}) and the Likelihood Ratio method (see Sect. \ref{sec:like});
\subitem $c$) coincident radio and X-ray sources selected by the Likelihood Ratio method (see Sect \ref{sec:like}).
\item AGN -- this class is more generic than AGU. These sources are not confirmed blazars nor blazar candidates (such as AGU).  Although they may have had evidence for their flatness in radio emission or broad-band emission, our intensive optical follow-up program did not provide a clear evidence for optical blazar characteristics.
\end{itemize}

As compared to the 1LAC, the classification scheme in the 2LAC has improved thanks to the two additional association methods, allowing for two more types of AGUs (classes {\it b} and {\it c} in the above description). With the previous association procedure, only about 50\% of the current AGUs would have been included in the 2LAC.

In addition to the optical classifications, sources have also been classified according to their SEDs using the scheme detailed in \S \ref{sec:sedclass}. 

\subsection{\label{sec:optclass} Follow-up Optical Program for Redshift and Optical Classification}

A large fraction ($\sim$ 60\%) of the redshifts and optical classifications presented in Table \ref{tab:clean} are derived from dedicated optical follow-up campaigns and specifically from spectroscopic observations performed with the Marcario Low-Resolution Spectrograph \citep{lrs} on the 9.2~m Hobby-Eberly Telescope (HET) at McDonald Observatory. Other spectroscopic facilities used for these optical results include the 3.6 m New Technology Telescope at La Silla, the 5 m Hale Telescope at Palomar, the 8.2 m Very Large Telescope at Paranal, the 10 m Keck I Telescope at Mauna Kea and the  DOLORES spectrograph  at 3.6 m Telescopio Nazionale Galileo at La Palma.
Our spectroscopic campaigns first considered all the sources which were statistically associated (probability larger than 90\%) with one of the still unclassified $\gamma$-ray sources in the 1LAC  which have X-ray, radio and optical counterparts within their error boxes. We then consider all sources with a flat radio spectrum.
This work will be detailed in two upcoming publications (M.~S. Shaw et al., 2011, in preparation and S. Piranomonte et al., 2011, in preparation). Overall, about 67 1LAC sources have gained a measured redshift between the 1LAC and the 2LAC.

\subsection{\label{sec:sedclass}SED Classification}

As in 1LAC, we classify blazars also based on the synchrotron peak frequency of the broadband SED \citep{SEDpaper}. This scheme extends to all blazars the standard classification system introduced by \cite{pg95} for BL~Lacs. We estimate the synchrotron peak frequency $\nu^S_\mathrm{peak}$, using the broadband indices $\alpha_{ro}$  (between 5 GHz and 5000 \AA) and $\alpha_{ox}$  (between 5000 \AA~and 1 keV).  The analytic relationship $\nu^S_\mathrm{peak}=f(\alpha_{ro},\alpha_{ox})$ was calibrated  with 48 SEDs in \citet{SEDpaper}. We use the estimated value of $\nu_\mathrm{peak}^\mathrm{S}$ to classify the source as either a low-synchrotron-peaked blazar (LSP, for sources with $\nu_\mathrm{peak}^\mathrm{S} < 10^{14}$~Hz), an intermediate-synchrotron-peaked blazar (ISP, for $10^{14}$~Hz~$< \nu_\mathrm{peak}^\mathrm{S} < 10^{15}$~Hz), or a high-synchrotron-peaked blazar (HSP, if $\nu_\mathrm{peak}^\mathrm{S} > 10^{15}$~Hz).

In this work the broad-band spectral indices are calculated from data in the radio, optical and X-ray bands. The radio flux measurements are obtained mainly from the GB6 \citep{GB6cat} and PMN  catalogs. The optical fluxes are taken mainly from the USNO-B1.0 \citep{USNOcat}   and SDSS \citep{SDSS} catalogs. For BL~Lac objects we applied  a correction to the optical flux assuming a giant elliptical galaxy with absolute magnitude M$_r$= $-$23.7 as the host galaxy of the blazar \citep[see ][]{2000Urry}. In the case of FSRQs we neglected the dilution of non-thermal light by the host galaxy. Finally, the X-ray fluxes are derived from the  RASS \citep{RASSbright},  {\it Swift}-XRT, WGA \citep{WGAcat}, {\it XMM} \citep{XMMcat} and BMW \citep{BMWcat} catalogs.

We express the value of $\nu_\mathrm{peak}^\mathrm{S}$ in the rest frame. BL~Lacs without known redshifts were assigned the median BL~Lac redshift,  z=0.27. The same redshift was assigned to AGU without measured redshifts, except for those with FSRQ-like properties ($\nu_\mathrm{peak}^\mathrm{S}<10^{15}$ Hz in the observer frame and $\Gamma\ge 2.2$, corresponding to the approximate dividing line between FSRQs and BL~Lacs found in 1LAC), which were given the FSRQ redshift median, z=1.12.

We note that the SED classification method assumes that the optical and \mbox{X-ray} fluxes come exclusively from  non-thermal emission. Recently, using simultaneous {\it Planck}, {\it Swift} and {\it Fermi} data, \cite{paperPLANCK} found that the optical/UV emission was significantly contaminated by thermal/disk radiation (known as the big blue bump).  FSRQs (and the AGUs which we assumed to be FSRQ like) are most affected by this contamination. To account for this, we systematically reduce  $\nu_\mathrm{peak}^\mathrm{S}$ by 0.5 in logarithmic space for these sources as suggested by \cite{paperPLANCK}.

\begin{figure}
\centering
\resizebox{16cm}{!}{\rotatebox[]{0}{\includegraphics{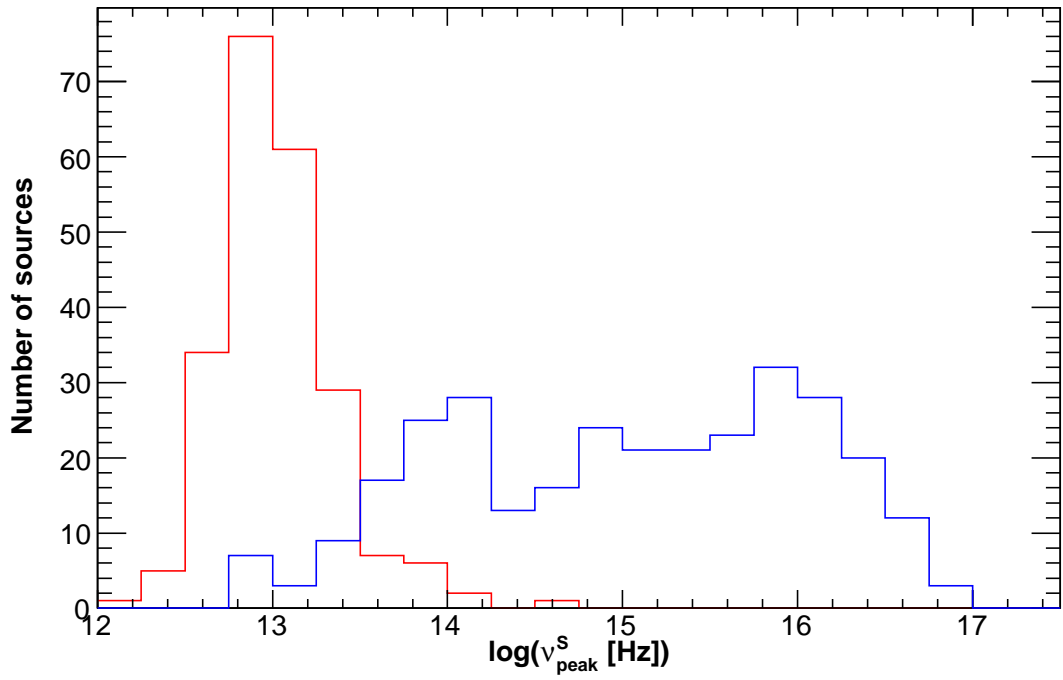}}}
\caption{Distributions of the synchrotron peak frequency $\nu^S_{peak}$ for FSRQs (red) and BL~Lacs (blue) in the Clean Sample.}
\label{fig:syn_hist}
\end{figure}

The $\nu_\mathrm{peak}^\mathrm{S}$  distributions for FSRQs and BL Lacs are displayed in Figure \ref{fig:syn_hist}.
Some individual sources can differ from the general behavior of their class, e.g.,  2FGL J0747.7+4501 seems to be an ISP-FSRQ  with $\log \nu_\mathrm{peak}^\mathrm{S}=  14.66$. Inspection of the SED reveals that this high peak value is partly due to the blue bump (thermal emission in the optical band). The same feature is found in the other ISP-FSRQs.
Indeed,  we can conclude that even with the applied corrections  this method may lead to a significant overestimation of the position of $\nu_\mathrm{peak}^\mathrm{S}$ for some sources  where the thermal components are non-negligible. 

However, looking at the whole sample we can see that the two classes of objects have different distributions.
For FSRQs, the  average  $\langle \log \nu^S_\mathrm{peak}\rangle$ obtained in the 2LAC Clean Sample is 13.02 $\pm 0.35 $ while BL~Lacs are spread over the whole parameter space from low (LSP) to the highest frequencies (HSP). These results are consistent with those presented in \cite{1LAC} and in \cite{paperPLANCK}.

\begin{figure}
\centering
\resizebox{16cm}{!}{\rotatebox[]{0}{\includegraphics{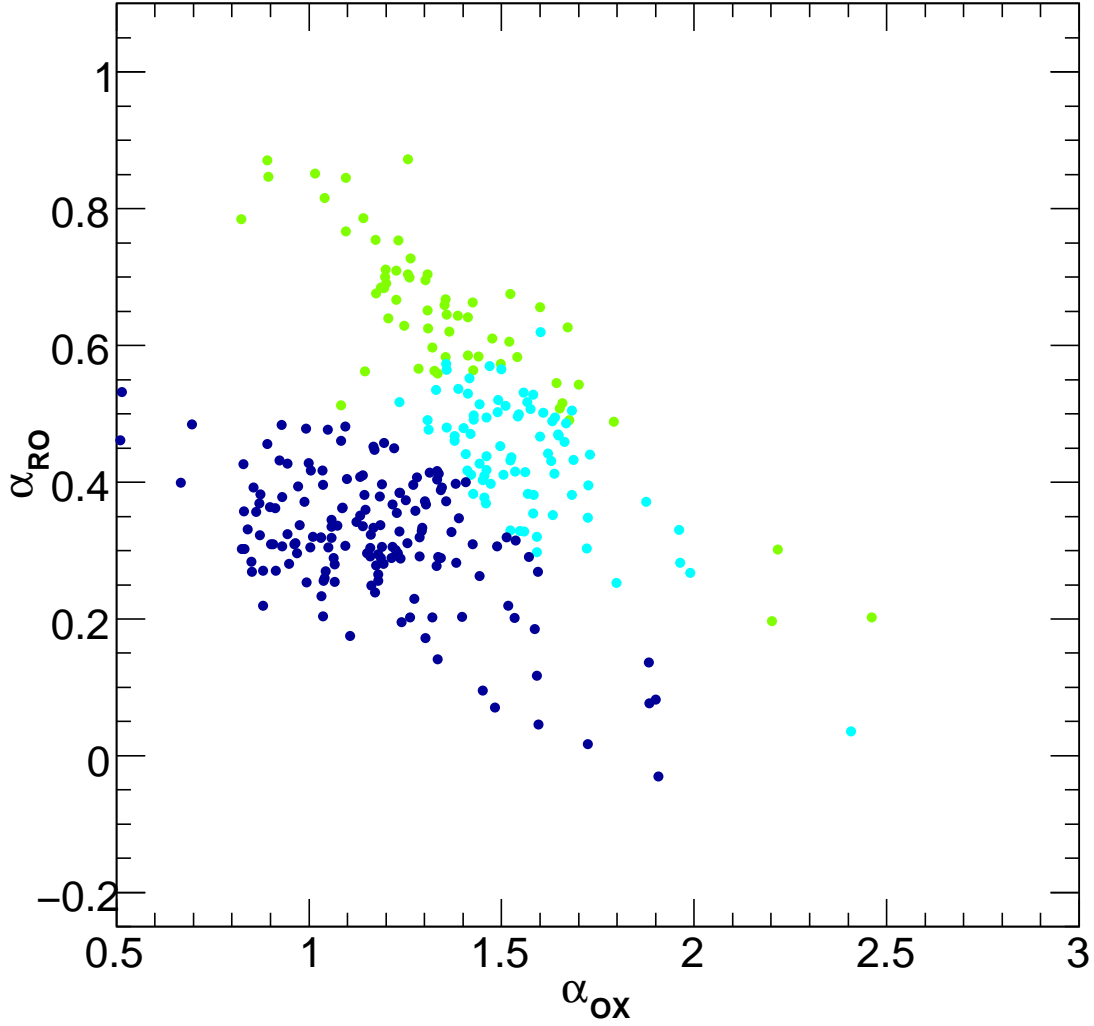}}}
\caption{$\alpha_\mathrm{ro}$ plotted against $\alpha_\mathrm{ox}$ for BL~Lacs. Green: LSPs, light blue: ISPs,  dark blue: HSPs. The overlap of sources with different classes in this plane is due to the redshift correction applied to $\nu^S_{peak}$ (determined in the rest frame).}
\label{fig:alpha_alpha}
\end{figure}

Figure \ref{fig:alpha_alpha} displays $\alpha_\mathrm{ro}$ versus  $\alpha_\mathrm{ox}$. Some sources, filling the bottom part of the $\alpha_\mathrm{ox} - \alpha_\mathrm{ro}$ plane, have much greater contamination by the host galaxy than the average assumed in our estimate.  Other outliers can be found in the upper part of the plane especially  for some extreme HSP sources including 2FGL J2343.6+3437, 2FGL J0304.5$-$2836, 2FGL J2139.1$-$2054, 2FGL J0227.3+0203 have a very low value of $\alpha_\mathrm{ox}$. This is probably due their being in high states in the X-ray band during the ROSAT observations. However, the SEDs built from archival data do point to a HSP classification.

\begin{figure}
\centering
\resizebox{16cm}{!}{\rotatebox[]{0}{\includegraphics{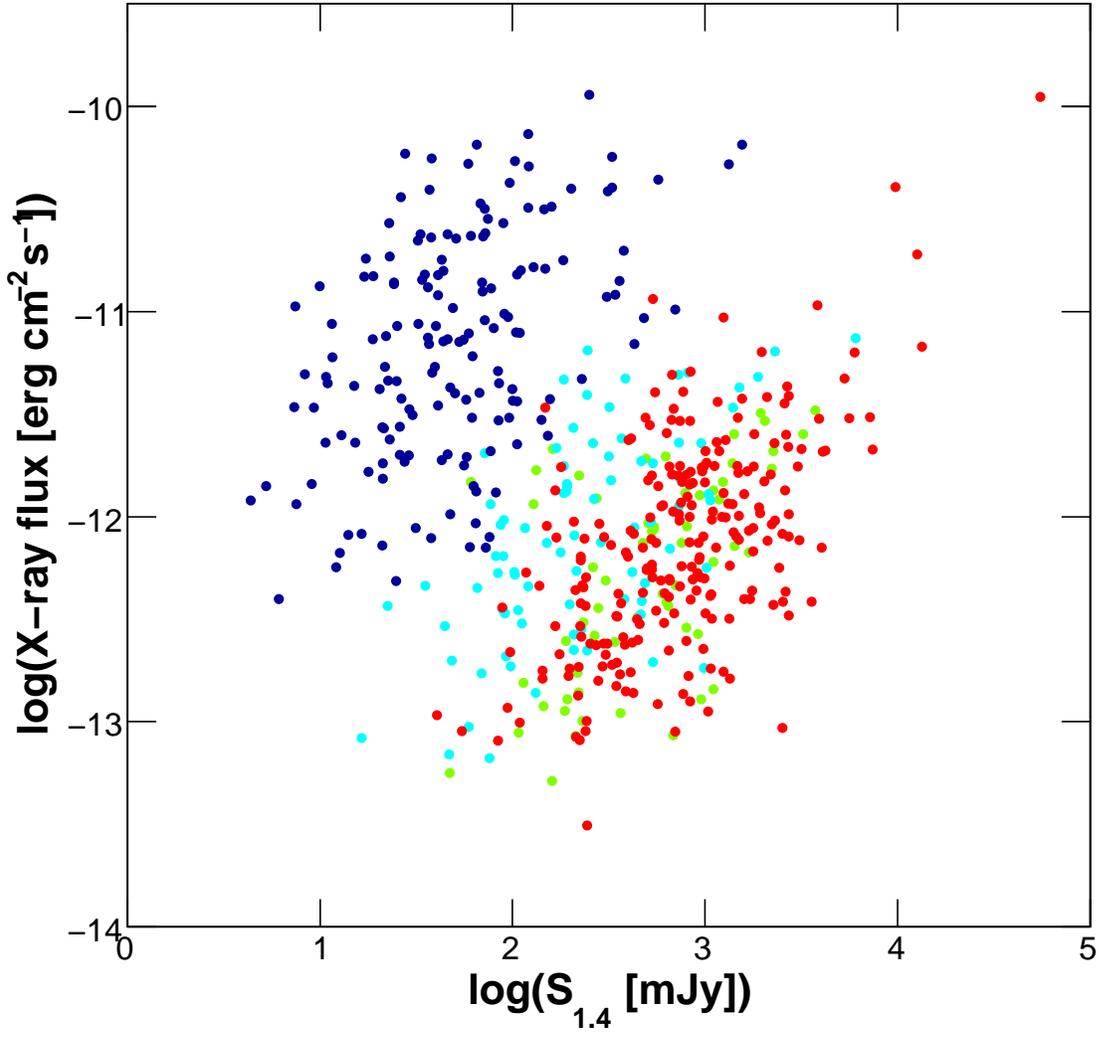}}}
\caption{X-ray flux  versus radio flux for blazars in the Clean Sample. Red: FSRQs,  green: LSP-BL Lacs, light blue: ISP-BL Lacs,  dark blue: HSP-BL Lacs.}
\label{fig:Fr_Fx}
\end{figure}

The X-ray flux is plotted against the radio flux in Figure \ref{fig:Fr_Fx}.  As in 1LAC, we see that the FSRQs (essentially all of the LSP type) and HSPs (all BL Lacs) are clearly divided. This plot  supports our method to classify the sources using multifrequency properties to estimate synchrotron peak frequency.

\section{\label{sec:cat} The Second LAT AGN Catalog (2LAC) }

The 2LAC catalog includes all sources with a significant detection over the two-year time period. Sources with only sporadic activity will be missing if they do not make the $TS>25$ cut as computed over the full time span.

\subsection{\label{sec:census}2LAC Population Census}

Table~\ref{tab:census} presents the breakdown of sources by type for
the entire 2LAC, the Clean Sample, and the low-latitude sample. The entire 2LAC includes 360 FSRQs, 423 BL~Lacs, 204 blazars of  unknown type and 30 other AGNs.  Of the 373 unassociated 1FGL sources located at $|b|>10\arcdeg$, 107 are now firmly associated with AGNs and listed in the 2LAC. Interestingly, 84 of these were predicted to be AGNs in \cite{1FGL_un}.  In the following only the Clean Sample  is considered in tallies and figures.  The Clean Sample comprises 886 sources in total, 395 BL~Lacs, 310 FSRQs, 157 sources of unknown type, 22 other AGNs, and 2 starburst galaxies. For BL~Lacs, 302 (76\% of the total) have an SED classification (i.e., 93 sources cannot be classified for lack of archival data), with HSPs representing the largest subclass (53\% of SED-classified sources), ISPs the second largest (27\%) and LSPs the smallest subclass (20\%, see Figure \ref{fig:syn_hist}). FSRQs with SED classification (224/310=72\%) are essentially all LSPs (99\%). 

\begin{figure}
\centering
\resizebox{16cm}{!}{\rotatebox[]{0}{\includegraphics{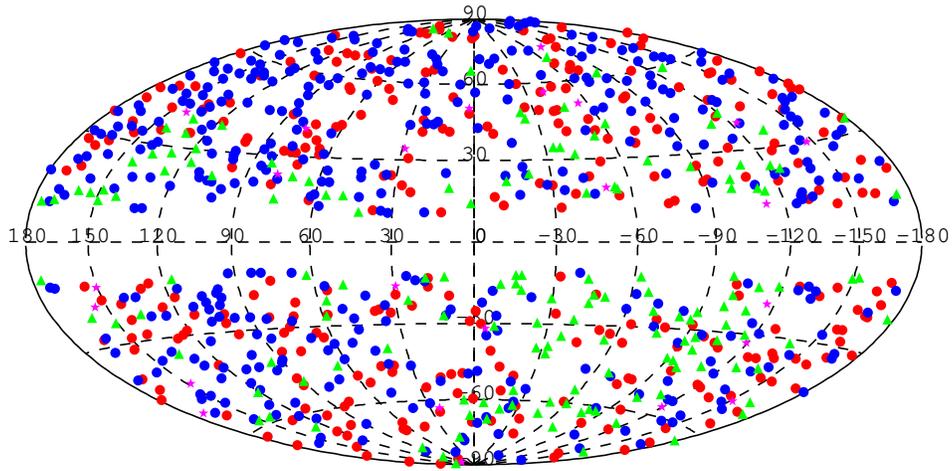}}}
\caption{Locations of the sources in the Clean Sample.  Red:\ FSRQs, blue:\ BL~Lacs, magenta:\ non-blazar AGNs, green:\ AGNs of unknown type.}
\label{fig:sky_map}
\end{figure}

\begin{figure}
\centering
\resizebox{16cm}{!}{\rotatebox[]{0}{\includegraphics{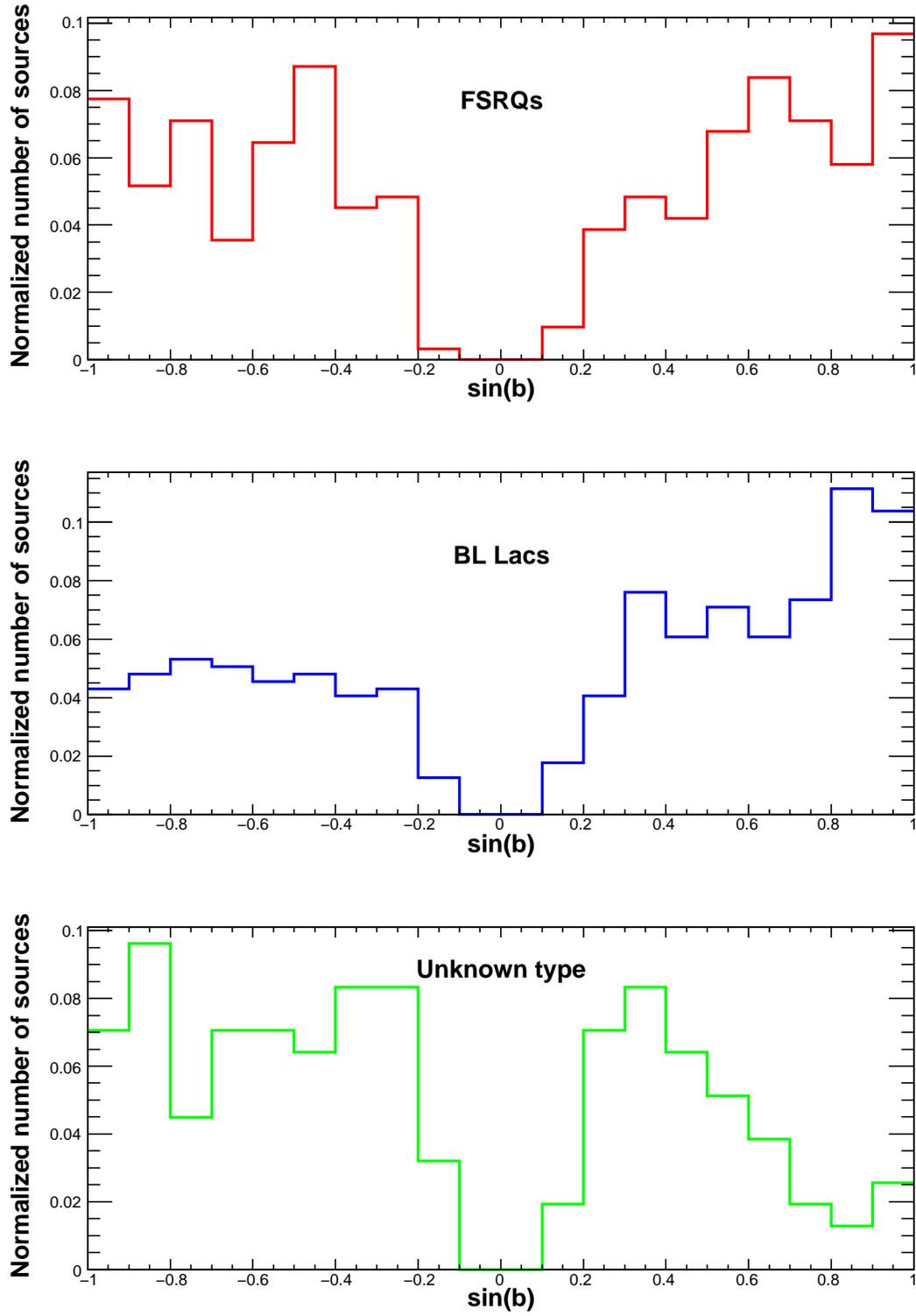}}}
\caption{Galactic latitude distributions of FSRQs (top) and BL Lacs (middle) and sources of unknown type (bottom)  from the Clean Sample.}
\label{fig:gal_lat}
\end{figure}

Figure \ref{fig:sky_map} shows the locations of the 2LAC sources. Some relative voids are present, the most prominent centered on ($l$,$b$)=($-$45$\arcdeg$,$-$45$\arcdeg$) reflecting a relative lack of  counterparts in the BZCAT catalog at that location. More generally, the observed anisotropy is mainly governed by the  non-uniformity of the counterpart catalogs.  A difference in the numbers of sources between the northern and the southern Galactic hemispheres is clearly visible for BL~Lacs in Figure \ref{fig:sky_map}. This conclusion is confirmed in Figure \ref{fig:gal_lat} displaying the Galactic-latitude distributions for FSRQs and BL~Lacs and blazars of unknown type. While the FSRQs show an approximately isotropic distribution\footnote{Although a relative deficit exists at intermediate northern Galactic latitudes, this is somewhat offset by blazars of unknown type.}, only 40\% of the total number of BL~Lacs are found in the southern Galactic hemisphere (152 at $b<-10\arcdeg$, 243 at $b>10\arcdeg$). At least approximately 100 other 2FGL sources at $b<-10\arcdeg$ are thus expected to be BL~Lac blazars. Some of them fall into the category blazars of unknown type, which are indeed found to be more numerous at $b<-10\arcdeg$ than at $b>10\arcdeg$ (97 versus 60), but a large fraction of these  BL~Lacs obviously remain unassociated 2FGL sources.

\begin{figure}
\centering
\resizebox{14cm}{!}{\rotatebox[]{0}{\includegraphics{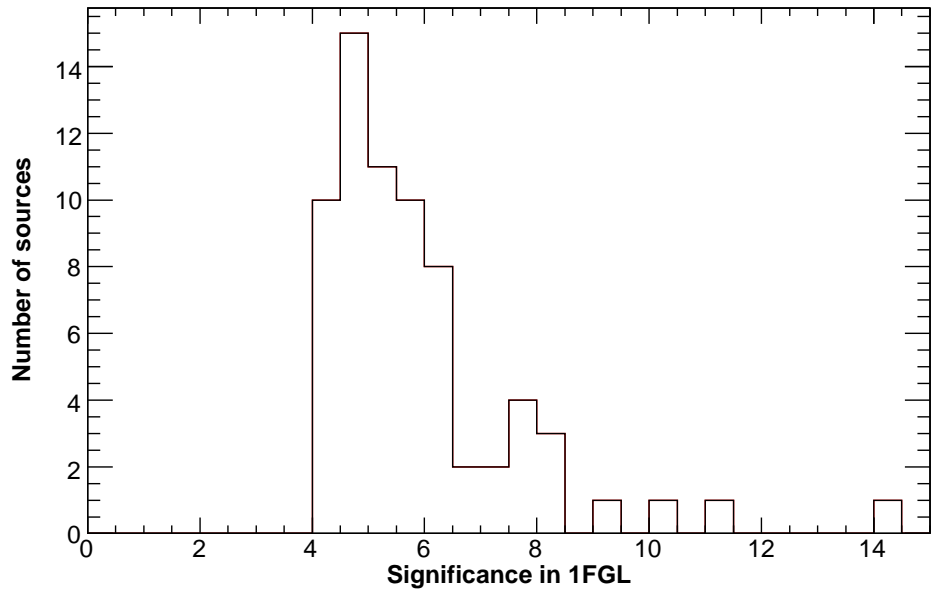}}}
\caption{Significance reported in the 1FGL for 1LAC sources missing in the 2LAC.The 1FGL detection threshold is 4.05, corresponding to $TS$=25.}
\label{fig:TS_gone}
\end{figure}

The comparison of the results inferred from the 1LAC and 2LAC enables the following observations:
\begin{itemize}
\item The 2LAC Clean Sample includes 287 more sources than the 1LAC Clean Sample, i.e., a  48\% increase. Of these, 234 were not present in 1FGL (58 FSRQs, 65  BL~Lacs, 108 blazars of  unknown type, 3 non-blazar objects); a total of 116 sources were present in 1FGL but not included in the 1LAC Clean Sample for various reasons (their associations were not firm enough, they had more than one counterpart or were flagged in the analysis).     
\item The fraction of FSRQs has dropped from 41\% to 35\% between the 1LAC and the 2LAC. The number of 2LAC Clean-Sample FSRQs has increased by 22\% relative to the  1LAC Clean Sample.
\item The fraction of BL~Lacs has remained about constant ($\sim$45\%  for both 1LAC and 2LAC). The number of 2LAC Clean-Sample BL~Lacs has increased by 42\% relative to the  1LAC Clean Sample.
\item The fraction of sources with unknown type has increased fairly dramatically between the two catalogs (from 8\% to 18\%), in part due to the improved association procedure. The number of these sources in the 2LAC Clean Sample has increased by more than a factor of 3  relative to that in the 1LAC Clean Sample.
\item The overall fraction of FSRQs and BL~Lacs without SED classification has increased from 25\% to 32\%: 155 sources in the Clean Sample are without optical magnitude while 227 are without X-ray flux. 
\item Out of 599 sources in the 1LAC Clean Sample, a total of 45 sources (listed in Table \ref{tab:1LAC}) are missing in the full 2LAC sample, most of them due to variability effects.  A few others are present in 2FGL but with shifted positions, ruling out the association with their former counterparts. The significances reported in the 1LAC for these 45 sources are relatively low (Figure \ref{fig:TS_gone}).  
\end{itemize}
These findings point to a need for more multiwavelength data, in particular in the optical and X-ray bands, enabling better classification  and characterization of the $\gamma$-ray loud blazars. 

\subsection{\label{sec:RG} Non-Blazar Objects and Misaligned AGNs}
Non-blazar $\gamma$-ray AGN are those not classified as FSRQs, BL Lacs, or as 
blazars of unknown/uncertain type, and constituted a small fraction of sources in 
the 1LAC ($\sim$4$\%$ in the Clean Sample). In the 2LAC, 
this fraction is similarly small ($\sim$3$\%$). 
Amongst these AGN are radio galaxies, which have emerged as a $\gamma$-ray source 
population due to the {\it Fermi}-LAT \citep[e.g.,][]{ngc1275,m87,magn}. The 2LAC 
contains in particular two new radio galaxies -- Centaurus~B and Fornax~A, 
associated with 2FGL~J1346.6$-$6027 and 2FGL~J0322.4$-$3717, respectively. 
The LAT detects extended emission from Centaurus A \citep{cena}, and this source is modeled with a extended spatial template in 2FGL. \citet{che07} and \citet{geo08} predicted that the radio lobes of Fornax A might be seen as extended sources in the LAT, though to date no extension has been detected. In this context we also note that the position of the 2FGL source associated with the large radio galaxy NGC 6251 ($\sim1\fdg2$ in angular extent), 2FGL J1629.4$+$8236, is shifted toward the western radio lobe with respect to the 1FGL source position (1FGL J1635.4$+$8228).

The source 2FGL~J0316.6+4119 is associated with the 
head-tail radio galaxy IC~310, whose spectrum extends up to TeV energies and was discovered with 
the LAT \citep{ner10} and with MAGIC \citep{ale10ic}. Missing from the 
2LAC/2FGL are three radio galaxies reported previously -- 1FGL~J0308.3+0403 and 
1FGL~J0419.0+3811, associated with 3C~78 (NGC~1218) and 3C~111, respectively \citep{1LAC}, and 
3C~120 \citep{magn}. In the cases of 3C~111 and 3C~120 this may be due to the $\gamma$-ray 
emission being variable \citep{kat11} and the analysis being complicated by their relatively low Galactic latitudes ($b=-8.8\arcdeg$ and $b=-27.4\arcdeg$ respectively). The 1FGL~J0308.3+0403/3C~78 source is confirmed but at a significance level lower than the $TS = 25$ threshold for inclusion in the 2FGL catalog \citep[see Table 7 of ][]{2FGL}.

Nearby AGN with dominant $\gamma$-ray emitting starburst components were detected in the first year of LAT observations: M~82 and NGC~253 \citep{starburst1} and NGC~1068 and NGC~4945 \citep{starburst2}. A study on star-forming galaxies observed with the LAT has been carried out \citep{sta11}.  The low-probability association of 1FGL J1307.0$-$4030 with the nearby Seyfert galaxy ESO 323$-$G77 is confirmed with 2FGL~J1306.9$-$4028, with a probability of 0.8, just above the threshold. The low-probability (65$\%$) association of 1FGL~J2038.1+6552 
with NGC~6951 in the 1LAC is not confirmed -- instead, the $\gamma$-ray source in 
this vicinity, 2FGL~J2036.6+6551, is now associated with the blazar CLASS J2036+6553. Finally, one new 
Seyfert association of note is NGC 6814 to 2FGL~J1942.5$-$1024 with a probability of 0.91 for its radio-$\gamma$-ray match. LAT studies of 
other nearby Seyfert galaxies have so far resulted only in upper limits \citep{seyfert}.  We conclude that such radio-quiet sources do not emit strongly in $\gamma$-rays.

No new radio-loud narrow-line Seyfert I galaxies beyond those four detected in the first year \citep{pmnj0948,rlnlsy1} were found, although such objects can be highly variable in
$\gamma$-rays and one such example (SBS 0846+513) has been recently detected while flaring \citep{don11}, though it does not make it into 2FGL/2LAC as it was too faint during the first 24 months of LAT operation.

\subsection{\label{sec:lowlat}Low-Latitude AGNs}

Diffuse radio emission, Galactic point sources, and heavy optical extinction make the low-latitude sky a difficult region for AGN studies, and catalogs of AGNs and AGN candidates often avoid it partially or entirely.  However, we are able to make associations with 104 low-latitude AGNs (while about 210 AGNs would be expected in this region from the high-latitude observations if the LAT sensitivity remained the same); these are presented in Table~\ref{tab:lowlat}.  Although the associations are considered valid, these sources have, in general, been studied much less uniformly and much less thoroughly than the high-latitude sources at virtually all wavelengths, so we do not include them as part of the Clean Sample in order to keep them from skewing any of our analyses of the overall $\gamma$-ray AGN population.

\subsection{\label{sec:indiv} Notes on Individual Sources}

As in the 1LAC, we provide additional notes on selected sources. Associations
discussed in the previous subsection (\S \ref{sec:RG}) on non-blazars and misaligned AGNs are not repeated.

\noindent {\bf 2FGL~J0319.8+4130:} This is the LAT source associated with the radio galaxy NGC~1275 discovered early in
the \textit{Fermi} mission \citep{ngc1275}. During the first two years of LAT operation, the MeV/GeV
emission is variable with significant spectral changes at $>$GeV energies \citep{Kat10,Bro11}.

\noindent {\bf 2FGL~J0339.2$-$1734:} As noted in the 1LAC, the optical spectrum of the associated AGN source PKS~0336$-$177 is not easily classified as BL Lac or FSRQ.

\noindent {\bf 2FGL~J0523.0$-$3628:} The radio source associated with this EGRET $\gamma$-ray source is PKS 0521$-$36, which has historically been classified as a BL~Lac object because of its optically variable continuum \citep{Dan79}. However, its spectrum obtained in our optical follow-up program did not enable a clear classification. It is thus flagged as a generic AGN.

\noindent {\bf 2FGL~J0627.1$-$3528:} This LAT source was associated with PKS~0625$-$35, classified as a radio galaxy, but with BL Lac
characteristics in the optical as discussed in \cite{magn}.

\noindent {\bf 2FGL~J0840.7+1310:} This LAT source was associated with 3C~207, classified as a SSRQ, and was analyzed in more detail in \cite{magn}.

\noindent {\bf 2FGL~J0847.0$-$2334:} This source is associated with CRATES J0847$-$2337 and has been classified as a ``galaxy'' in our optical follow-up program. 

\noindent {\bf 2FGL~J0903.6+4238:} This radio source, S4 0900+42 was selected by \citet{Fan01} in a search for candidate Compact Steep Spectrum radio sources. It was then rejected because -- interestingly -- deeper observations revealed an extended ($>$40 kpc) low frequency radio structure. In the lack of an optical spectrum, this source could then be considered as a candidate misaligned AGN.

\noindent {\bf 2FGL~J0904.9$-$5735:} The associated radio source, PKS 0903$-$57, was classified as a Seyfert-I galaxy at $z=0.695$ by \cite{Tho90}. Its spectrum obtained in our optical follow-up program did not enable a clear classification.

\noindent {\bf 2FGL~J0942.8$-$7558:} The LAT source was associated with the radio source, PKS~0943$-$76, and studied in \cite{magn}. The photometric redshift of the radio source is $z=0.26$ and it appears to have an
FR~II morphology \citep{Bur06}.

\noindent {\bf 2FGL~J1230.8+1224:} This LAT source is associated with the radio galaxy M87, discovered initially in the first year
LAT data \citep{m87}. No significant variability is observed with the LAT within the first two years of observations \citep[see ][]{M87a}.

\noindent {\bf 2FGL~J1256.5$-$1145:} The associated source is CRATES J1256$-$1146 ($z=0.058$) whose spectrum obtained in our optical follow-up program did not enable a clear classification.  

\noindent {\bf 2FGL~J1329.3$-$0528:} The associated AGN, 1RXS~132928.0$-$053132, is not a known radio emitter (e.g., in the NVSS survey).

\noindent {\bf 2FGL~J1641.0+1141:} The associated AGN, CRATES~J1640+1144, was noted in the 1LAC as simply a ``galaxy.''  Its spectrum obtained in our optical follow-up program did not enable a clear classification. 

\noindent {\bf 2FGL~J1647.5+4950:} The associated AGN is SBS~1646+499, already noted in the 1LAC as characterized as a nearby ($z=0.047$) late-type galaxy. It is a BZU type in BZCAT. Its spectrum obtained in our optical follow-up program did not enable a clear classification. 

\noindent {\bf 2FGL~J1829.7+4846:} This LAT source was associated with 3C~380, classified as a SSRQ,  and was analyzed in more detail in \cite{magn}.

\noindent {\bf 2FGL~J2250.8$-$2808:} The LAT detected a flare from this object in 2009 March \citep{Koe09}.
The associated flat spectrum radio source, PMN~J2250$-$2806, has a redshift  $z=0.525$.  Its spectrum obtained in our optical follow-up program did not enable a clear classification. 

\section{Properties of the 2LAC Sources}

\subsection{\label{sec:z}Redshift Distributions}

\begin{figure}
\centering
\resizebox{16cm}{!}{\rotatebox[]{0}{\includegraphics{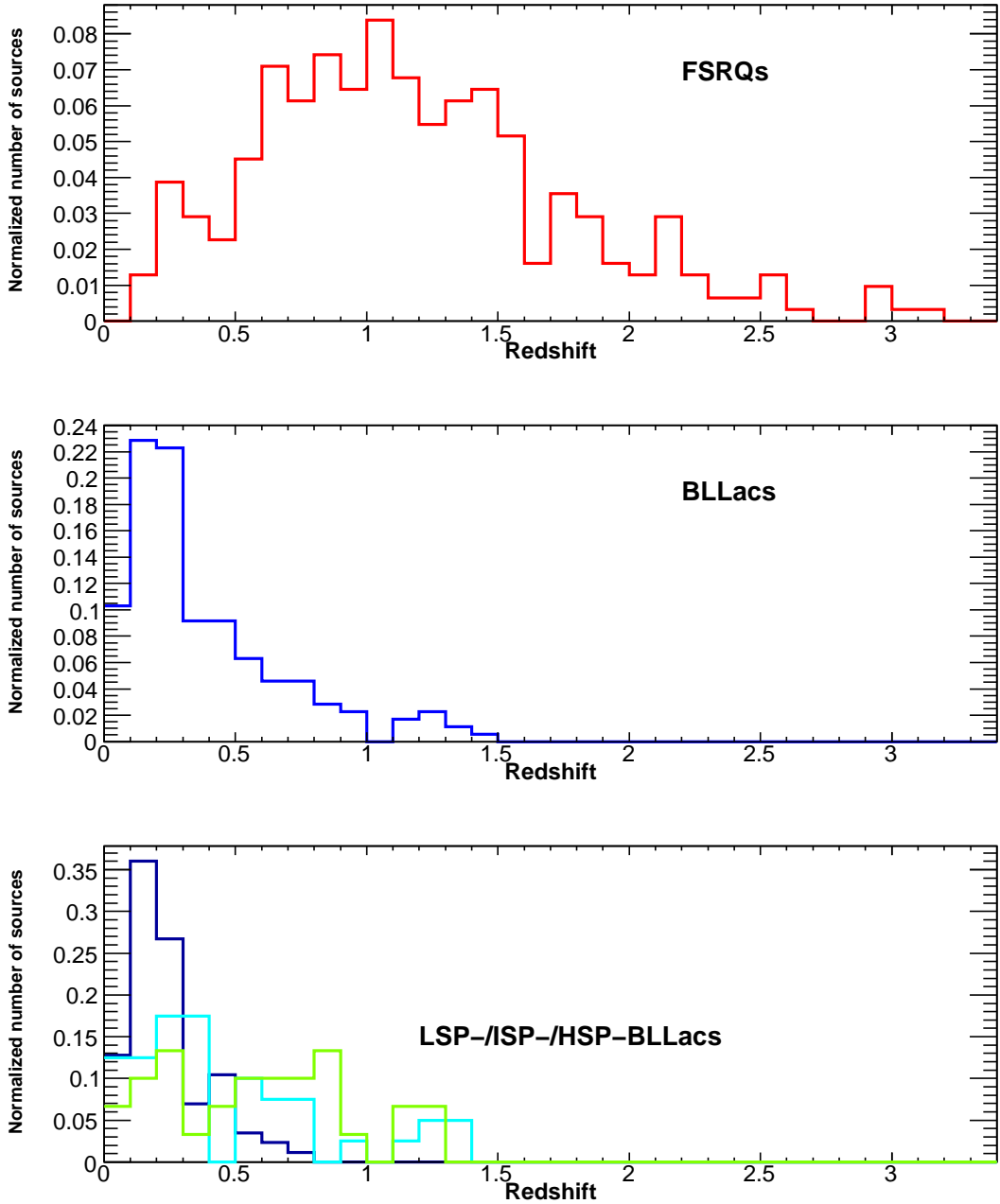}}}
\caption{Redshift distributions for FSRQs (top), BL~Lacs (middle), LSP-BL~Lacs (bottom, green), ISP-BL~Lacs (bottom, light blue), HSP-BL~Lacs (bottom, dark blue)}
\label{fig:redshift}
\end{figure}

The redshift distributions of the various classes are shown in Figure \ref{fig:redshift}.  They are very similar to those obtained with 1LAC. 
The distribution peaks around z=1 for FSRQs (Fig. \ref{fig:redshift} top) and extends to z=3.10. This distribution contrasts with that of  sources observed in the BAT catalog \citep{Aje09} where 40\%  of FSRQs have a redshift greater than 2. The distribution peaks at a lower redshift for BL~Lacs (Figure \ref{fig:redshift} middle). Note that 56\% of the BL~Lacs have no measured redshifts. The fraction of BL~Lacs having a measured redshift is higher for sources with a SED-based classification. This fraction is essentially constant for the different subclasses, (49\%, 49\%, 54\%) for (LSPs, ISPs, HSPs) respectively. Figure \ref{fig:redshift} bottom shows the redshift distributions for the different subclasses of BL~Lacs. These distributions gradually extend to lower redshifts as the location of the synchroton peak shifts to higher frequency, i.e., from LSPs to HSPs.

\begin{figure}
\centering
\resizebox{16cm}{!}{\rotatebox[]{0}{\includegraphics{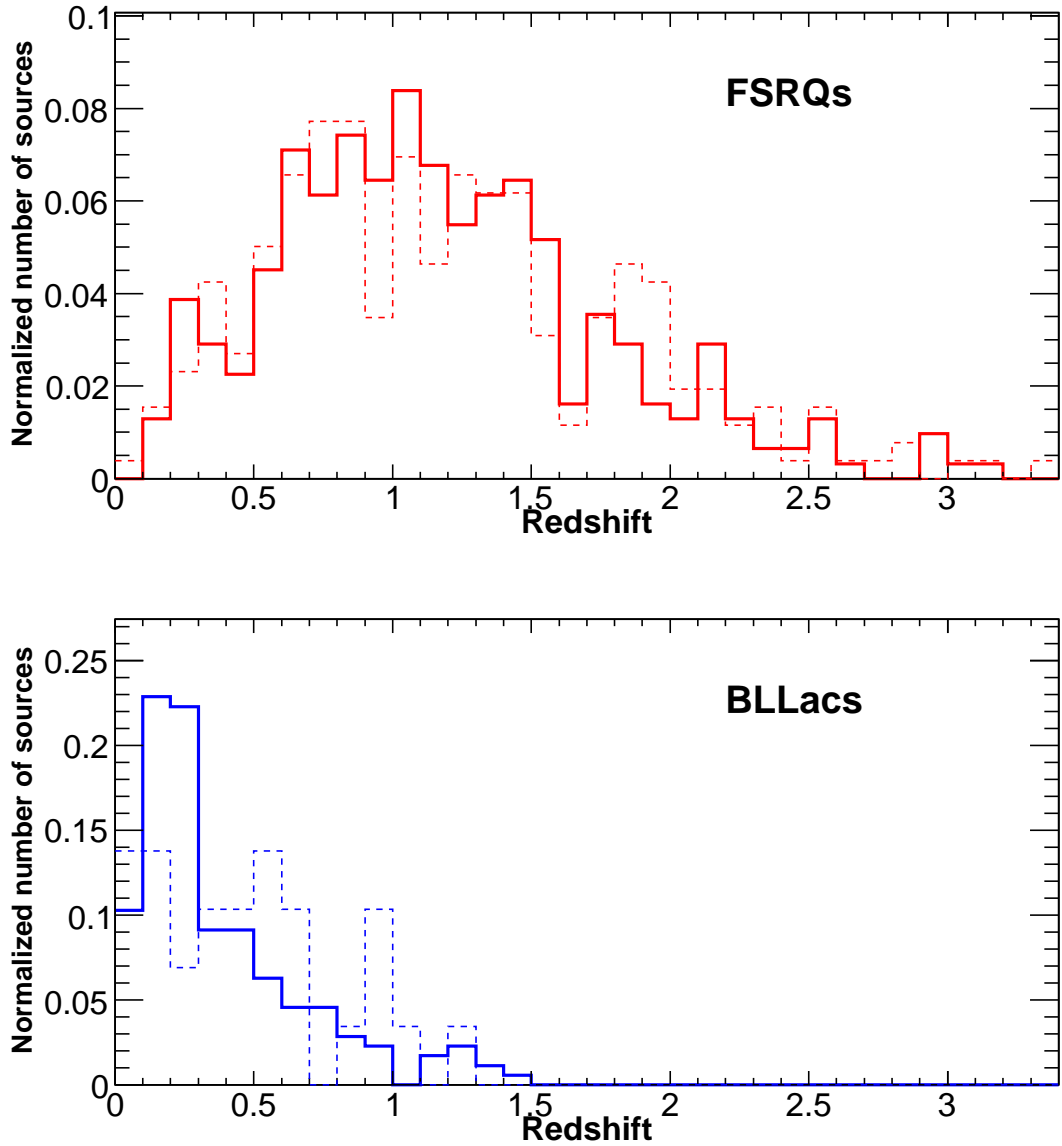}}}
\caption{Comparison between redshift distributions for blazars in the 2LAC Clean Sample (solid) and the 5-Year WMAP complete sample (dashed). Top: FSRQs. Bottom: BL~Lacs.}
\label{fig:redshift_w}
\end{figure}

The redshift distributions of FSRQs and BL~Lacs are compared in Figure \ref{fig:redshift_w} to the corresponding distributions for the sources obtained by cross correlating the seven-year WMAP catalog \citep{Gol11} with BZCat, using a correlation radius of 11$\arcmin$ (thus selecting 339 sources of a total of 471). Good agreement is observed for FSRQs. The agreement between the 2LAC and WMAP distributions of BL~Lacs is more marginal, but the low number of BL~Lacs with measured redshifts in the WMAP sample (29 sources) prevents us from drawing definite conclusions. Note that all BL~Lacs in the WMAP catalog are detected by the LAT, while only 50\% (130 of 260) of the WMAP FSRQs fulfill this condition.

\subsection{\label{sec:flux}Flux and Photon Spectral Index Distributions}

\begin{figure}
\centering
\resizebox{15cm}{!}{\rotatebox[]{0}{\includegraphics{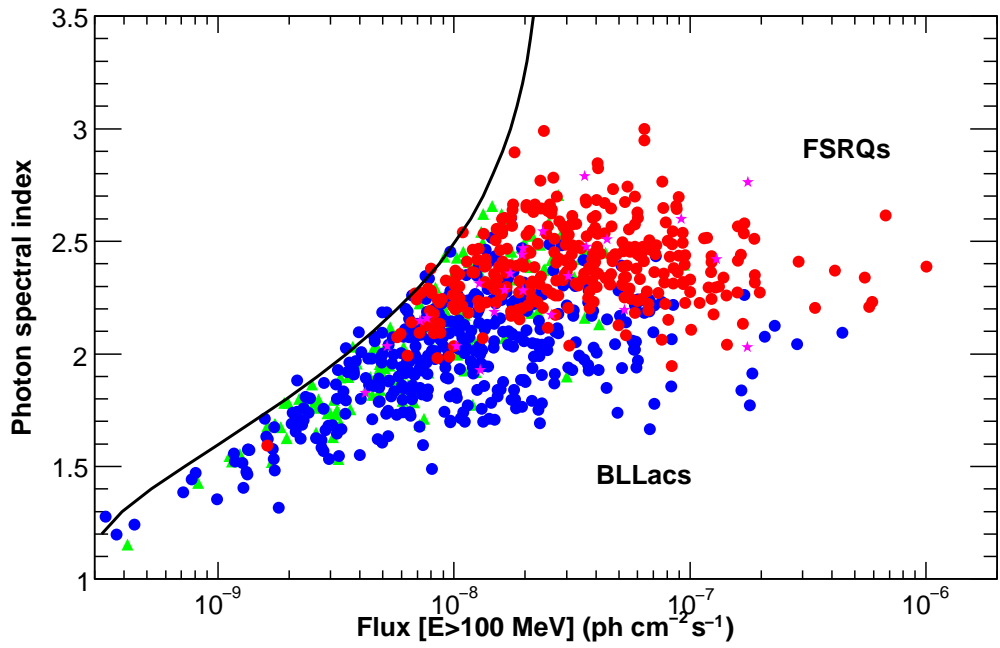}}}
\caption{Photon index versus flux above 100 MeV for blazars in the Clean Sample. Red:\ FSRQs, blue:\ BL~Lacs, magenta:\ non-blazar AGNs, green:\ AGNs of unknown type.}
\label{fig:index_flux}
\end{figure}

\begin{figure}
\centering
\resizebox{15cm}{!}{\rotatebox[]{0}{\includegraphics{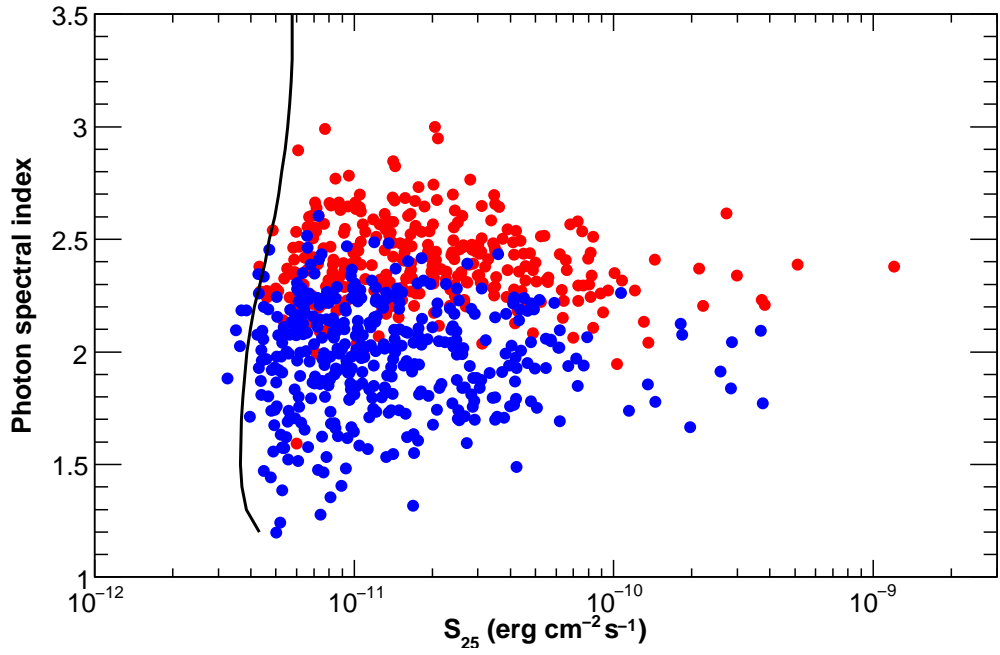}}}
\caption{Photon index versus energy flux above 100 MeV. Red:\ FSRQs, blue:\ BL~Lacs. The curve represents the approximate detection limit.}
\label{fig:index_S}
\end{figure}

The photon index is plotted versus the mean flux (E$>$100 MeV) in Figure \ref{fig:index_flux}, along with an estimate of the flux limit.  The flux limit strongly depends on the photon index as harder sources are easier to discriminate against the background,which is due to the narrowing of the point-spread function (PSF) of the LAT with increasing energy and to the relative softness of the diffuse Galactic $\gamma$-ray emission. In contrast, the limit in energy flux above 100 MeV is almost independent of the photon index as illustrated in  Figure \ref{fig:index_S}. 

\begin{figure}
\centering
\resizebox{14cm}{!}{\rotatebox[]{0}{\includegraphics{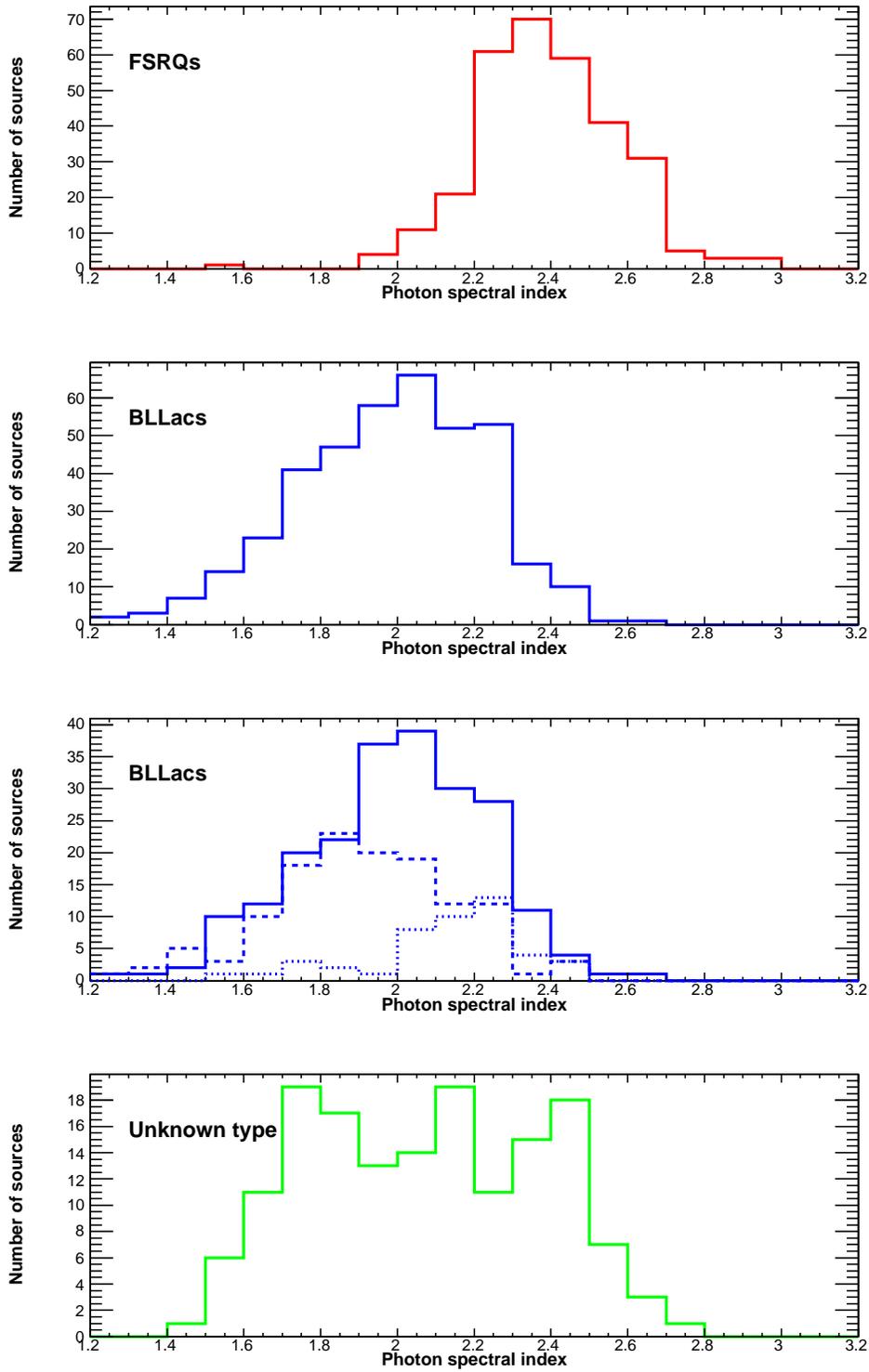}}}
\caption{Photon index distributions. Top: FSRQs. Second from top: BL~Lacs. Second from bottom: BL~Lacs without redshift (solid), BL~Lacs with z$<$0.5 (dashed), BL~Lacs with z$>$0.5 (dotted). Bottom:  blazars of unknown type.}
\label{fig:index}
\end{figure}

The photon index distributions are given in Figure \ref{fig:index} for the different classes of blazars. The now well-established spectral difference in the LAT energy range between FSRQs and BL~Lacs, with a moderate overlap between the distributions \citep{LBAS,1LAC} is still present.  The index distribution of sources with unknown types spans a wider range than those of FSRQs and BL~Lacs separately. Assuming that the class of sources with unknown types is entirely made up of FSRQs and BL~Lacs lacking classification, each with  the same photon index distributions as the classified sources,
FSRQs and BL~Lacs would contribute about equally to this component.   

\begin{figure}
\centering
\resizebox{14cm}{!}{\rotatebox[]{0}{\includegraphics{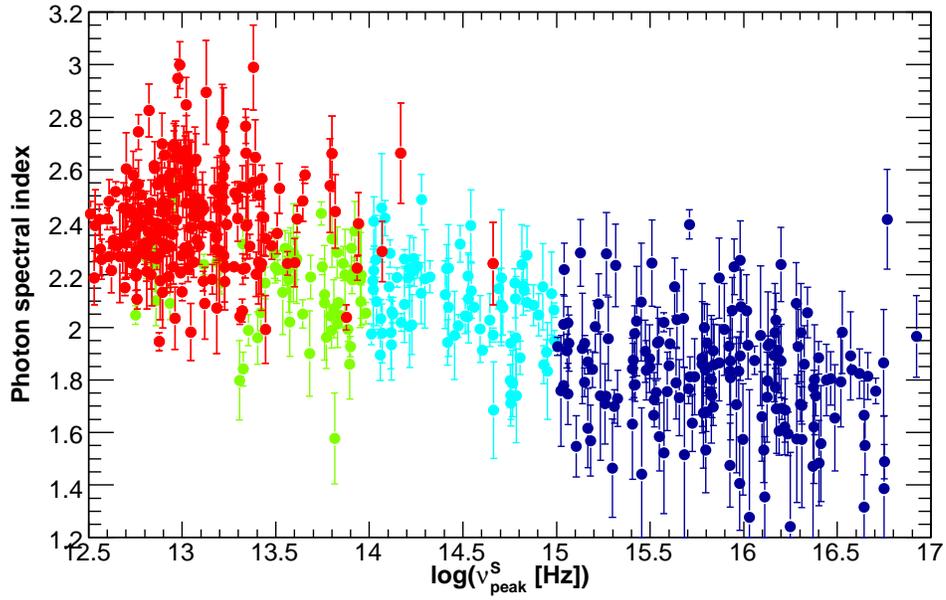}}}
\caption{Photon index versus frequency of the synchrotron peak $\nu^S_{peak}$. Red: FSRQs, green: LSP-BL Lacs, light blue: ISP-BL Lacs,  dark blue: HSP-BL Lacs.}
\label{fig:index_nu_syn}
\end{figure}

\begin{figure}
\centering
\resizebox{14cm}{!}{\rotatebox[]{0}{\includegraphics{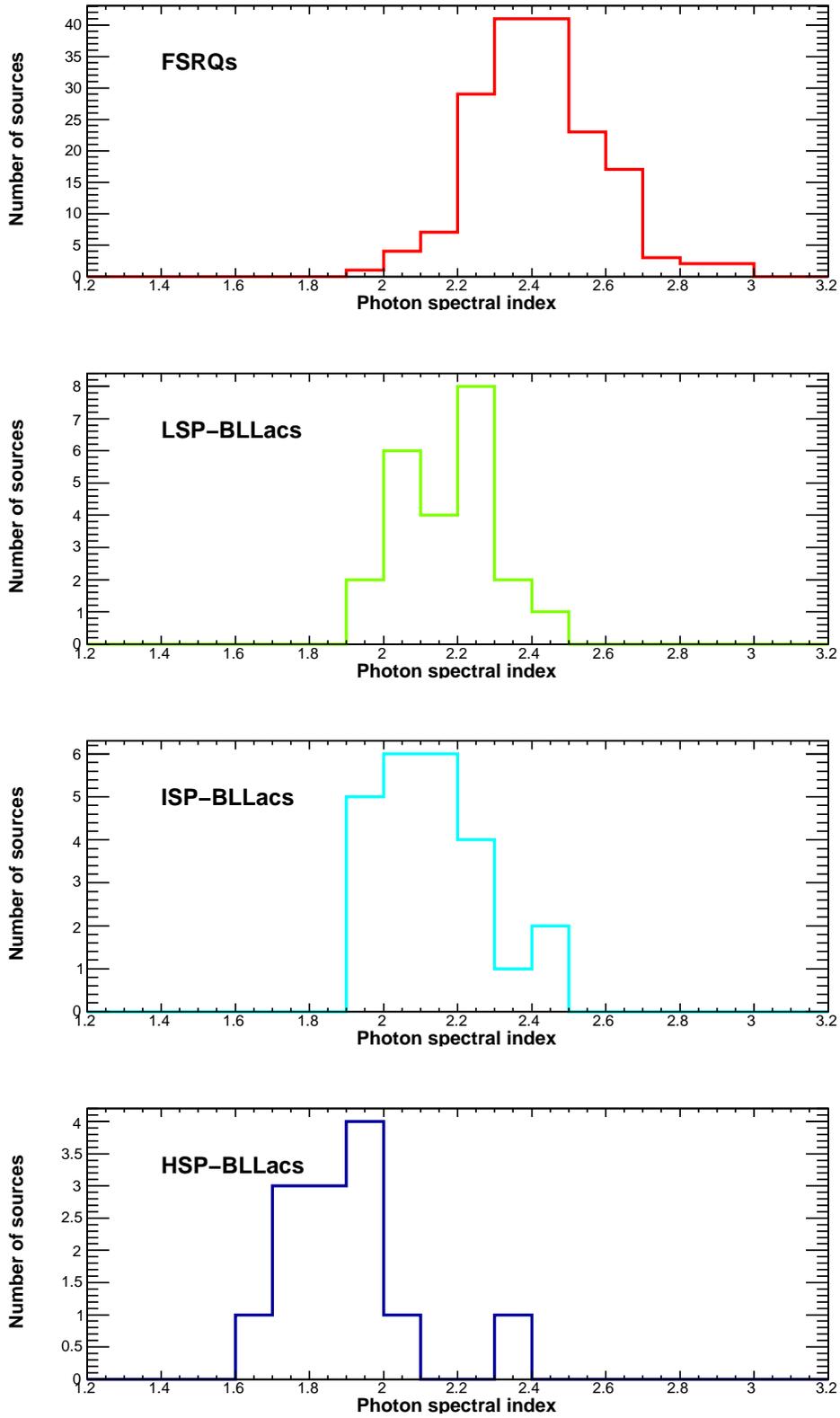}}}
\caption{Photon spectral index distributions for the different blazar classes for sources in the Clean Sample with $F$[${E>100}$~MeV]~$>1.5\times10^{-8}$ \pflux{}.}
\label{fig:index_c}
\end{figure}

\begin{figure}
\centering
\resizebox{14cm}{!}{\rotatebox[]{0}{\includegraphics{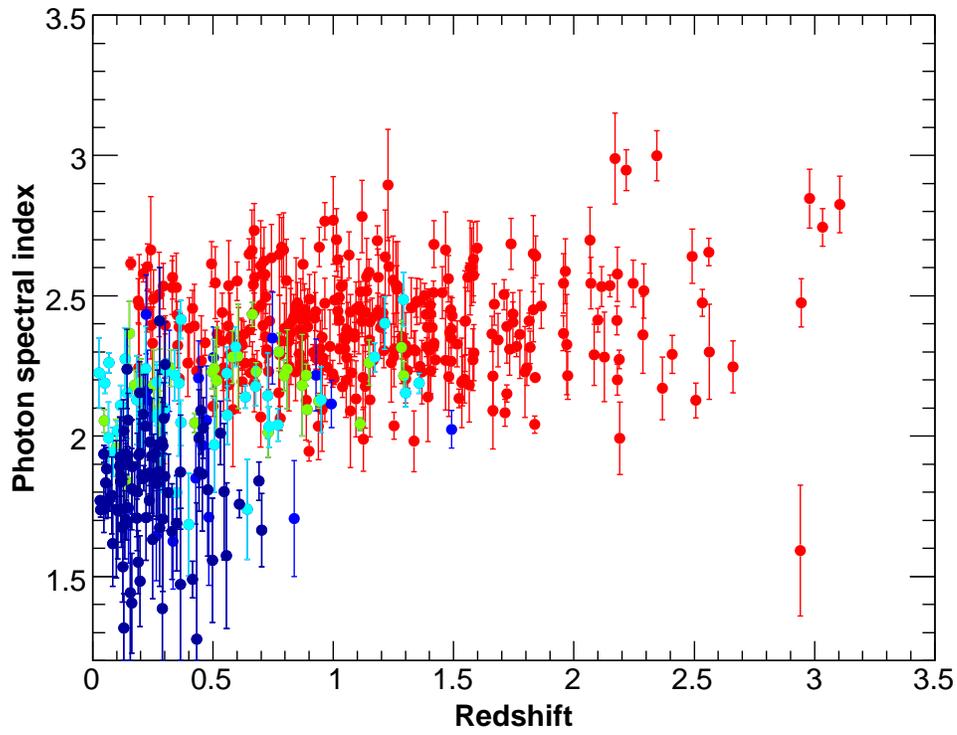}}}
\caption{Photon spectral index versus redshift. Red: FSRQs, green: LSP-BL Lacs, light blue: ISP-BL Lacs,  dark blue: HSP-BL Lacs.}
\label{fig:index_z}
\end{figure}

The photon index  is plotted versus the frequency of the synchrotron peak in Figure \ref{fig:index_nu_syn}. A relatively strong correlation between these two parameters, again reported earlier \citep{LBAS,1LAC} is observed. Strong conclusions regarding the  HSP-BL~Lac outliers (e.g., 2FGL~J1213.2$-$2616/ RBS~1080 and 2FGL~J1023.6+2959/RX~J1023.6+3001 with $\Gamma$=2.4 and $\Gamma$=1.2 respectively) should not be made as these sources are very faint and are significantly detected at best in only one energy band. In order to make a meaningful comparison between the photon index distributions for different classes, it is advantageous to use the flux-limited sample, i.e., sources with Flux[E$>$100 MeV]$>$1.5$\times$10$^{-8}$ \pflux{}, which is free of the bias arising from  the photon-index dependence of the flux limit (Figure \ref{fig:index_flux}). The resulting photon index distributions are shown in  Figure \ref{fig:index_c}. The distribution mean values and rms are (2.42$\pm$0.17, 2.17$\pm$0.12, 2.13$\pm$0.14, 1.90$\pm$0.17) for (FSRQs, LSP-BL~Lacs, ISP-BL~Lacs, HSP-BL~Lacs) respectively. For orientation, the mean values in the significance-limited sample are  (2.39, 2.14, 2.09, 1.81) for (FSRQs, LSP-BL~Lacs, ISP-BL~Lacs, HSP-BL~Lacs).
No significant dependence of the photon index on redshift is observed {\sl if blazar subclasses are considered separately}, as illustrated in Figure \ref{fig:index_z}, corroborating the conclusion drawn with 1LAC. Note that the region populated by LSP-BL~Lacs in the (redshift, $\Gamma$) plane overlaps but does not strictly coincide with that populated by FSRQs.   
The FSRQ with z=2.941 and $\Gamma= 1.59\pm0.23$ is  2FGL\,J0521.9+0108/CRATES J0522+0113, which, while having a definite classification, exhibits a complex optical spectrum. This source is located in the Orion region, where uncertainties in our knowledge of the Galactic diffuse emission  can affect the determination of the source photon spectral index.  The three photon index distributions for BL~Lacs with z$<$0.5 (mostly HSPs), with z$>$0.5 (mostly LSPs), and for  BL~Lacs without redshifts are compared in Figure \ref{fig:index}. The distribution of BL~Lacs without redshifts is markedly different from the two other distributions and thus does not favor any conclusions concerning the actual redshift distributions of these blazars.

\begin{figure}
\centering
\resizebox{16cm}{!}{\rotatebox[]{0}{\includegraphics{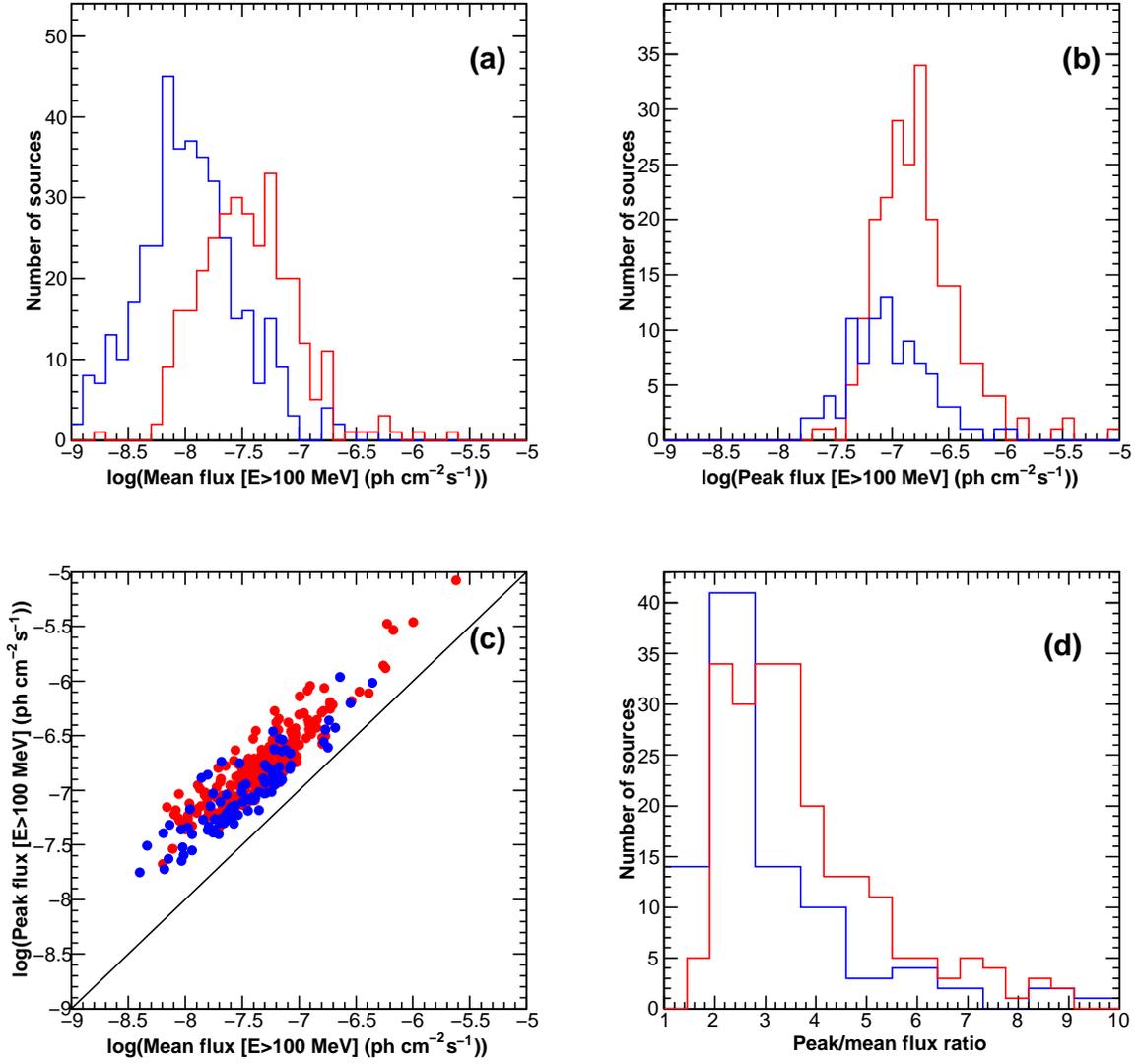}}}
\caption{(a) Mean flux distributions. Red:\ FSRQs, blue:\ BL~Lacs.(b) Peak flux distributions. (c) Peak flux vs mean flux. (d) Peak flux over mean flux ratio. } 
\label{fig:flux_mean_peak} 
\end{figure}

The time-averaged, mean flux distributions for FSRQs and BL~Lacs are compared in  Figure \ref{fig:flux_mean_peak}a. As suggested by Figure \ref{fig:index_flux}, the fluxes of the FSRQs extend to higher values than do BL~Lacs, but FSRQs have a higher detection flux limit due to their spectral softness. For sources showing significant variability, the monthly peak-flux distributions are compared in  Figure \ref{fig:flux_mean_peak}b. These distributions are more similar for the two blazar classes. The peak flux is plotted as a function of  mean flux in  Figure \ref{fig:flux_mean_peak}c, and the  distribution of peak flux over mean flux ratio is given in Figure \ref{fig:flux_mean_peak}d. Larger flux ratios are observed for FSRQs. Variability is discussed further in \S \ref{sec:var}.     

\subsection{Comparison of 2LAC and 1LAC fluxes}
\begin{figure}
\centering
\resizebox{13cm}{!}{\rotatebox[]{0}{\includegraphics{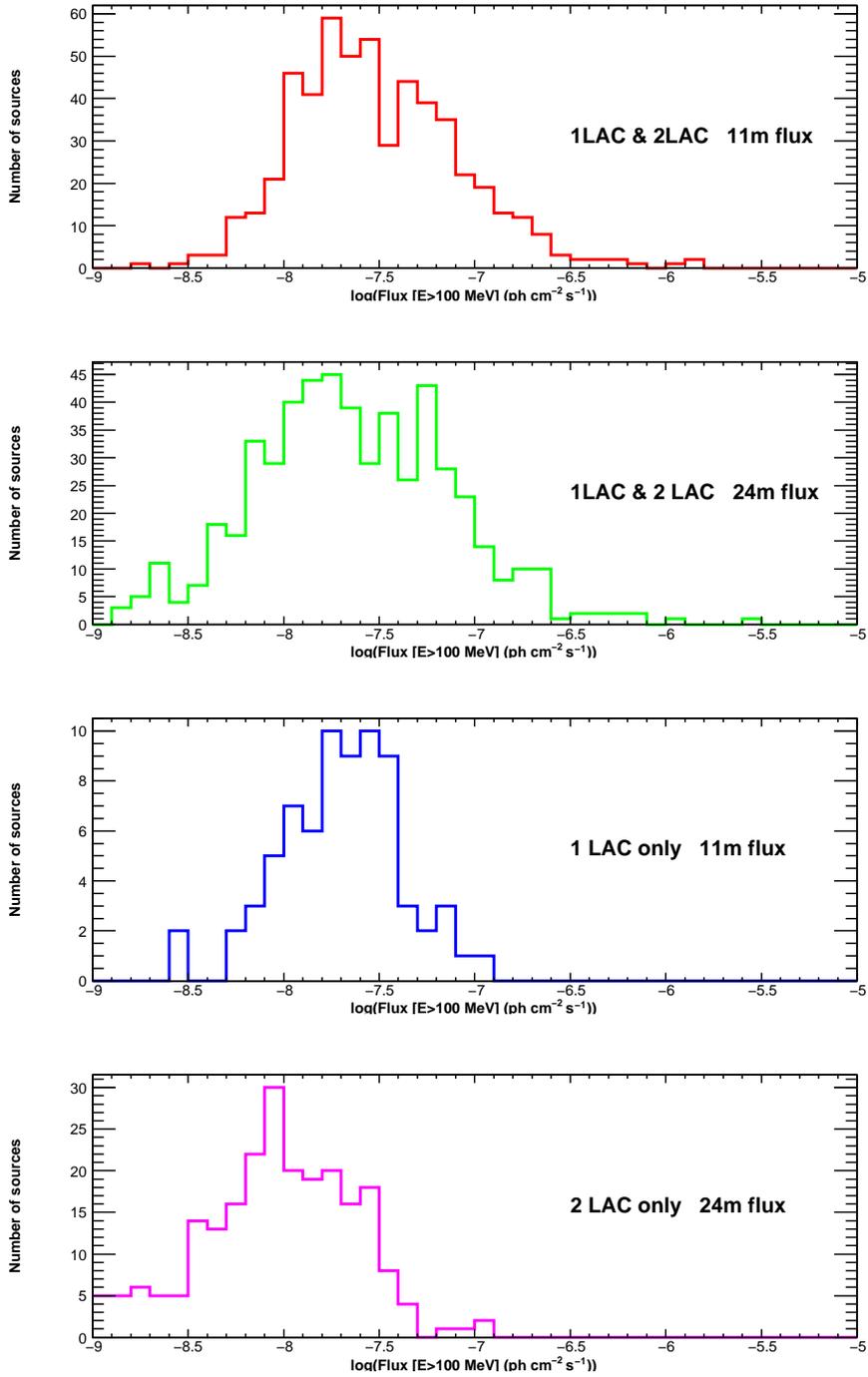}}}
\caption{Distributions of flux above 100 MeV. The top and second panels show the 1LAC fluxes and 2LAC fluxes for sources in both 1LAC and 2LAC respectively. The third panel shows 1LAC fluxes of 1LAC sources missing in the 2LAC. The bottom panel displays the 2LAC fluxes for new sources in the 2LAC absent in the 1LAC.} 
\label{fig:flux_11_24}
\end{figure}

 Photon flux distributions from 1LAC and 2LAC are displayed in  Figure \ref{fig:flux_11_24}. The top two panels show the  1LAC fluxes and  2LAC fluxes for sources present in both 1LAC and 2LAC. As expected the 2LAC distribution is broader than the 1LAC distribution, especially at the low-flux end.  The bottom two panels represent the 1LAC flux distribution for the 45 missing 1LAC sources and the 2LAC flux distribution for the 250 newly-detected 2LAC sources in the Clean Sample. The high-flux end of these distributions look alike, which can presumably arise from the facts that a similar pool of sources i) were comparatively bright during the first 11 months and then faded away, or   ii) have brightened during the last 13 months spanned by the 2LAC while being faint during the 1LAC period. Of course, the low-flux ends of the two distributions are different as the new 2LAC sources include sources fainter than the 1LAC detection limit.

\subsection{\label{sec:curv} Energy Spectra}

First observed for 3C~454.3 \citep{Abdo_3C} early in the {\it Fermi} mission, a significant curvature in the energy spectra of many bright FSRQs and some bright LSP-/ISP-BL~Lacs is now a well-established feature \citep{spec_an,1LAC}. The break energy obtained from a broken power-law fit has been found to be remarkably constant as a function of the flux, at least for 3C~454.3  \citep{Abdo_3C_11}. Several explanations  have been proposed to account for this feature,  including $\gamma\gamma$ attenuation from He\,{\sc ii} line photons \citep{Pou10},
intrinsic electron spectral breaks \citep{Abdo_3C}, Ly $\alpha$ scattering
\citep{Abdo_3C_10}, and hybrid scattering \citep{Fin10}. 

Although broken power-law  (BPL) functions have been found to better reproduce most curved blazar energy spectra, the LogParabola function (\S \ref{sec:obs}) has been selected here since it has only one more degree of freedom with respect to a power law, convergence of spectral fits is easier than for BPL and the function decreases more smoothly at high energy than a power law with exponential cutoff form. Physical arguments supporting the use of a LogParabola function have been presented in \cite{Tra11}.

\begin{figure}
\centering
\resizebox{16cm}{!}{\rotatebox[]{0}{\includegraphics{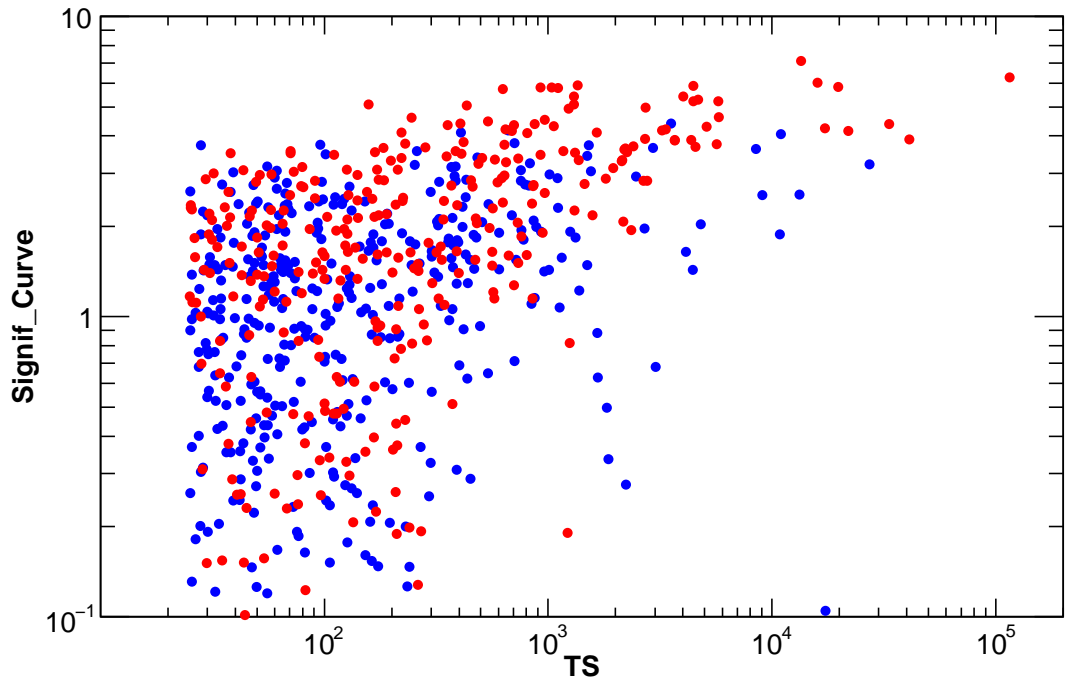}}}
\caption{$Signif\_Curve$, defined as the square-root of $TS_{curve}$ times a correction factor accounting for systematic effects,  versus $TS$. Red:\ FSRQs, blue:\ BL~Lacs.}
\label{fig:TSCurve_TS}
\end{figure}

\begin{figure}
\centering
\resizebox{16cm}{!}{\rotatebox[]{0}{\includegraphics{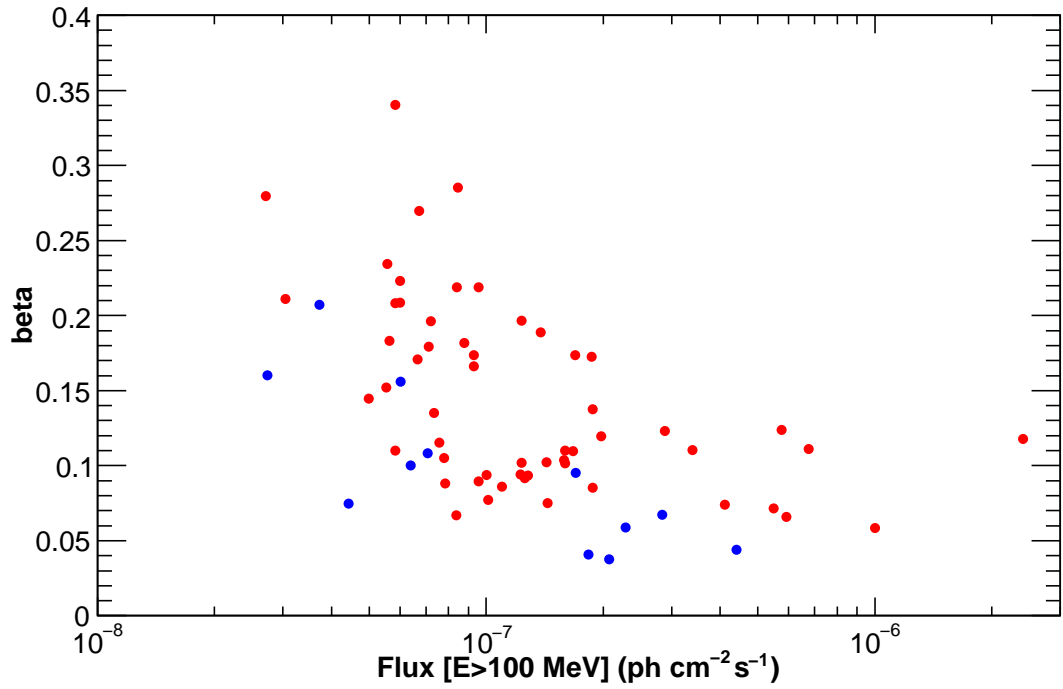}}}
\caption{LogParabola parameter $\beta$ plotted as a function of flux above 100 MeV. Red:\ FSRQs, blue:\ BL~Lacs.}
\label{fig:beta_flux}
\end{figure}

The spectral curvature is characterized by the parameter $Signif\_Curve$, equal to $\sqrt{c \times TS_{curve}}$, where $TS_{curve}$ is defined in \S \ref{sec:obs} and $c$ is a source-dependent correction factor accounting for systematic effects \citep[see ][for details]{2FGL}. $Signif\_Curve$ is plotted as a function of $TS$ in Figure \ref{fig:TSCurve_TS}. For $TS>$ 1000, most FSRQs have large  $Signif\_Curve$, while BL~Lacs exhibit a variety of behaviors. As mentioned earlier, LogParabola results were retained for sources with $TS_{curve}$$>$16 (corresponding to  $Signif\_Curve \simeq$  4).    
The LogParabola parameter $\beta$ is plotted as a function of the flux in Figure \ref{fig:beta_flux}
for the 57 FSRQs and 12 BL~Lacs in the Clean Sample with $TS_{curve}>$16. The average $\beta$ is significantly lower for BL~Lacs than for FSRQs (0.11$\pm$0.02 versus 0.18$\pm$0.02 respectively), possibly due to the fact that different regions of the Inverse-Compton peak (assuming a leptonic scenario) are probed in the LAT energy band. 

\begin{figure}
\centering
\resizebox{16cm}{!}{\rotatebox[]{0}{\includegraphics{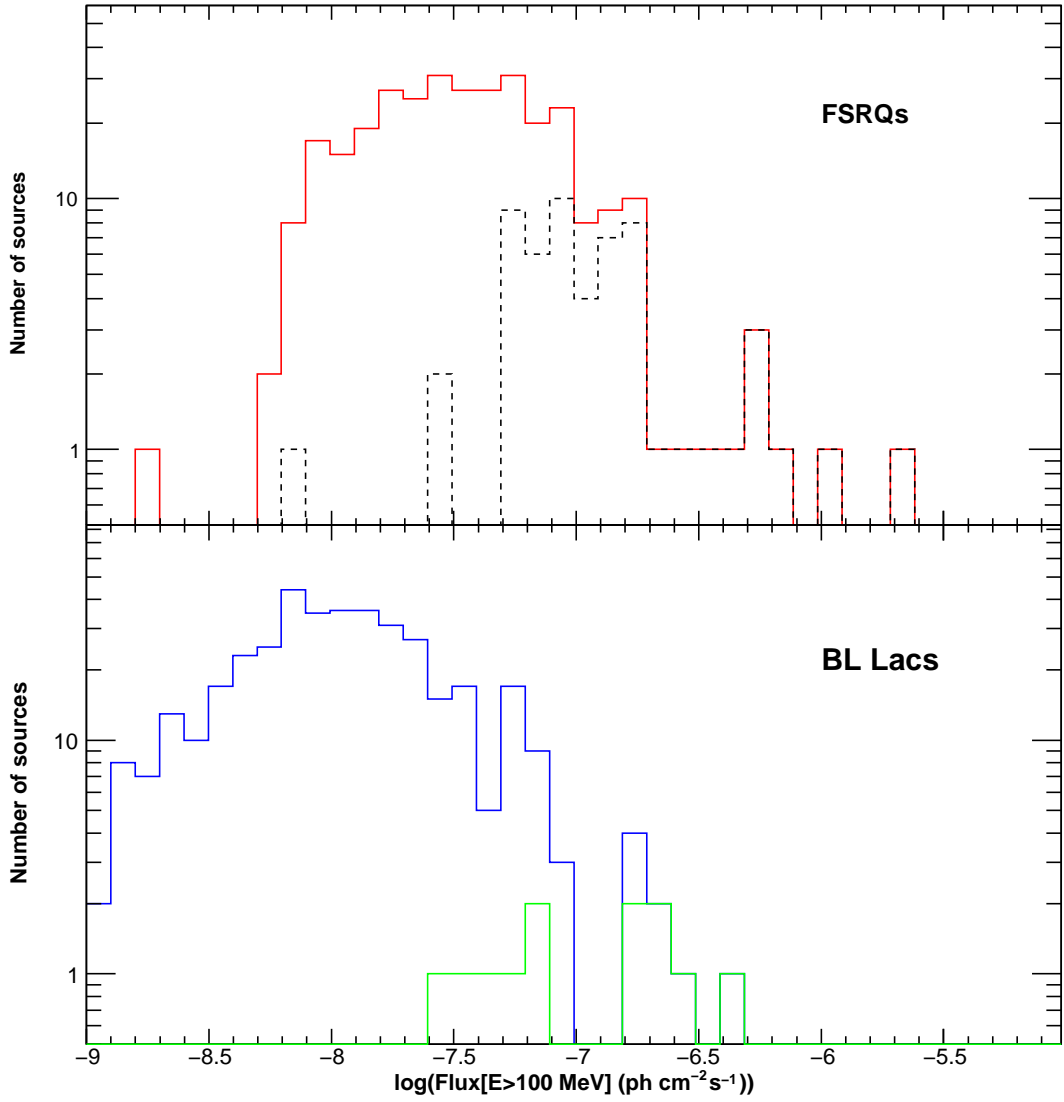}}}
\caption{Top: Flux distribution of sources exhibiting significant spectral curvature (black, dashed) compared to the full distribution (red) for FSRQs. Bottom: Flux distribution of sources exhibiting significant spectral curvature (green) compared to the full distribution (blue) for BL~Lacs. }
\label{fig:Flux_curv}
\end{figure}

The 12 BL~Lacs comprise 7 LSPs, 3 ISPs, 1 HSP  and 1 BL~Lac lacking SED classification. The HSP is BZB~J1015+4926 (GB~1011+496), the SED of which has a maximum at a few GeV. The flux distributions for these sources are compared to the overall distributions in Figure \ref{fig:Flux_curv}, and are seen to confirm the trend observed in Figure \ref{fig:TSCurve_TS}. 

\subsection{\label{sec:var}Variability}

Variability at all time scales is one of the distinctive properties of blazars. Since launch, detections by the {\it Fermi}-LAT of $\gamma$-ray activity from 81 flaring blazars have been reported in Astronomer's Telegrams (ATels). Four of them are not listed in the 2LAC since they did not pass the $TS=25$ cut for inclusion in the 2FGL: SBS~0846+513, PMN~J1123$-$6417 (at b=3.0$\arcdeg$), PMN~J1913$-$3630, PKS~1915$-$458. 

Two-year light curves with monthly binning were obtained as part
of the 2FGL catalog. The large bin width leads to a substantial smoothing of the light curves for the brightest blazars, for which peak fluxes may be much higher than the one-month average fluxes reported here.  A more extensive analysis using higher-resolution light curves, thus containing richer temporal information will be presented elsewhere. Nevertheless these light curves constitute the largest set ever produced in the $\gamma$-ray band, allowing variability analysis on a wide sample of blazars. 
In this section we will give an overview of the variability properties for the sources in the 2LAC Clean Sample. This includes the detection of variability via the
LAT $\gamma$-ray variability index, a measure of the $\gamma$-ray variability duty cycle and a derivation of population variability characteristics from the Discrete Auto Correlation Function, DACF, first order Structure Function, SF, and from Power Density Spectra, PDS. DACF \citep[see, e.g.,][]{ede88,huf92}, SF \citep[see, e.g.,][]{sim85,smi93,lai93,pal97}, and PDS \citep{Vaugh03} are methods providing insights into fluctuation modes, characteristic timescales and flavors of the variability modes in the $\gamma$-ray monthly-bin light curves. A short description of these three analysis methods are given in \citet{Abdo_var}.
 
\begin{figure}
\centering
\resizebox{16cm}{!}{\rotatebox[]{0}{\includegraphics{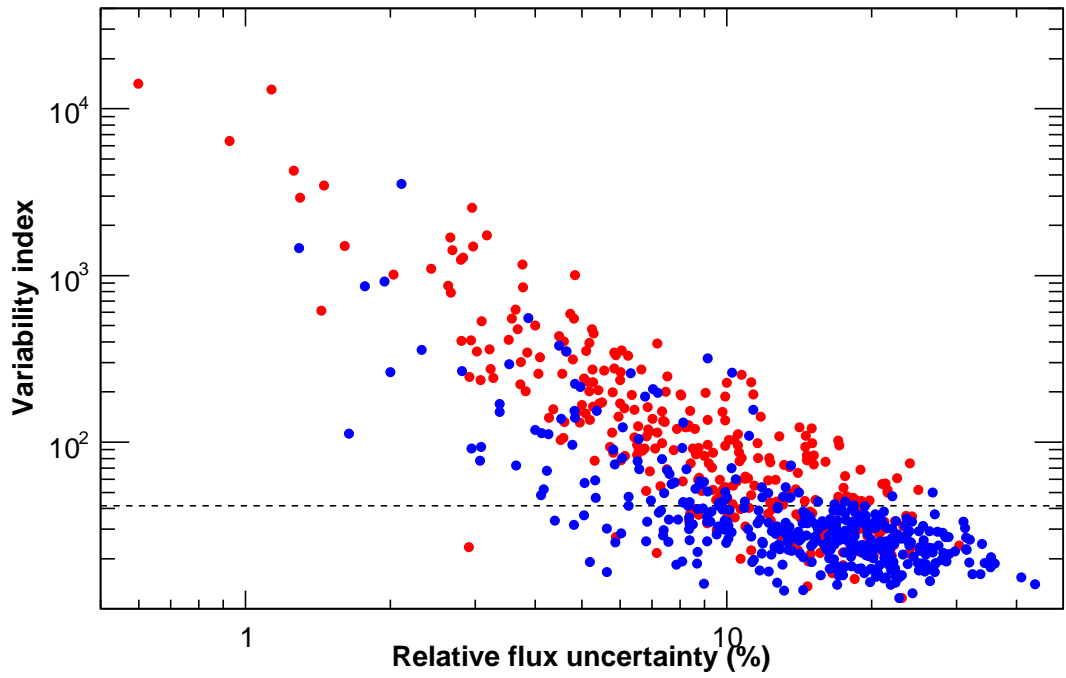}}}
\caption{Variability index versus relative flux uncertainty. Red:\ FSRQs, blue:\ BL~Lacs. The dashed line corresponds to the 99\% confidence level for a source to be variable. }
\label{fig:varind_relunc}
\end{figure}

The variability index $TS_{var}$, which is described in section \ref{sec:obs}, is plotted as a function of the relative flux uncertainty in Figure \ref{fig:varind_relunc}. The relative flux uncertainty, computed with a fixed photon index \citep[see section 3.6 of ][]{2FGL},  reflects the photon statistics. This parameter allows meaningful comparisons between sources with different fluxes and photon indices. Figure \ref{fig:varind_relunc} illustrates the fact that for a source to be labeled as variable on the basis of its variability index it must be both intrinsically variable and sufficiently bright. All very bright sources, including both FSRQs and BL~Lacs are found to be variable at a confidence level greater than 99\%, depicted by the line at $TS_{var}>$41.6 in Figure \ref{fig:varind_relunc}. At a given relative flux uncertainty, BL Lacs have on average lower $TS_{var}$ than FSRQs. 

\begin{figure}
\centering
\resizebox{16cm}{!}{\rotatebox[]{0}{\includegraphics{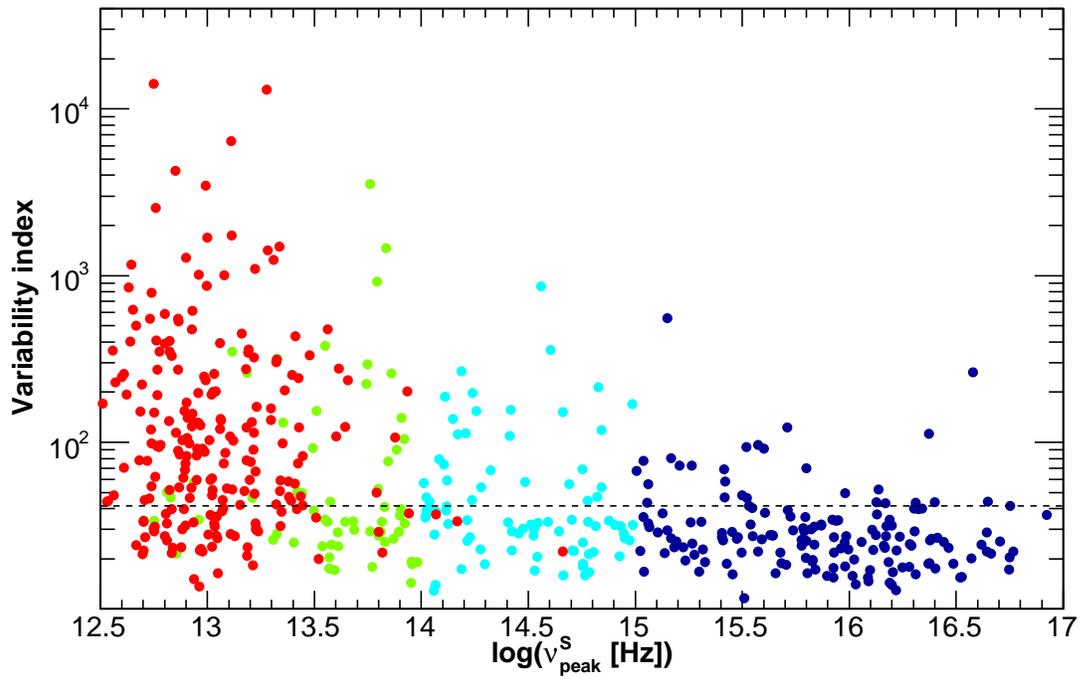}}}
\caption{Variability index versus synchrotron peak frequency. Red: FSRQs, green: LSP-BL Lacs, light blue: ISP-BL Lacs,  dark blue: HSP-BL Lacs. The dashed line corresponds to the 99\% confidence level for a source to be variable.}
\label{fig:varind_sync}
\end{figure}

\begin{figure}
\centering
\resizebox{16cm}{!}{\rotatebox[]{0}{\includegraphics{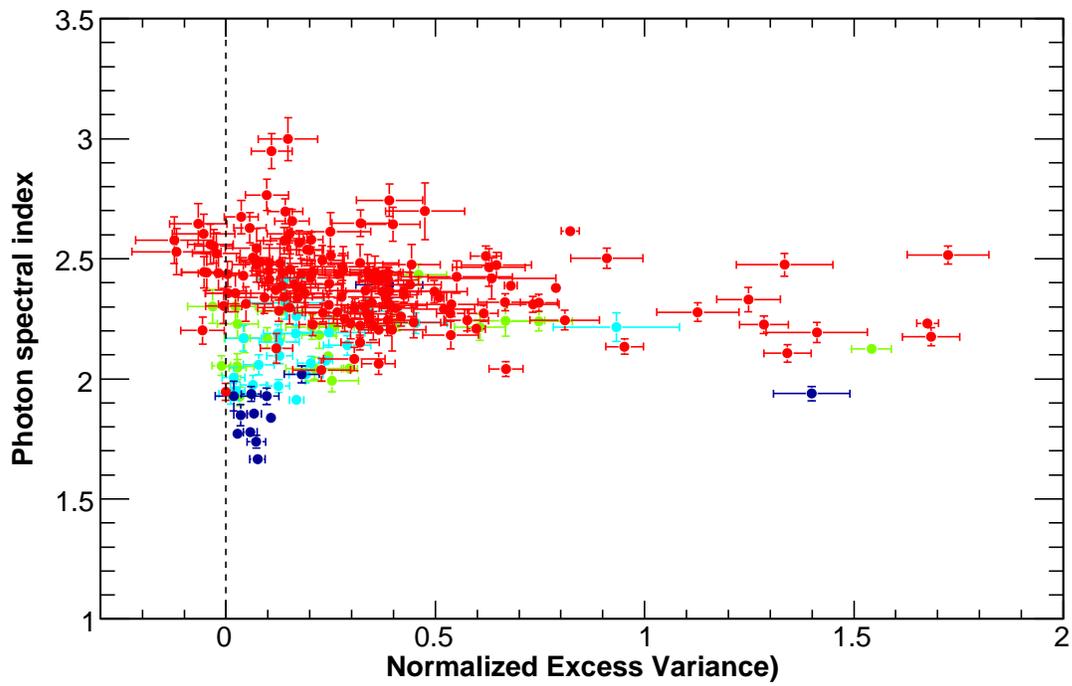}}}
\caption{Photon spectral index  versus normalized excess variance for sources with flux greater than $3 \times 10^{-8}$ \pflux{}.   Red: FSRQs, green: LSP-BL Lacs, light blue: ISP-BL Lacs,  dark blue: HSP-BL Lacs.}
\label{fig:index_var}
\end{figure}

\begin{figure}
\centering
\resizebox{16cm}{!}{\rotatebox[]{0}{\includegraphics{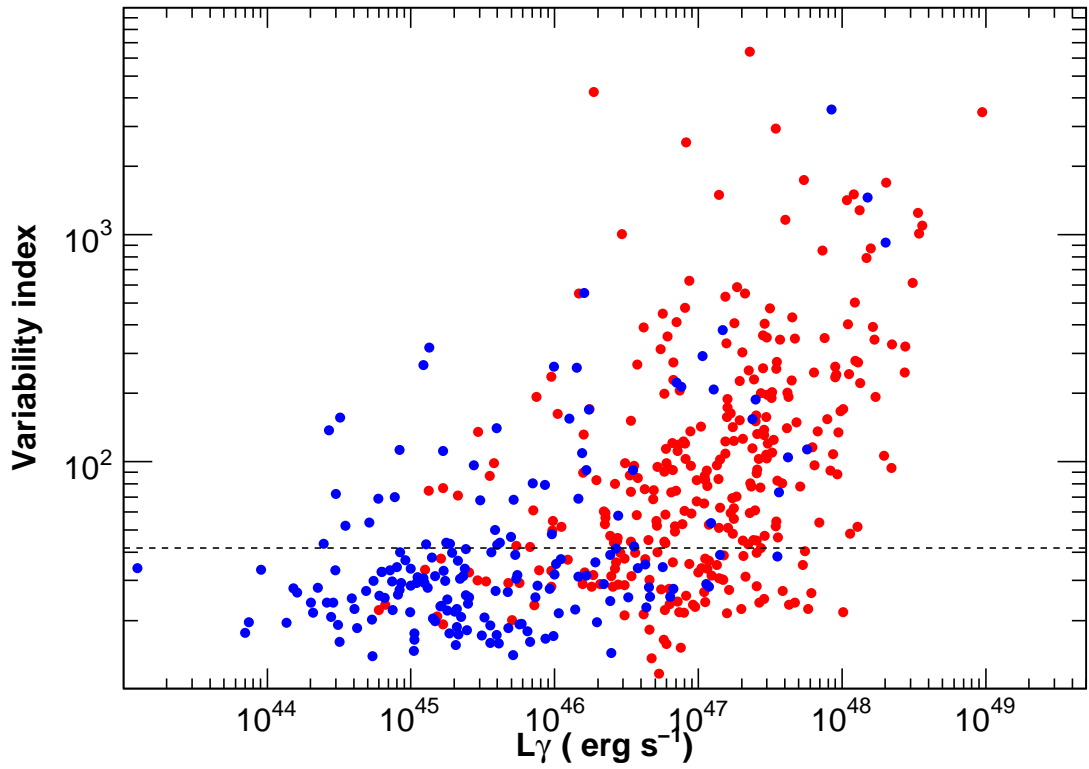}}}
\caption{Variability index versus $\gamma$-ray luminosity. Red: FSRQ, blue: BL~Lacs. The dashed line corresponds to the 99\% confidence level for a source to be variable.}
\label{fig:L_varind}
\end{figure}

\begin{figure}
\centering
\resizebox{16cm}{!}{\rotatebox[]{0}{\includegraphics{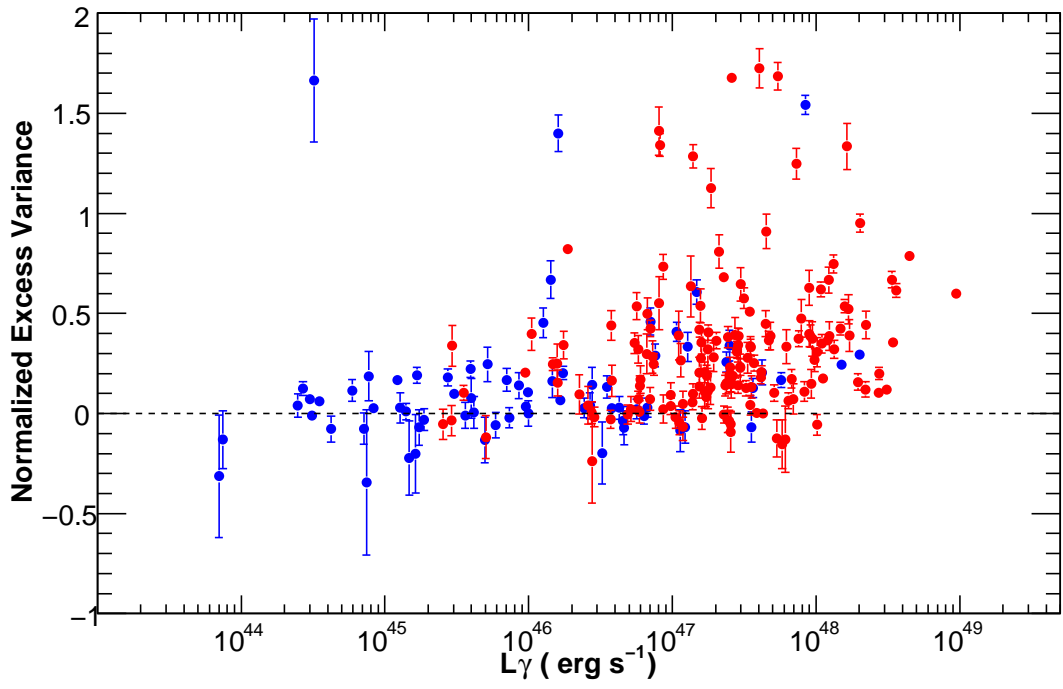}}}
\caption{Normalized excess variance versus $\gamma$-ray luminosity. Red: FSRQs, blue: BL~Lacs.}
\label{fig:L_sig}
\end{figure}

A total of 224 FSRQs (out of 310), 91 BL~Lacs (out of 395) and 33 sources of unknown type (out of 157) are variable at a confidence level greater than 99\%. Thus 348 blazars of the 2LAC Clean Sample fulfill this condition, while there were only 189 in the 1LAC Clean Sample. Figure \ref{fig:varind_sync} shows the variability index versus synchrotron peak position. Only a small fraction of the HSP-BL~Lacs detected by the LAT shows significant variability  (27 out of 160), substantially less than LSP-BL~Lacs (25 out of 61) and ISP-BL~Lacs (30 out of 81). The photon indices of variable FSRQs and BL~Lacs  are shown in Figure \ref{fig:index_var} versus the normalized excess variance \citep{Vaugh03}. The plot reveals a trend of variability with spectral index. Most variable sources have a photon index greater than 2.2.  These sources are observed at energies greater than the peak energies of their SEDS, where the variability amplitude tends to be larger. The harder sources, including all but one (PKS~0301$-$243) of the HSPs and ISPs have   normalized excess variance $<$ 0.5.
The average normalized excess variance for each of the blazar classes is 0.37 $\pm$0.03 (FSRQs), 0.28$\pm$0.07 (LSP-BL~Lacs), 0.19$\pm$0.04 (ISP-BL~Lacs) and 0.20$\pm$0.10 (HSP-BL~Lacs). Excluding the outlier (PKS~0301$-$243) the value for the HSP-BL~Lacs becomes 0.10$\pm$0.03 which implies that even if significant variability is detected only in a fraction of the individual HSPs, they do, as a class, exhibit variability but at a lower level than the other classes.
The variability index and normalized excess variance are also plotted against $\gamma$-ray luminosity. These are shown in Figure \ref{fig:L_varind} and \ref{fig:L_sig} respectively. The normalized excess variance does show a gradual increase with $\gamma$-ray luminosity for both BL Lacs and FSRQs. The BL Lac with low luminosity and high normalized excess variance ($>$1.5) is 2FGL J0217.4+0836, which underwent a flare with a Flux[E$>$100 MeV]=1.3$\times$10$^{-7}$ \pflux{} flare in January 2010.  

\begin{figure}
\centering
\resizebox{16cm}{!}{\rotatebox[]{0}{\includegraphics{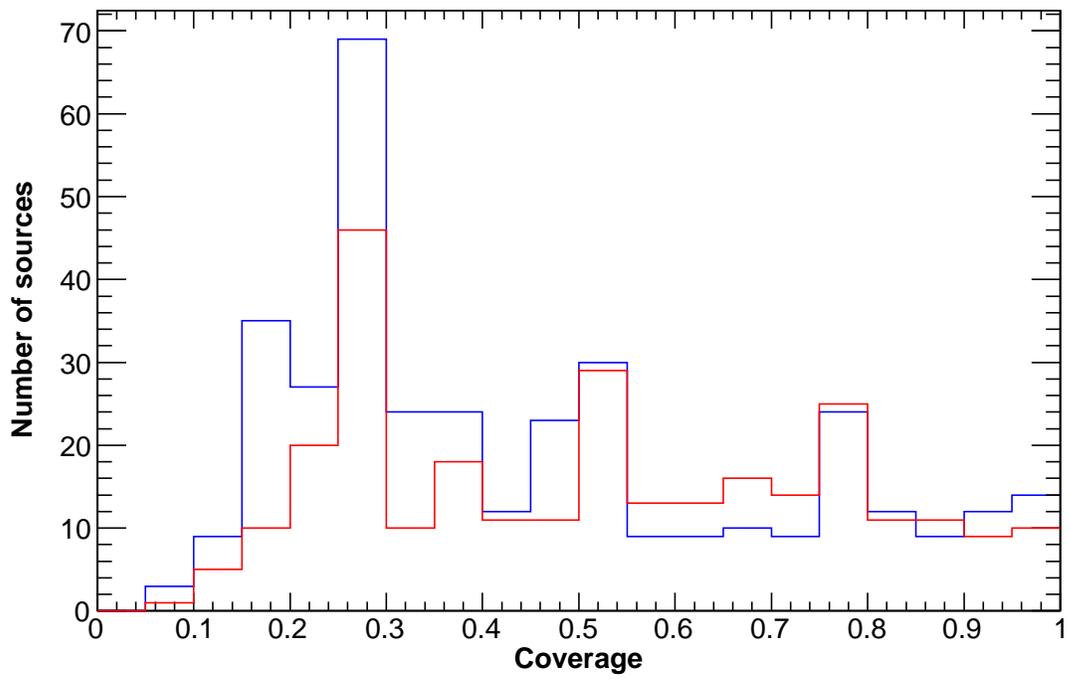}}}
\caption{Coverage distributions for BL~Lacs (blue) and FSRQs (red) in the Clean Sample.}
\label{fig:coverage_lc}
\end{figure}

\begin{figure}
\centering
\resizebox{16cm}{!}{\rotatebox[]{0}{\includegraphics{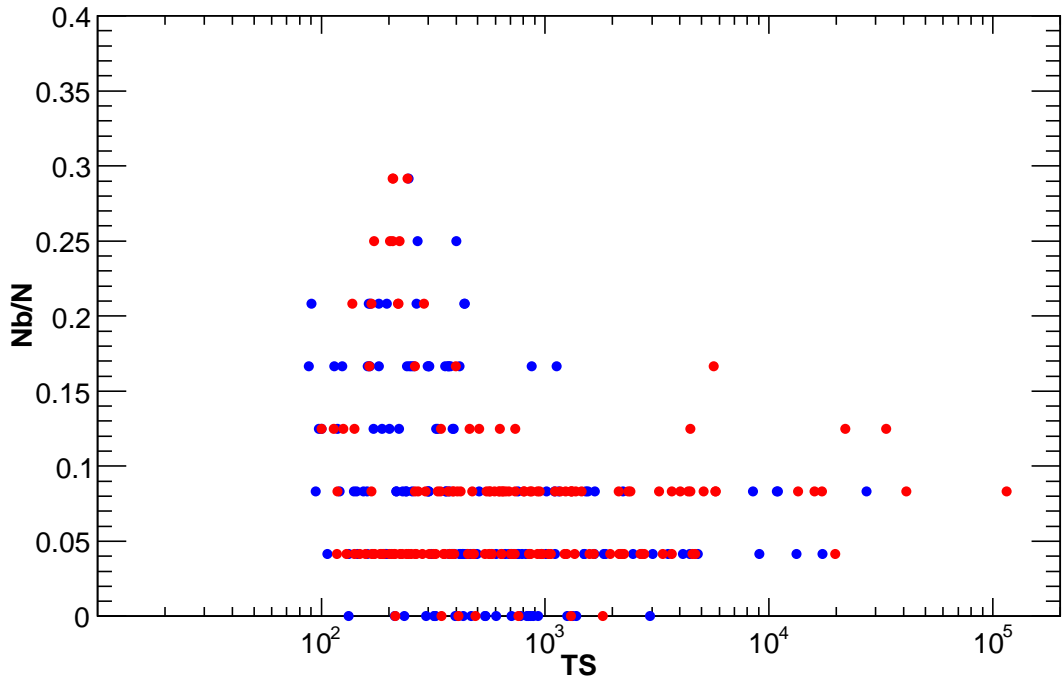}}}
\caption{Duty cycle (defined as the fraction of monthly periods where the flux exceeds $<F>+1.5\:S + \sigma_i$) versus $TS$ for FSRQs (red) and BL~Lacs (blue).  }
\label{fig:duty_cycle}
\end{figure}

The monthly-binned light curves also provide information about the duty cycle of blazars at $\gamma$-ray energies. Sources are in general not detected in all 1-month-bins. This is illustrated in Figure \ref{fig:coverage_lc}, which shows the distribution of {\it coverage}, i.e., the fraction 
of months where the source was detected with  $TS>$4. Not surprisingly, the coverage distribution is skewed toward low values. We find that 161 FSRQs and 152 BL~Lacs have a coverage greater than 0.5. Only these sources will be considered in the variability studies presented below. We define the {\it duty cycle} as the fraction of monthly periods N$_b$/N$_{tot}$ where the flux exceeds $<$F$>+1.5 S + \sigma_i$, where $<F>$ is the average flux, $S$ is the total standard deviation and $\sigma_i$ is the flux uncertainty of month $i$ \citep{Abdo_var}. These duty cycle values are shown as a function of $TS$ in Figure \ref{fig:duty_cycle}. 
Bright sources with $TS >$ 1000 essentially have all N$_b$/N$_{tot}\ge$ 0.05. Simulations considering the actual $TS$ distributions of both blazar classes were performed and showed that the measurement of N$_b$/N$_{tot}$ for these sources was not significantly affected by measurement noise. The wider distribution in N$_b$/N$_{tot}$ for sources with $TS <$ 1000 is consistent with these sources having similar duty cycle as the brighter ones and only results from a lower signal-to-noise ratio. 

\begin{figure}
\centering
\resizebox{16cm}{!}{\rotatebox[]{0}{\includegraphics{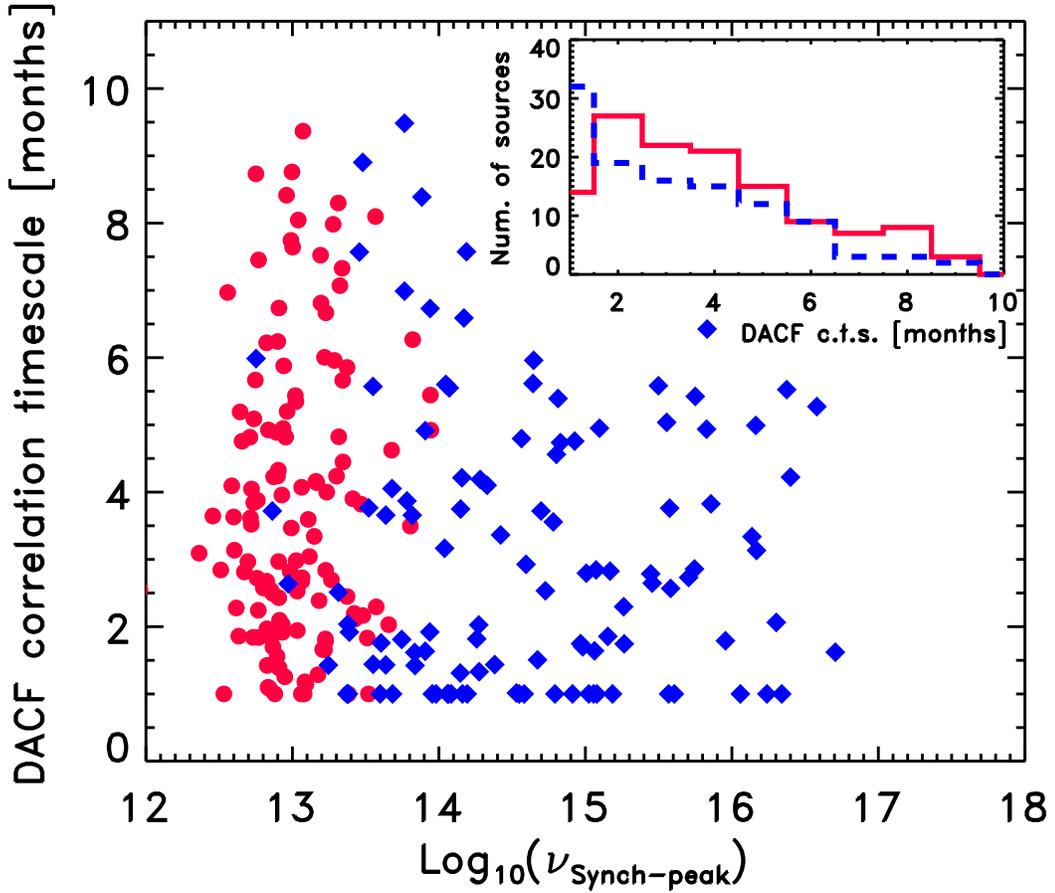}}}
\caption{ The observed discrete auto-correlation function (DACF) $\gamma$-ray correlation timescales versus the source-frame synchrotron peak frequency for the monthly light curves of the 2LAC sources having at least 50\% of the 24 bins with flux detections of $TS \geq 4$. Red circles: FSRQs, blue diamonds: BL Lacs. Inset panel: distribution of DACF $\gamma$-ray correlation timescales. Red/continuous line: FSRQs, blue/dashed line: BL Lacs.}
\label{fig:ACF}
\end{figure}

DACF and PDS were calculated for all sources with coverage larger than 0.5 and mean flux above 100 MeV exceeding $3 \times 10^{-8}$ \pflux{} (156 FSRQs and 59 BL~Lacs), while the SF analysis was applied to the whole Clean Sample.  From each DACF a correlation timescale was estimated
as the time lag of the first zero crossing of the function, computed by linear interpolation between the lag points. These observer-frame timescale estimates for both FSRQs and BL~Lacs are plotted in Figure \ref{fig:ACF} as a function of synchrotron peak frequency for the selected sources. The timescale distribution is shown in the inset plot. Interestingly the observation that FSRQs have $\gamma$-ray correlation extending to longer timescales than BL Lacs confirms the trend found for the LBAS sample \citep{LBAS} using  weekly light curves obtained over the first 11 months of observation \citep{Abdo_var}.

\begin{figure}
\centering
\resizebox{16cm}{!}{\rotatebox[]{0}{\includegraphics{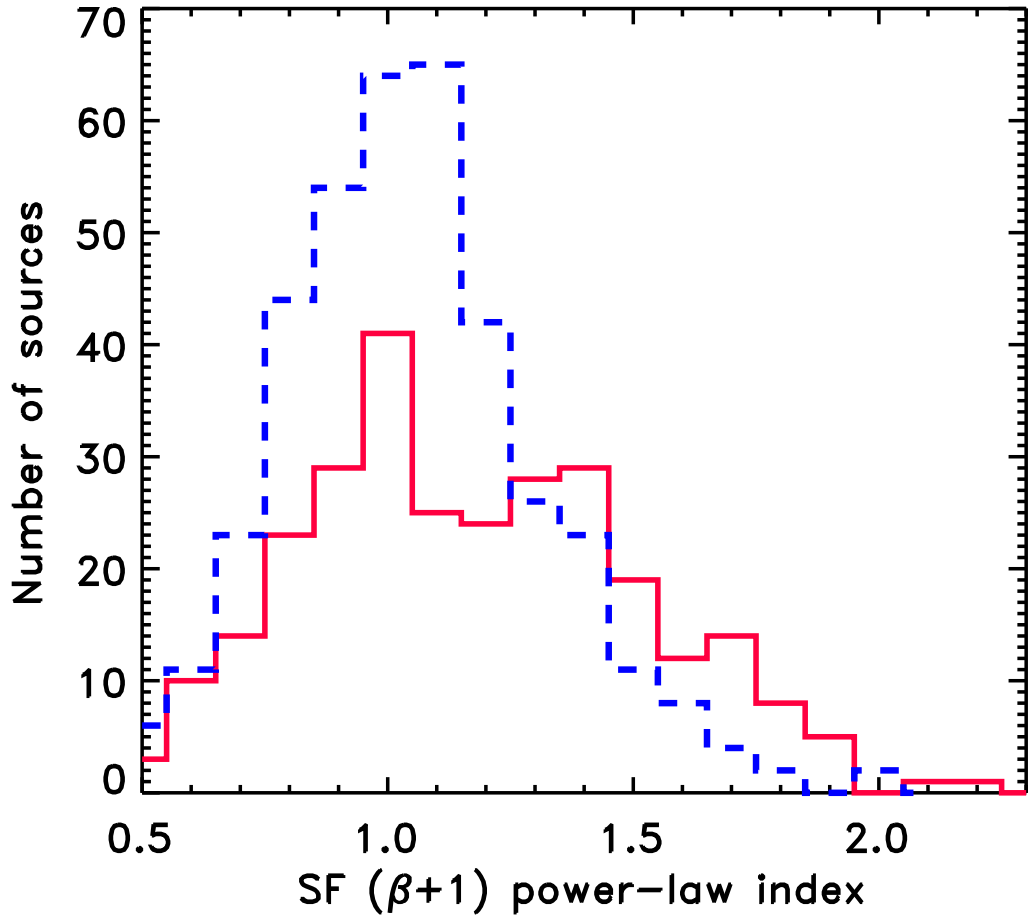}}}
\caption{Distribution of the temporal PDS power-law indexes ($\beta+1$) for the FSRQs (red) and BL Lacs (blue) of the 2LAC Clean Sample, evaluated in time domain using a first order structure function (SF) analysis with blind power-law $\beta$ slope estimation.}
\label{fig:SF}
\end{figure}

The SF, which is equivalent to the PDS of the signal but calculated in the time domain, which makes it less subject to irregular sampling, low significance bins and upper limit problems, was applied to the light curves of the entire 2LAC Clean Sample sources. Results are shown in Figure \ref{fig:SF} where the 
distribution of the PDS power-law indexes evaluated in the time domain ($\beta + 1$, where $\beta$ is the blind power-law slope estimated from the $SF$ of each light curve) are reported for the FSRQs and BL Lacs. The resulting distributions of the power-law indices appear whitened (i.e., closer to white noise with flatter $SF$ power-law indices) because of the short extent of the time lag range investigated (from 1 to 24 months) and of the fact that a consistent subset of the 2LAC Clean Sample showed low-flux, noisy and non-variable monthly-bin light curves, when compared with the same analysis performed on the brightest and better sampled light curves of the LBAS sample \citep{Abdo_var}. Again the distribution shows FSRQs with slightly more Brownian-like (steeper) and more scattered SF indexes, with respect to the more flicker-like (flatter) ones for BL Lacs in agreement with what was already found for the LBAS sample \citep{Abdo_var}.

\begin{figure}
\centering
\resizebox{16cm}{!}{\rotatebox[]{0}{\includegraphics{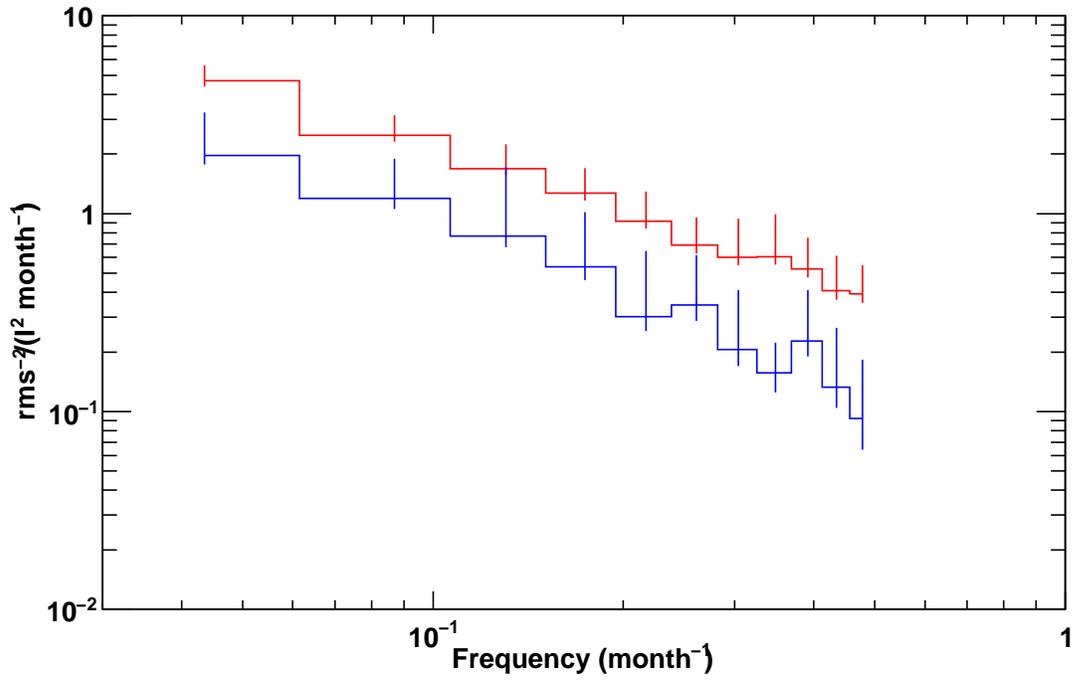}}}
\caption{Power density spectrum (PDS) for bright FSRQs (red) and BL~Lacs (blue).}
\label{fig:PDS}
\end{figure}

In Figure \ref{fig:PDS} we have plotted the average PDS for FSRQs and BL~Lacs. The power density is normalized to fractional variance per frequency unit (\mbox{\,rms$^{2}$~I$^{-2}$~day$^{-1}$}, where I is the average flux) and the PDS points are averaged in logarithmic frequency bins. The white noise level was estimated from the rms of the flux errors and was subtracted for each PDS. The error bars were computed as the standard error of the mean for each frequency bin. The PDS slope (power-law index)
is similar for the two groups, $\sim$ 1.15 $\pm$0.10. This is somewhat flatter than was deduced for the very brightest sources in the LBAS sample \citep{Abdo_var}. The difference in
the height of the PDS means that the fractional variability
of BL~Lacs is lower than that of FSRQs. This is in line with the
LBAS results. With the PDS normalization used here, we can compute a normalized excess variance
by integrating the PDS over frequency. To limit the effect of statistical noise this integration was done for frequencies up to 0.2 month$^{-1}$, which also contains most of the variance. 
The resulting normalized excess variance for the different blazar classes is 0.44 $\pm$0.04 (FSRQs), 0.27 $\pm$0.10 (LSP-BL~Lacs), 0.19 $\pm$0.04 (ISP-BL~Lacs) and 0.14 $\pm$0.07 (HSP-BL~Lacs). The trend and values are consistent with the normalized excess variance calculated directly from the light curves as described above.

\subsection{\label{sec:hep} Highest-energy photons}

\begin{figure}
\centering
\resizebox{16cm}{!}{\rotatebox[]{0}{\includegraphics{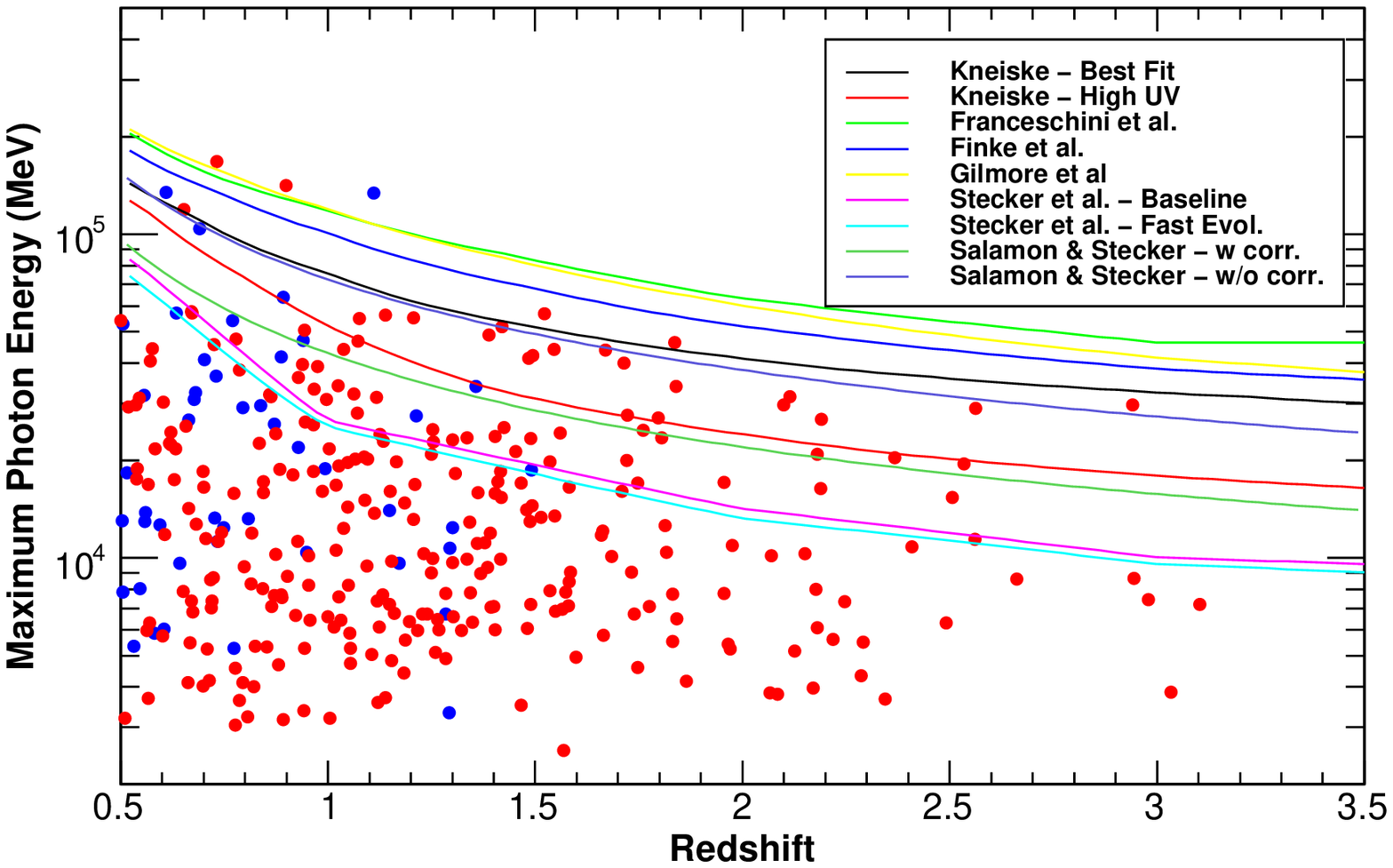}}}
\resizebox{16cm}{!}{\rotatebox[]{0}{\includegraphics{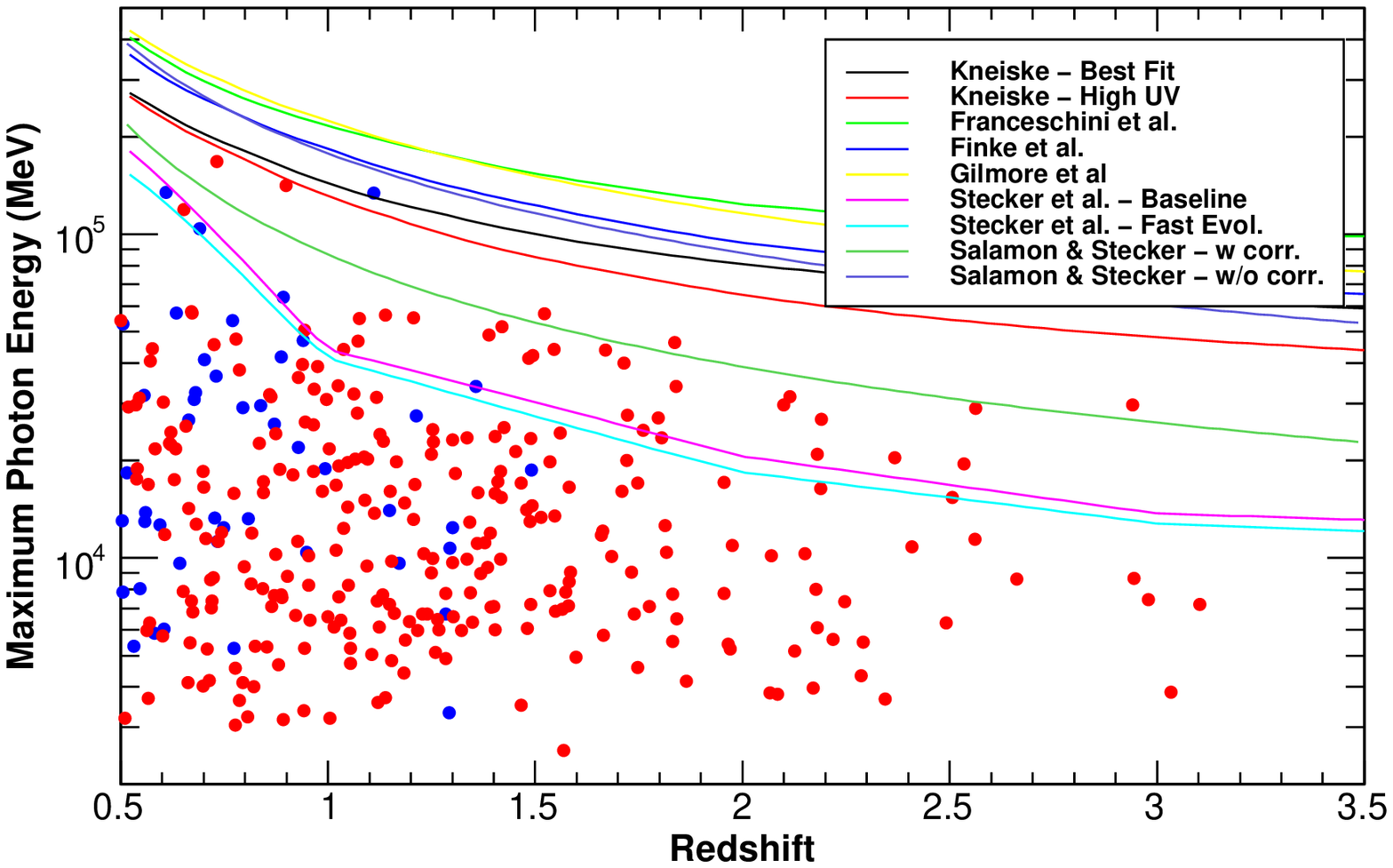}}}
\caption{Top: Maximum photon energy versus redshift. Red: FSRQs, blue: BL~Lacs. The curves correspond to predictions for $\tau$=1 for different models. Bottom: Same but the curves correspond to predictions for $\tau$=3 for different models.}
\label{fig:redshift_he}
\end{figure}

Figure \ref{fig:redshift_he} displays, as a function of redshift,  the highest energy photon (HEP) detected by the LAT from the 2LAC AGN sample using the Pass 7\_V6 Ultraclean  event selection and that is associated with the source within the 68\% containment radius. Further work is being carried out to improve the capability to reconstruct event tracks and reject background at high energy \citep{Baldini}.  In comparison to the corresponding sample based on 11 months of LAT operation \citep{Abdo_EBL} we find about a factor $\sim 2$ more candidate photon events coming from sufficiently high redshift ($z>0.5$) to probe the models of the extragalactic background light (EBL). 

Predictions of $\gamma\gamma$ opacity curves, $\tau_{\gamma\gamma} = 1$ (top panel) and $\tau_{\gamma\gamma} = 3$ (bottom panel), for different EBL models are also shown in Figure \ref{fig:redshift_he}. Detection of HEPs above the opacity curve predicted by a given model  makes the model less likely. In the new 2LAC AGN sample, we find 30 HEP events from $z>0.5$ sources beyond the $\tau_{\gamma\gamma} = 3$ regime of the \cite{Ste06} ``baseline model'', which is already severely constrained by the LAT 11 month data set \citep{Abdo_EBL}. Only one event appears beyond $\tau_{\gamma\gamma} = 3$ of the \cite{Kne04}  ``best-fit'' and ``high-UV'' models.

None of the HEP events seems to be in strong contradiction with EBL models that are of lower photon density \citep[e.g., ][]{Fran08, Fin10b,Gil09}. Note, however, that  we don't have redshift information for more than 50\%  of the 36 sources with  HEPs at energies greater than 100 GeV, which can therefore not be tested against any EBL models. Apparent in Figure \ref{fig:redshift} is the clustering of HSPs at low redshifts ($z\le 0.2$) while LSPs cover a broad redshift range up to $z = 3.1$. Because HSPs are intrinsically hard sources, and LSPs intrinsically soft (see Figure \ref{fig:index_nu_syn}) any systematic trend between redshift and spectral properties (spectral index, HEP) is unlikely to be caused by EBL absorption only. For the $> 500$ events without an assigned source redshift, the HEP is located above $\sim 10$~GeV in more than $\sim 70\%$ of all cases. Interestingly, we found $\sim 4$ FSRQs with HEPs that reach energies $> 100$ GeV \citep[4C~+55.17, see ][ 4C~+21.35,  PKS~1958$-$179, BZQ~J1722+1013]{McC11} with the latter two (at redshifts $z = 0.652$ and $z = 0.732$ respectively) displaying no significant deviation from a power-law spectrum (with indices $\Gamma \sim 2.4$ and $\Gamma \sim 2.2$, respectively) in the energy range of the LAT. One BL~Lac (2FGL J0428.6$-$3756, PKS 0426$-$380) at redshift $z =1.10$  of LSP spectral type has also been detected at $> 100$ GeV.

\subsection{Luminosity Distributions}

\begin{figure}
\centering
\resizebox{16cm}{!}{\rotatebox[]{0}{\includegraphics{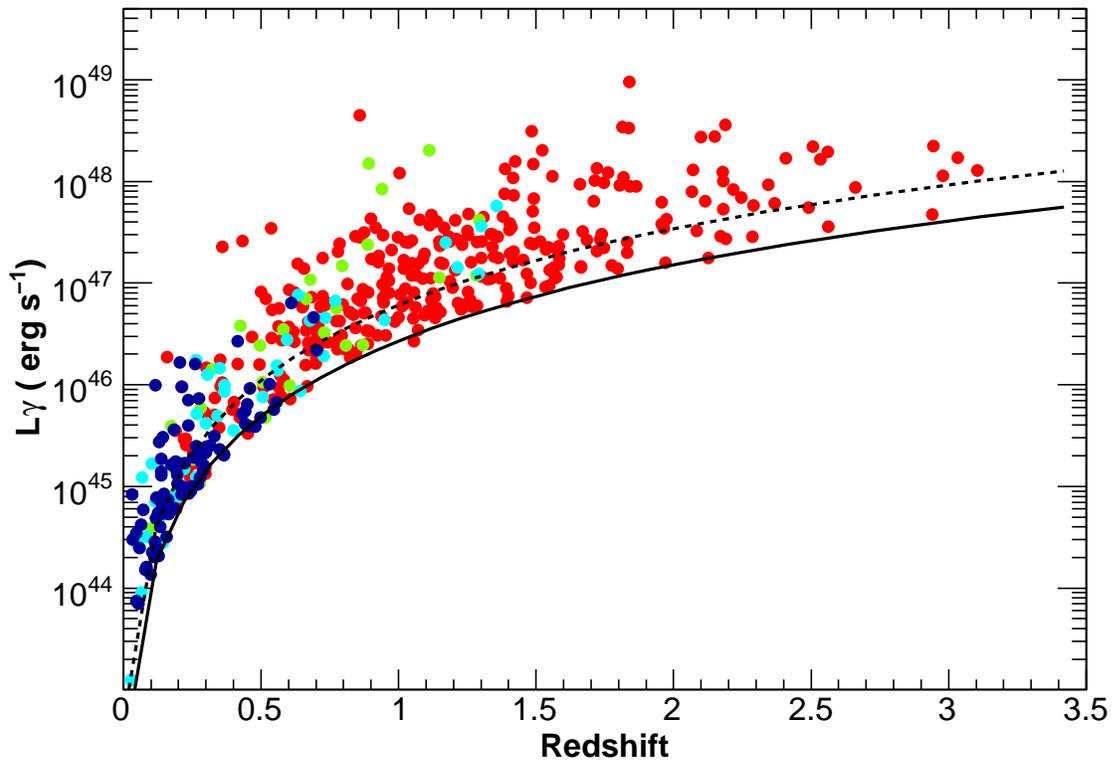}}}
\caption{Gamma-ray luminosity versus redshift. Red: FSRQs, blue: BL~Lacs. The solid (dashed) curve represents the approximate detection limit for $\Gamma$=1.8 ($\Gamma$=2.2). }
\label{fig:L_redshift}
\end{figure}

The $\gamma$-ray luminosity is plotted as a function of redshift in Figure \ref{fig:L_redshift}. A Malmquist bias is readily apparent in this figure as only high-luminosity sources (mostly FSRQs) are detected at large distances. Given their $\gamma$-ray luminosity distribution, most BL~Lacs could not be detected if they were located at redshifts greater than 1. 

\begin{figure}
\centering
\resizebox{16cm}{!}{\rotatebox[]{0}{\includegraphics{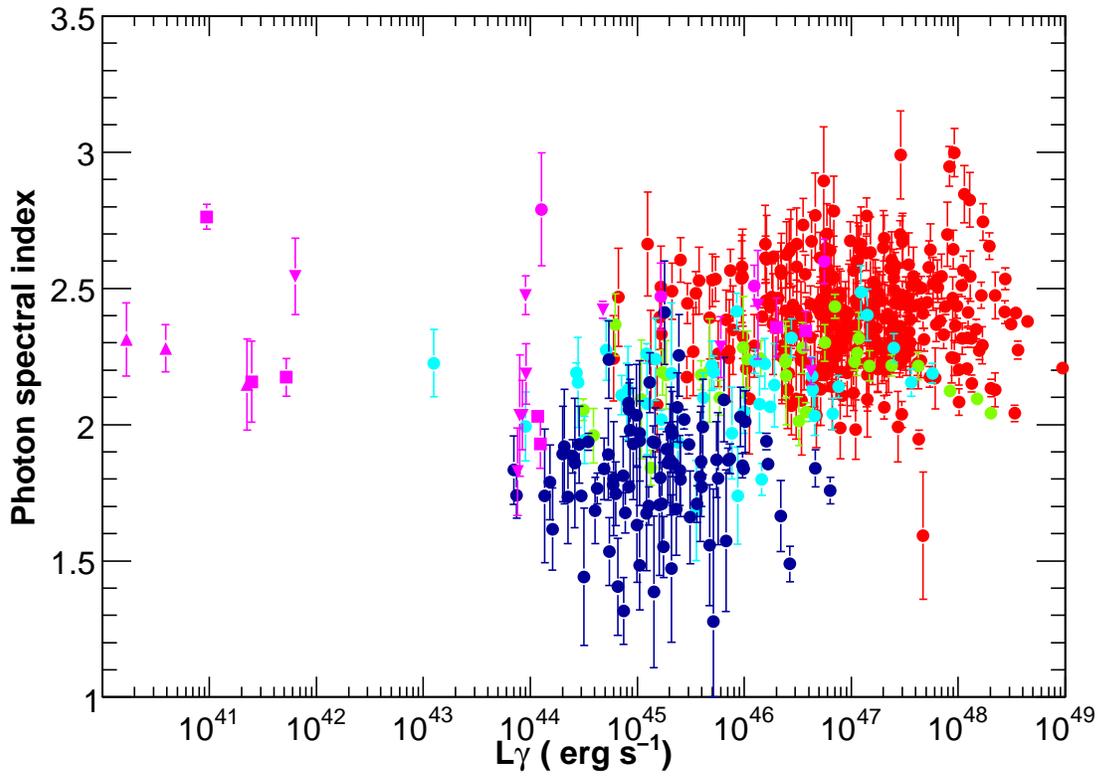}}}
\caption{Photon index versus $\gamma$-ray luminosity. Red: FSRQs, green: LSP-BL Lacs, light blue: ISP-BL Lacs,  dark blue: HSP-BL Lacs, magenta: non-blazar AGNs (circles: NLS1s, squares: misaligned AGNs, up triangles: starbursts, down triangles: other AGNs).}
\label{fig:index_L}
\end{figure}

Figure \ref{fig:index_L} shows  photon index versus $\gamma$-ray luminosity. This correlation has been discussed in detail in the context of the ``blazar divide'' \citep{Ghisellini09}. Note that since the $\gamma$-ray luminosity is derived from the energy flux and that the detection limit in energy flux is essentially independent of the photon index (Figure \ref{fig:index_S}), no significant LAT-related detection bias is expected to affect this correlation. The ISP-BL~Lac outlier at L$_\gamma \simeq$ 3$\times$ 10$^{43}$ erg cm$^{-2}$ s$^{-1}$ is 4C~04.77 (2FGL J2204.6+0442) at z=0.027, which was classified as an AGN in 1LAC.

\begin{figure}
\centering
\resizebox{16cm}{!}{\rotatebox[]{0}{\includegraphics{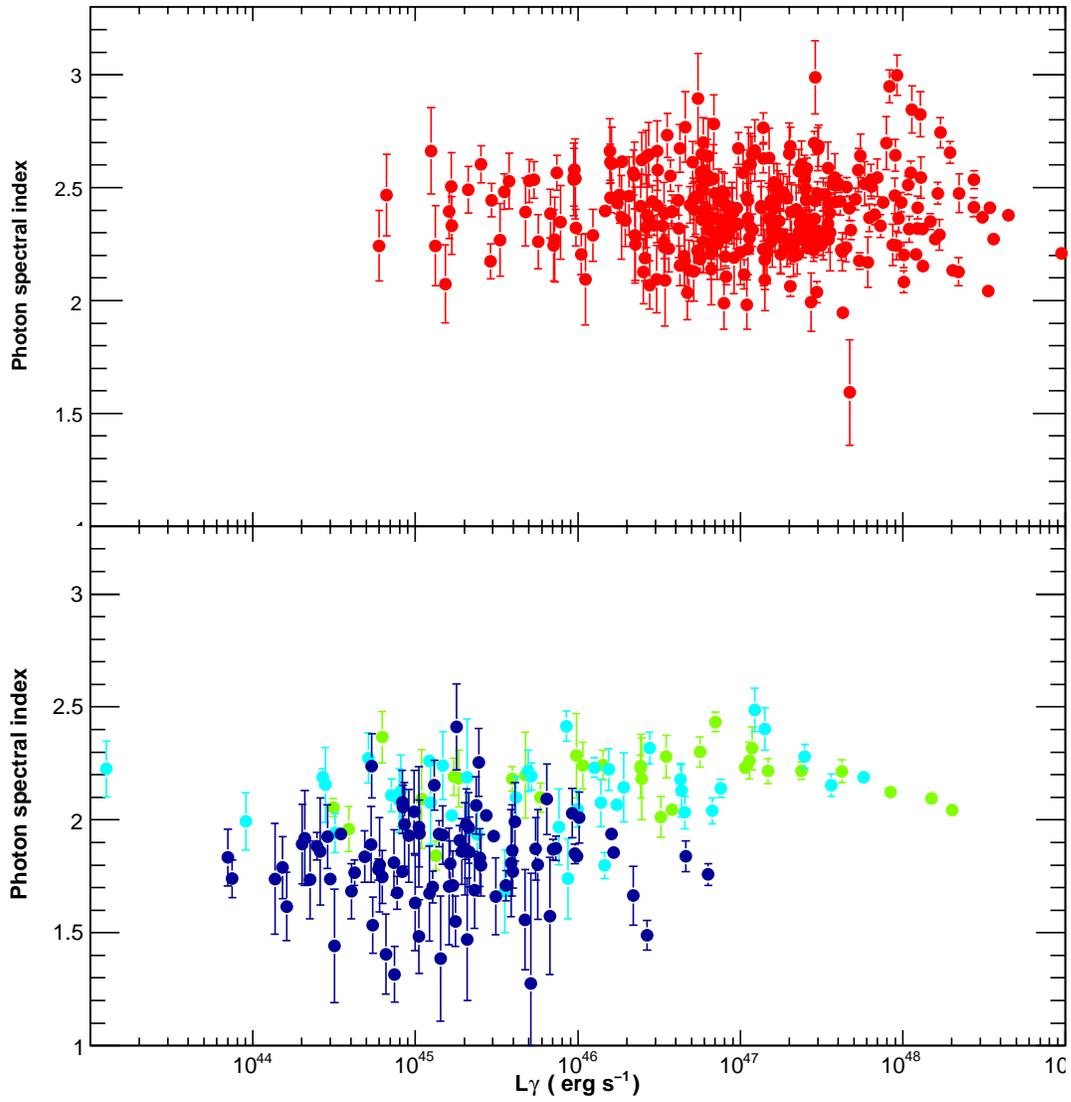}}}
\caption{Photon index versus $\gamma$-ray luminosity. Top: FSRQs. Bottom:green: LSP-BL Lacs, light blue: ISP-BL Lacs,  dark blue: HSP-BL Lacs.}
\label{fig:index_L_2}
\end{figure}

Figure \ref{fig:index_L_2} shows  photon index versus  $\gamma$-ray luminosity for FSRQs (top) and BL~Lacs (bottom) separately. The Pearson correlation coefficients are $-$0.04 and 0.14 for FSRQs and BL~Lacs respectively. For a given class, the correlation is very weak.

\section{Multiwavelength properties of the 2LAC sample} 

In this section, we explore the properties of the 2LAC sample in the radio, optical, X-ray and TeV bands. Table \ref{tab:prob1} gives archival fluxes in different bands for these sources. For completeness, Table \ref{tab:prob2} provides the corresponding fluxes for the low-latitude sources. 

\subsection{Radio Properties}

The 2LAC sources are associated with a population of radio sources, whose flux density distribution spans the range between a few mJy and several tens of Jy. This is rather typical for blazars, whose radio emission has often been found to be correlated  with the $\gamma$-ray activity \citep{Kovalev2009,Ghirlanda2010,Ghirlanda2011,Mahony2010,radiogamma}. In particular, \citet{radiogamma} have shown a highly significant correlation (chance probability $<10^{-7}$) between the radio and $\gamma$-ray fluxes for both FSRQ and BL Lacs in the 1LAC, although with a large scatter.

\begin{figure}
\plotone{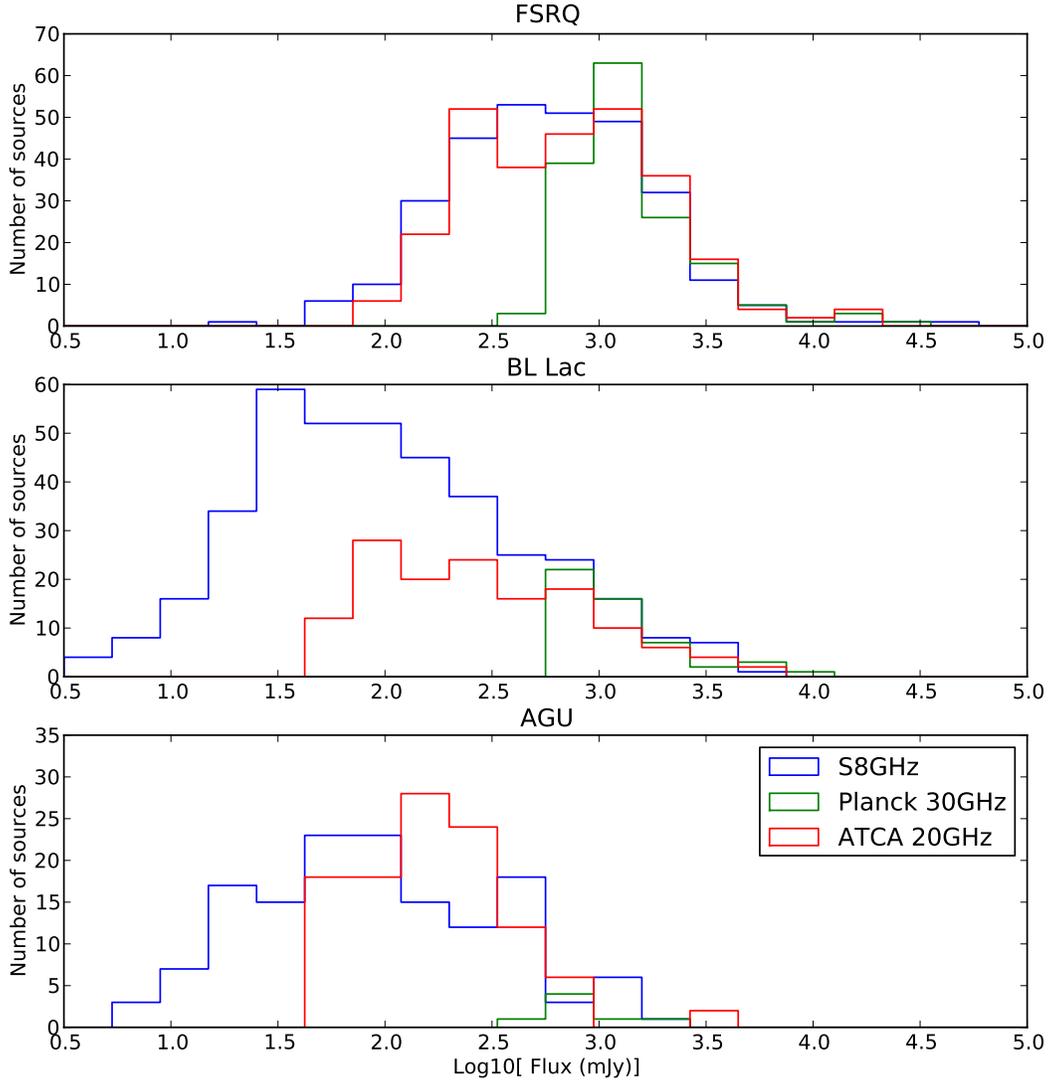}
\caption{Radio flux density distributions of the 2LAC counterparts: FSRQs (top),
 BL Lacs (middle), blazars of unknown type (bottom). For each panel we show the counts at 8 GHz (blue line, from CRATES or similar surveys), at 20 GHz (red line, obtained from the AT20G survey and multiplied by 2 to normalize for the sky coverage) and at 30 GHz (green line, from the Planck ERCSC). \label{fig:fluxhisto}}
\end{figure}

In Figure ~\ref{fig:fluxhisto} we plot the
radio flux density distributions for sources in the 2LAC, divided according to the optical type.  For all sources, we plot the radio flux density at 8 GHz, obtained either using interferometric data from CRATES \citep[][or similar surveys, when available]{crates}, or extrapolated from  low frequency (NVSS or SUMSS)
measurements assuming $\alpha=0.0$; we also plot the distribution of the radio flux density at higher frequency, i.e., at 20 GHz as obtained from the AT20G survey  and at 30 GHz as obtained from the {\it Planck} ERCSC \citep{Planckcatalog}. Since AT20G only covers half of the sky, we multiply the counts by 2 to have a consistent normalization (2LAC and {\it Planck} are all-sky surveys).

The distributions for BL Lacs and FSRQs are quite broad, with well separated peaks, FSRQs being on average significantly brighter radio sources. The median flux densities of the two distributions at 8 GHz are 86 and 581 mJy for BL Lacs and FSRQ, respectively. 
In the highest flux density bins, the various surveys are all basically complete. The distributions are similar for the three frequencies (8 GHz, 20 GHz, 30 GHz), confirming that the 2LAC sources have flat radio spectra. Below 1 Jy, {\it Planck} counts drop rapidly owing to sensitivity limits, while AT20G becomes less and less complete below 100 mJy.  Interestingly, AT20G shows a deficit of BL Lac sources in the 100--300 mJy range, which cannot be attributed to sensitivity limits; this is most likely to arise from the lack of spectroscopic information for sources in the Southern hemisphere (see Fig.\ \ref{fig:sky_map}), where the  AT20G survey was carried out. 

As shown in the 2FGL paper, the radio flux density distribution of the {\it Fermi} sources accounts for nearly all the brightest radio sources in CRATES, while a significant fraction of lower flux density sources have not been detected by {\it Fermi} so far. One viable possibility is that the $\gamma$-ray duty cycle of FSRQs (which is the dominant population in CRATES) is quite low, so these sources have not yet gone through a phase of activity during the {\it Fermi} lifetime; combined with the typically soft $\gamma$-ray spectra of FSRQs and the lower sensitivity and broader PSF of the LAT at low energy, this could account for the lack of such sources. 

On the other hand, the BL~Lac population extends to lower flux densities (even below the CRATES sensitivity) and is more consistently detected by the LAT. For example, the $\gamma$-ray detection rate in the VIPS survey established with the 1LAC sample, was $\sim 2/3$ for BL Lacs and only  9\% (50/529) for the FSRQs \citep{Linford2011}. In particular, a large number of BL~Lacs have now been detected and associated thanks to the extension to lower flux density of the association methods, which is essential for the radio-weak HSP sources, and their more persistent (less dramatically variable) $\gamma$-ray emission.

\begin{figure}
\resizebox{16cm}{!}{\rotatebox[]{0}{\includegraphics{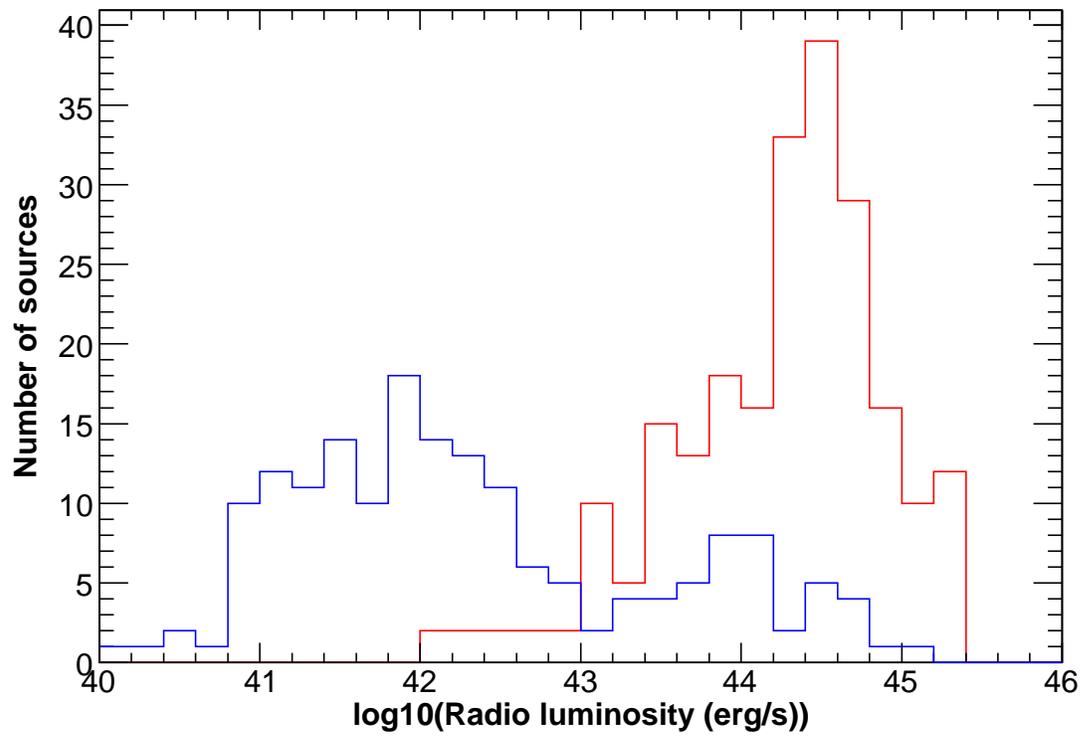}}}
\caption{Radio luminosity distribution of the 2LAC counterparts for FSRQs (red) and BL~Lacs (blue). \label{fig:lumhisto}}
\end{figure}

When combined with the different redshift distributions (see
Sect.~\ref{sec:z}), the different flux density distributions result in markedly distinct radio luminosity distributions, as shown by Figure ~\ref{fig:lumhisto}. The overall luminosity
interval spans the range between $10^{40}$ and $10^{45}$ erg s$^{-1}$, with
FSRQs more clustered at high luminosity ($\log L_{\rm r, FSRQ} [$erg\,s$^{-1}] = 44.1 \pm
0.7$), while the BL Lacs span a broader interval, down to lower luminosities
($\log L_{\rm r, BL} [$erg\,s$^{-1}]= 42.3 \pm 1.1$).

Not unexpectedly, given the large overlap between the two samples, these
properties are entirely consistent with those of the sources in the 1LAC. Also
the radio spectral index distribution for sources with data at both 8 GHz and $\sim
1$ GHz remains consistent with a flat value, with $\langle \alpha \rangle
= 0.08 \pm 0.30$. This is also suggestive that our extrapolation of the low
frequency data is solid, as confirmed by the similar distributions of the 8 GHz, 20 GHz, and 30 GHz flux densities in the range where the three surveys are complete.

\subsection{Properties in the optical/infrared and hard X-ray bands }

Optical and infrared bands are important for our understanding of GeV $\gamma$-ray blazars. For LSPs, the peak of the synchrotron emission is located in these bands and significant correlation with the GeV emission has been observed. Both synchrotron and thermal emission components can contribute in these bands, creating a complex spectral-temporal behavior. On the other hand, our limited knowledge about their host galaxy, nucleus and stellar core profiles hamper studies in these bands, as do difficulties in measuring line widths, ratios, and fluxes.

Correlated variability between optical-infrared and $\gamma$-ray variability points to a common population of electrons producing non-thermal emission through synchrotron and Inverse Compton processes. High-quality  data (GeV and optical/NIR) obtained on flaring sources thanks to intensive multifrequency campaigns \citep[e.g., ][]{Abdo_1502,Abdo_3C279}, have already revealed the existence of correlated flares, with no true orphan flares \citep[as sometimes observed in the X-ray band, e.g., ][]{Abdo_3C279}.

Our 2LAC sample is characterized by different optical spectra, with a number of BL Lac - FSRQ transition objects. Those include BL Lacertae itself, the prototype of the class displaying at times moderately strong, broad lines and a complex SED \citep{Abdo_BLLacertae}, and 3C 279, one of the prototypes of the FSRQ class, which can appear nearly featureless in the optical band in a bright state \citep{Abdo_3C279}. 
The four NLS1 sources in 2LAC have flat radio spectra and strong but narrow emission
lines, interpreted as the apparent luminosity of the jets compared to the line luminosity being
lower, possibly because of lower intrinsic jet power, or slight misalignment of the jet with respect to our
line of sight.

\begin{figure}
\centering
\resizebox{16cm}{!}{\rotatebox[]{0}{\includegraphics{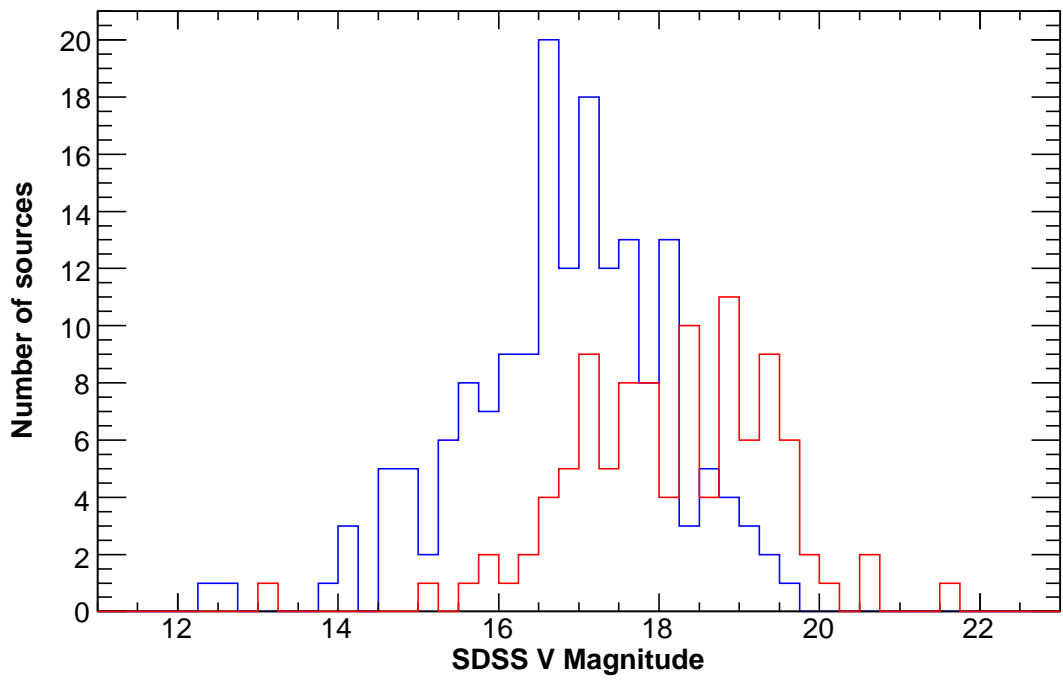}}}
\caption{SDSS magnitude for blazars in the Clean Sample. Red: FSRQs,  blue: BL Lacs.}
\label{fig:Magu}
\end{figure}

Figure \ref{fig:Magu} shows the V magnitude reported in SDSS for the FSRQs and BL~Lacs of the Clean Sample.  The BL Lacs are associated with brighter galaxies relative to the FSRQs, although the sources are all relatively bright. This brightness enables the monitoring of all Clean Sample sources with small optical telescopes to study correlated variability.

\begin{figure}
\centering
\resizebox{16cm}{!}{\rotatebox[]{0}{\includegraphics{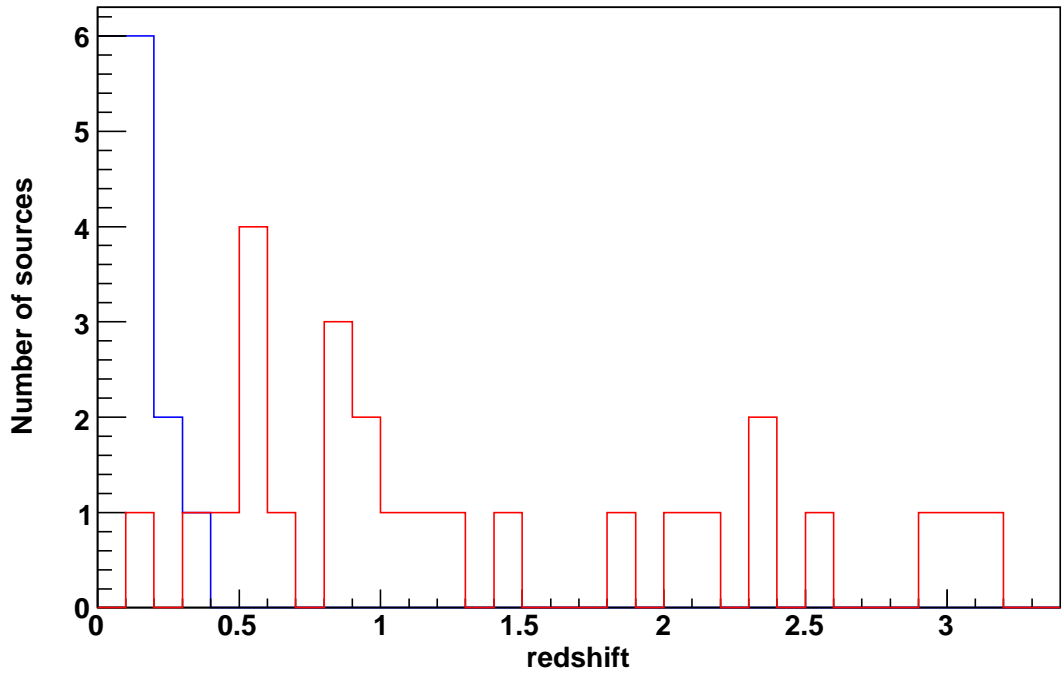}}}
\caption{Redshift distributions of blazars detected by both BAT and LAT.  Red: FSRQs,  blue: BL Lacs.}
\label{fig:BAT_z}
\end{figure}

\begin{figure}
\centering
\resizebox{16cm}{!}{\rotatebox[]{0}{\includegraphics{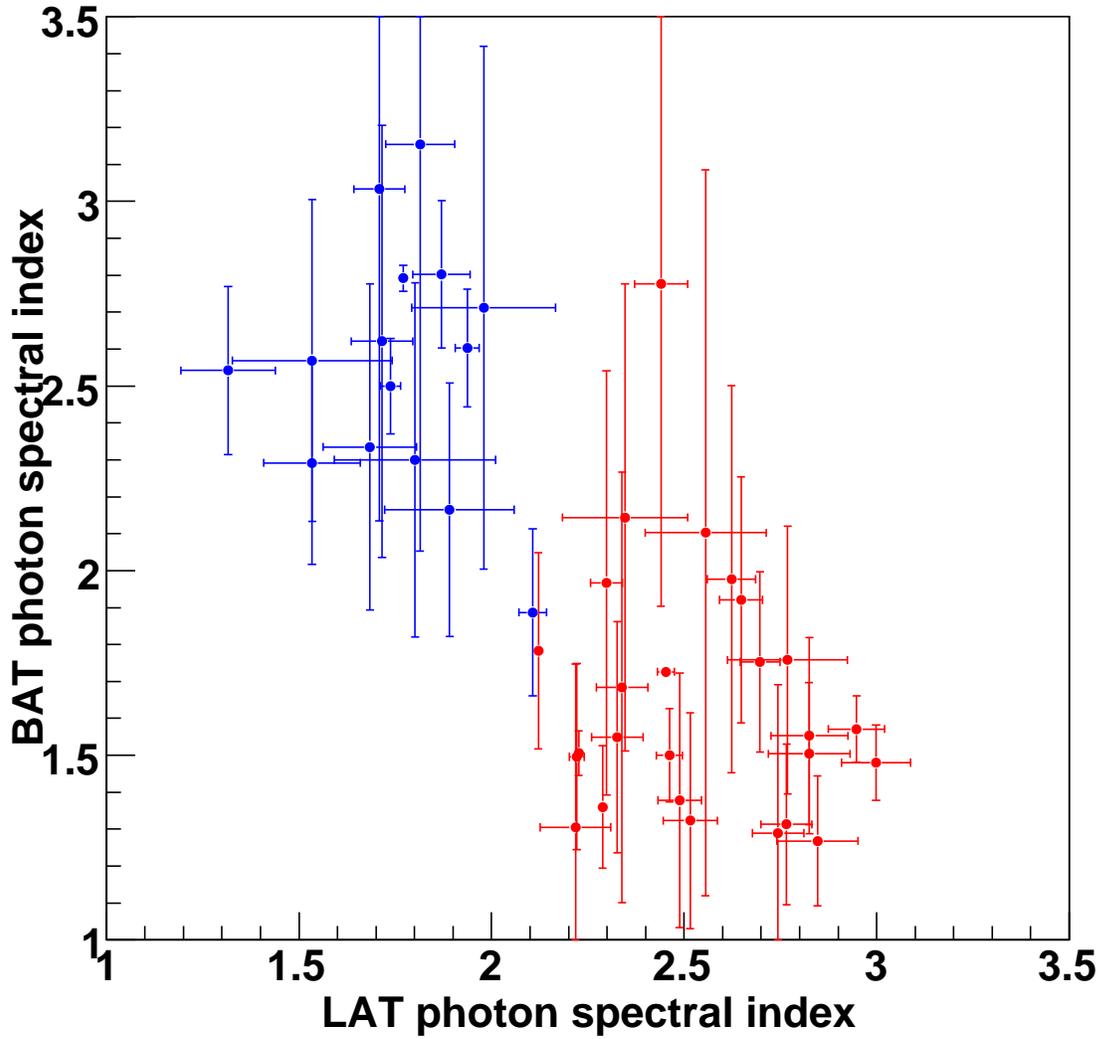}}}
\caption{Photon spectral index in the BAT band (14--195 keV) versus photon spectral index in the LAT band.  Red: FSRQs,  blue: BL Lacs.}
\label{fig:BAT_index}
\end{figure}

Cross-correlating the 2LAC with the {\it Swift} BAT 58-month survey \citep{BATcatalog} yields a total of 47 sources present in both catalogs.   
The redshift distributions of the FSRQs and BL~Lacs from this subset are given in Figure \ref{fig:BAT_z}. All 15 BL~Lacs are of the HSP type, except one,
which is an ISP. These distributions are very similar to those of the LAT blazars not detected by BAT. The photon spectral index measured in the BAT band is plotted against the  photon spectral index in the LAT band in Figure \ref{fig:BAT_index}.  A clear anticorrelation is visible in this Figure  (Pearson correlation factor=$-$0.73).  For the HSP-BL~Lacs considered here, BAT probe the high-frequency (falling) part of the $\nu F_{\nu}$ synchrotron peak while the LAT probes the rising side of the Inverse Compton peak (assuming a leptonic scenario). For FSRQs, which are all LSPs, BAT and LAT probe the rising and falling parts of the Inverse-Compton peak respectively.  Note that for this subset of sources which are quite distinct in properties, the LAT spectral indices for FSRQs and BL~Lacs do not overlap. The Pearson correlation factor is only $-0.15$ and $-0.17$ for FSRQs and BL~Lacs considered independently, respectively.

\subsection{GeV-TeV connection}

At the time of publication of 1LAC \citep{1LAC}, 32 AGNs had been
detected in the ``TeV'' or very high energy (VHE; E\,$\ge$ 100\,GeV)
regime \citep{TevCat}. All but four of these (RGB\,J0152$+$017,
1ES\,0347$-$121, PKS\,0548$-$322 and 1ES\,0229$+$200) were in
1LAC. Since then, an additional 13 AGNs (14 if we include the
unidentified, but likely AGN, VER\,J0648$+$152 that is discussed below)
have been detected at TeV energies, which brings the total number of
TeV AGNs to 45, 39 of which are in 2FGL. Just one of the TeV AGNs,
RGB\,J0152$+$017, that was not in 1LAC is in 2LAC. The clean 2LAC
sample contains 34 of the TeV AGNs, which we will refer to as the
GeV-TeV AGNs. The five TeV AGNs that are in 2FGL but not the clean
sample are: VER\,J0521+211, MAGIC\,J2001+435 and 1ES\,2344+514 (due to
their low Galactic latitudes) and IC\,310 and 1RXS\,J101015.9$-$311909
(due to their flags\footnote{IC\,310 has two flags indicating that its
  $TS$ changed from $TS$ $>$ 35 to $TS$ $<$ 25 when the diffuse model
  was changed and that it lies on top of an interstellar gas clump or
  small-scale defect in the model of the diffuse
  emission. 1RXS\,J101015.9$-$311909 has one flag indicating that when
  the diffuse model was changed, its position moved beyond the 95\%
  error ellipse; see \citet{2FGL} for more details on flagged
  sources.}). All of the TeV AGNs that were in 1LAC remained
significant LAT sources and are thus in the 2LAC Clean Sample. As can
be seen in Table~\ref{TAB:GeVTeV}, the largest subclass in the GeV-TeV
AGNs (18) is the HSPs but there also 6 ISPs, 5 LSPs and 5 AGNs whose
SED class remains unclassified using the technique described in
\S~\ref{sec:sedclass}. The mean photon index of the 2LAC sources
associated with the TeV AGNs is $1.87\,\pm\,0.27$, while the mean
photon index of the clean 2LAC sample is $2.13\,\pm\,0.30$, indicating
that those AGNs which are detected at TeV are, in general, harder than
the majority of the 2LAC sources at {\it{Fermi}}-LAT energies.

Since the launch of {\it{Fermi}}, 22 AGNs and one TeV source that was
classified as unidentified when discovered,
VER\,J0648$+$152,\footnote{VER\,J0648$+$152 is spatially coincident
  with 1FGL\,J0648.8$+$1516 and 2FGL\,J0648.9$+$1516, and seems likely
  to be an AGN. It is not in the 2LAC Clean Sample due to its low
  Galactic latitude.} have been discovered in the VHE regime.
{\it{Fermi}}-LAT was implicated in the detection of nine of these
objects \citep{ATel2486,1ES0502+675:ATel2301,PKS1424+240:ATel2084,
VERJ0521+211:ATel2260,MAGICJ2001+435:ATel2753,NGC1275:ATel2916,
4C+21.35:ATel2684,RBS0413:ATel2272,APLib:ATel2743},
a significant percentage of the entire catalog of TeV AGNs ($20\%$).  This demonstrates the close ties between these energy regimes
and also the unique capability of the LAT to provide the Cherenkov
telescopes with prime TeV candidates, which is especially valuable
input for these instruments since they have small fields of view and
low duty cycles ($\sim$10\%). These sources are flagged with asterisks
in Table~\ref{TAB:GeVTeV}.

\begin{figure}
\plotone{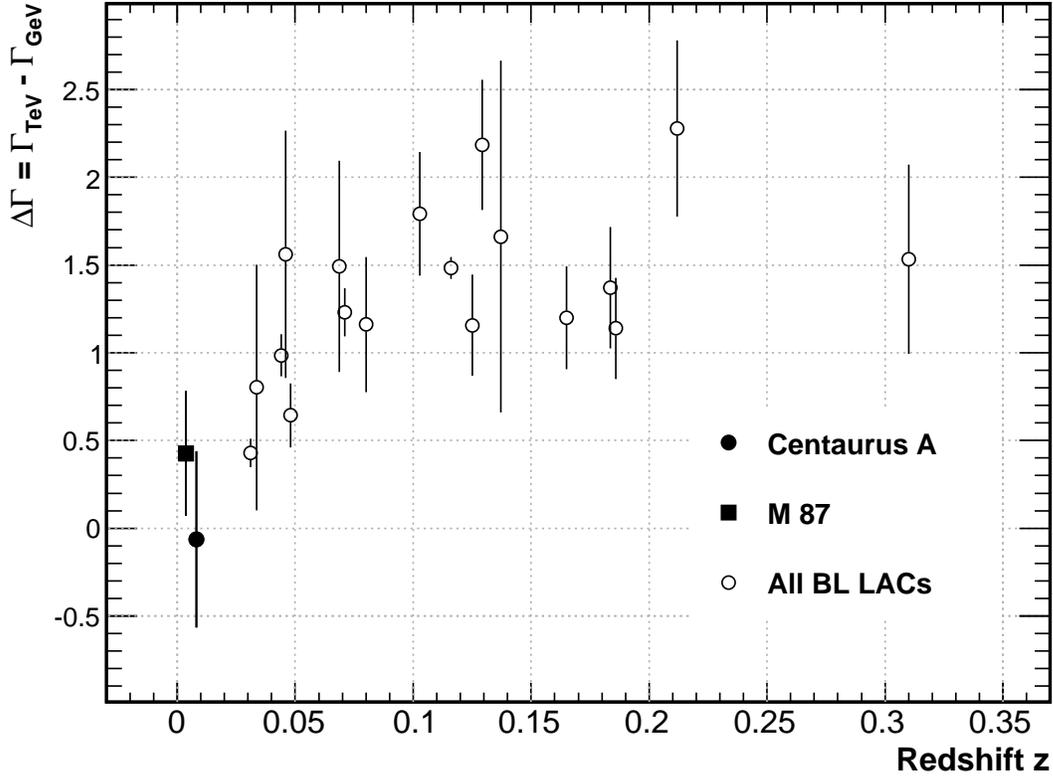}
\caption{The difference in photon index between that measured by {\it{Fermi}}
LAT and that reported in the TeV regime,
$\Delta\Gamma\,\equiv\,\Gamma_{TeV}\,-\,\Gamma_{GeV}$, for the 19
GeV-TeV AGN with reliable redshifts and reported TeV spectra (flagged
in Table~\ref{TAB:GeVTeV}), as a function of their redshift. In
addition to these, 1ES\,2344+514, which is not in the 2LAC clean
sample due to its low Galactic latitude ($b = -9.86$), has also been
included. Its photon index, $\Gamma_{GeV}$, as quoted in 2FGL
\citep{2FGL}, is used. The other two 2FGL TeV AGN that are not
in the Clean Sample due to their low Galactic latitudes (VER\,
J0521+211 and MAGIC\,J2001+435) are not included since no photon index
has yet been reported for them at TeV energies.}
\label{fig:GeVTeVEBL}
\end{figure}

As discussed in \citet{1LAC}, the majority of the GeV-TeV AGNs can be
well fit with power-law (PL) spectra in both $\gamma$-ray energy regimes
although, as detailed below, sometimes a LogParabola spectrum was the
preferred fit in the GeV regime. In many cases, there is a significant
difference between the PL spectral indices measured by {\it{Fermi}}
LAT, $\Gamma_{GeV}$, and by the Cherenkov telescopes, $\Gamma_{TeV}$,
indicating that the spectrum undergoes a break somewhere in the
$\gamma$-ray regime. In the same manner as described in \citet{1LAC},
the difference in photon index between that measured by {\it{Fermi}}
LAT and that reported in the TeV regime,
$\Delta\Gamma\,\equiv\,\Gamma_{TeV}\,-\,\Gamma_{GeV}$, for the GeV-TeV
AGNs with reliable redshifts and reported TeV spectra (flagged in
Table~\ref{TAB:GeVTeV}), are plotted as a function of the redshift in
Figure~\ref{fig:GeVTeVEBL}. It should be noted that the data used to
measure the spectral indices in question were not necessarily
simultaneous. It can be seen that there is a deficit of distant
sources with small values of $\Delta\Gamma$, confirming the trend
previously reported \citep{pg1553,1LAC}. One possible explanation for
this is the effect of the EBL: the $\gamma$-ray photons pair produce
with the photons of the EBL, softening the spectrum in the VHE band in
a redshift-dependent way.

As can be seen in the 2LAC, most of the GeV-TeV AGNs, 26 out of 34,
are best fit with power-law spectra in the {\it{Fermi}}-LAT band
pass. Of these sources, 17 are HSPs, two are ISPs, two are LSPs and
five are GeV-TeV AGN that are unclassified. Out of the three SED
classes, the HSPs have, by definition, their synchrotron peak
frequencies at the highest energies. Thus, in many emission model
scenarios, it is expected that their second SED peak would also occur
at the highest energies. For sources not subject to significant
absorption by the EBL, this means that their spectral turn-over may
occur at higher energies than covered by the 2LAC. Following these
arguments, it is not surprising then that most of the GeV-TeV sources
with power-law spectra in the LAT bandpass are HSPs.

By extension, it would seem likely that at least some of the five
GeV-TeV AGN that were not assigned SED classes by the procedure
described in \S~\ref{sec:sedclass} are HSPs. An examination of the
literature reveals three of them (2FGL\,J0416.8$+$0105/1H\,0413$+$009,
2FGL\,J1101$-$2330/1H\,1100$-$230 and
2FGL\,J2009.5$-$4850/PKS\,2009$-$489) to have been classified as
high-frequency peaked BL Lacs (\citealp{2011arXiv1105.5114V,
  2007A&A...470..475A, 2005A&A...436L..17A}). One of the remaining
sources (2FGL\,J1325.6$-$4300) is associated with the Centaurus\,A
core, a Fanaroff-Riley Type I galaxy. We note that these are all
Southern Hemisphere sources and that, typically, this hemisphere is
not as well surveyed at radio and optical wavelengths. This could be a
factor in the non-classification of their SEDs. The two ISPs that are
best fit by power laws are W\,Comae (2FGL\,J1221.4$+$2814; $z=0.103$)
and PG\,1424$+$240 (2FGL\,J1427.0+2347; the redshift is unknown but
upper limits of $z<1.19$ and $z<0.66$ have been derived by
\citet{2010arXiv1006.4401Y} and \citet{2010ApJ...708L.100A} while
\citet{2011arXiv1101.4098P} estimate $z=0.24\pm0.05$). The two LSPs
that were best fit by power laws are among the closest known GeV-TeV
AGN: AP\,Lib (2FGL\,J1517.7$-$2421; $z=0.048$) and M\,87
(2FGL\,J1230.8$-$1224, z=0.0036), a Fanaroff-Riley Type I galaxy.

Of the eight GeV-TeV whose {\it{Fermi}} LAT spectra are best fit by a
LogParabola, only one, 1H\,1013$+$498 (2FGL\,J1015.1+4925), is
classified as an HSP. With a redshift of $z\,=\,0.212$, this object is
less distant than many of the other sources (both those best-fit by
LogParabolas and by power laws) so the curvature in its spectrum is
not likely to be solely attributable to absorption from the photons of
the EBL. The remaining GeV-TeV sources with LogParabola spectra,
comprise four ISPs and 3 LSPs.

The six TeV AGNs that are not in 2FGL (SHBL\,J001355.9$-$185406,
1ES\,0229+200, 1ES\,0347$-$121, PKS\,0548$-$322, 1ES\,1312$-$423 and
HESS\,J1943+213\footnote{The subclass of this source has not been
  confirmed but all available observations favor its classification as
  a HBL \citep{2011A&A...529A..49H}}) are all high-frequency-peaked BL
Lacs, and are amongst the weakest extragalactic TeV sources detected
to date, with fluxes ranging between 0.4\% and 2\% that of the Crab
Nebula in that energy regime. The fact that it is the weakest TeV HBL
that remain below the 2LAC detection threshold is compatible with the
characteristics of this subclass of AGN, namely, that their second
emission peak occurs at high frequencies and that they have low
bolometric luminosities (when compared to that of the other blazar
subclasses).

\section{Discussion and Summary}

The 2FGL catalog contains 1319 sources at $|b| > 10\arcdeg$, of which 1017 sources are associated at high confidence with AGNs. These constitute the 2LAC. The 2LAC Clean Sample consists of 886 sources (see Table \ref{tab:census}), and is defined by requiring that sources have only one counterpart each and no analysis flags. It includes 395 BL~Lacs, 310 FSRQs, 157 blazars of unknown type, 8 misaligned AGNs, 4 NLSy1 galaxies, 10 AGNs of other types, and 2 starburst galaxies. The 2LAC Clean Sample represents a 48\% increase over the 599 high-latitude AGNs in the 1LAC Clean Sample. This reflects not only the increased exposure, but also follow-up campaigns on individual targets and the availability of more extensive catalogs.  

\subsection{Unassociated Sources and Redshift Incompleteness}

The observed deficit of BL~Lac objects at negative Galactic latitudes compared to positive latitudes (Figure \ref{fig:gal_lat}) is not fully accounted for by blazars of unknown type, suggesting that a significant number of blazars (at least 60) are present in the unassociated sample of 2FGL sources. This deficit results primarily from the greater incompleteness of the current counterpart catalogs at Southern declinations, in particular, the BZCAT \citep{bzcat}, which is biased by the greater number of Northern hemisphere arrays that have better exposure to positive Galactic latitudes. There is furthermore a modest anisotropy in LAT exposure favoring positive Galactic latitudes (Figure \ref{fig:sens}). The lack of extensive archival multiwavelength data also leads to an incomplete characterization of the 2LAC Clean Sample. Consequently we find that
\begin{enumerate}
\item 157 of the 862 blazars in the 2LAC ($\sim 18$\%, referred to as ``of unknown type'') lack firm optical classification. Their photon index distribution (Figure \ref{fig:index} bottom) suggests that they comprise roughly equal numbers of BL~Lacs and FSRQs.      
\item 220 of the 395 BL~Lac objects ($\sim 55$\%) lack measured redshifts, and this fraction is roughly the same for LSP, ISP, and HSP BL Lac objects;
\item 93 of the 395 BL~Lac objects ($\sim 23$\%), and 86 of the 310 FSRQs ($\sim  28$\%), lack SED-based classifications.
\end{enumerate} 
Despite the fact that intensive optical follow-up programs are underway (M.~S. Shaw et al., 2011, in preparation and S. Piranomonte et al., 2011, in preparation), these limitations, as was also the case for the 1LAC, hamper interpretation.  

The smaller error boxes that result from longer exposure fortunately result in fewer multiple associations in 2LAC than in 1LAC. Only 26 2LAC sources have more than one counterpart, whereas 33 sources had more than one counterpart in the 1LAC. Moreover, 2LAC sources have at most two counterparts, while there were cases of three counterparts in 1LAC.  Besides the difference in exposure, comparisons between 1LAC and 2LAC must take several other factors into account \citep[a full description is given in][]{2FGL}: i) the switch from unbinned to binned likelihood analysis; ii) the use of different instrument response functions (``P7\_V6 SOURCE" instead of ``P6\_V3 DIFFUSE"); and  iii) the use of different association methods. None of these changes is, however, expected to affect the number of overall associated sources by more than $\sim  10$\% (the former change leads to a lower 2LAC/1LAC count ratio, while the latter two have the opposite effect). 

Comparisons between the properties of BL Lac objects and FSRQs must carefully take into account the redshift incompleteness, given that more than half of the BL Lac objects in the 2LAC lack redshifts. Because the photon spectral-index distribution of blazars of unknown type differs from both those of BL Lac objects or FSRQs (Fig.\ \ref{fig:index}), the sample of blazars lacking redshifts therefore does not, apparently, represent a uniform subsample of any one class of objects with measured redshift. This incompleteness influences any conclusions concerning luminosity or other properties that depend on knowledge of redshift \citep{1LAC}. For example, strongly beamed emission can overwhelm the atomic line radiation flux and might preferentially arise from high luminosity, high redshift BL Lac objects \citep{Gio11}. These would then be absent in the spectral index/luminosity diagram (Fig.\ \ref{fig:index_L}) and skew the correlation.  Until the redshift incompleteness, the nature of the unassociated sources in the 2LAC, and underlying biases introduced by using different source catalogs \citep{Gio99,Padovani03,Gio11} are resolved, conclusions about the blazar sequence \citep{Fossati98,Ghi98} and the blazar divide \citep{Ghisellini09} remain tentative.
 
The GeV spectra of most FSRQs are softer than those of BL~Lac objects, suggesting that the strength of the emission lines is connected with and possibly determines the position of the  external Compton scattering peak, as would be expected in leptonic scenarios for blazar jets \citep[e.g.,][]{Ghi98,Boe02}. From their general properties, in particular in the $\gamma$-ray band,  LSP BL~Lacs appear to be transitional objects between FSRQs and the general BL~Lac population, confirming the trend established from their broadband SEDs \citep{Ghi11}. Clarifying the relationship between the line luminosities and the broadband SEDs of blazars is crucial to determine the evolutionary connection between various classes of $\gamma$-ray emitting blazars, and whether this is reflected in the blazar sequence.

\subsection{$\log N - \log S$  Distribution}

A complete analysis of the $\log N - \log S$ distribution requires a dedicated study ({\it Fermi}-LAT Collaboration, in preparation). Assuming, however, that the sources at high Galactic latitude are dominated by blazars, and furthermore neglecting the aforementioned effects of different analysis procedures, then the observed increase in the detected number of $|b| > 10^\circ$ sources between 1FGL and 2FGL is roughly compatible with the extrapolation of the integral $\log N - \log S$  derived from the 1LAC to lower fluxes, which exhibits a slope of $\sim -0.6$ at the low-flux end of the distribution \citep{Abdo_EDB}. The roles of source confusion, flux limits of the cataloged AGN data used to make AGN associations, and intrinsic AGN variability must be carefully considered, however.  With respect to the first issue, approximately 8\% of the $|b|>10^\circ$, $TS > 25$ sources were missing because of source confusion in 1FGL \citep{1FGL}, but this fraction went down to $\sim  3.3$\% in the 2FGL due to improved analysis techniques \citep{2FGL}. Source confusion is, of course, even more important for soft sources due to the larger PSF and the lower effective area for detection of lower energy photons that leads to poorer position determination, but should not strongly affect the results presented here. 

Regarding the flux limits of the cataloged sources, Figure \ref{fig:fluxhisto} shows that BL Lac objects are on average much fainter radio sources, with median 8 GHz fluxes nearly an order of magnitude fainter than for FSRQs ($\sim  80$ mJy for BL Lac objects and $\sim  500$ mJy for FSRQs). Incompleteness in radio catalogs therefore would likely be more important for BL Lac objects and especially the HSP BL Lac objects which, if this selection bias were not present, would further increase the fractional number of BL Lacs compared to FSRQs. Finally, concerning the issue of variability, we note that the averaging of fluxes over 2 years will dilute the presence of blazars with small duty cycles on monthly and yearly timescales. 

\begin{figure}
\plotone{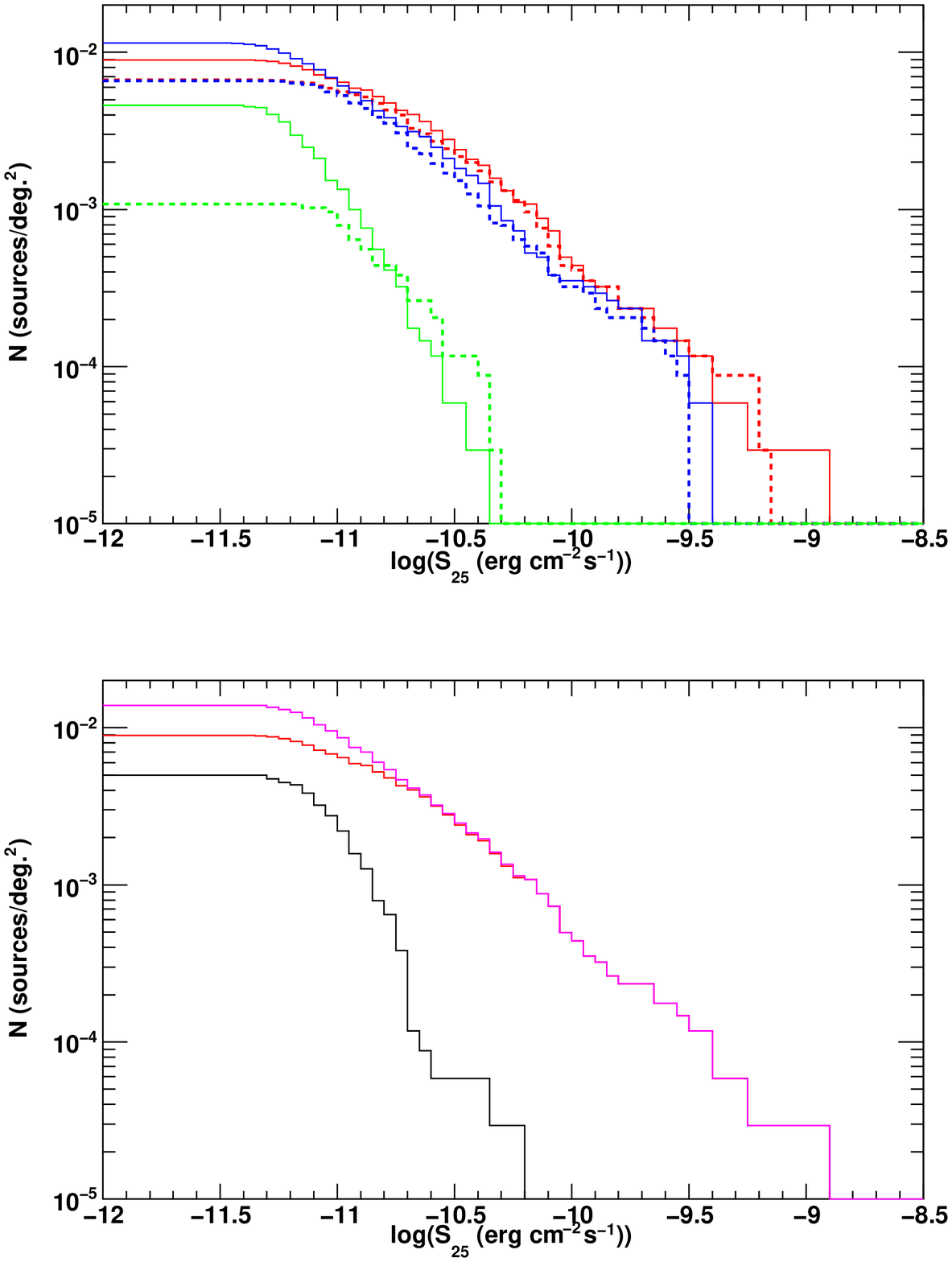}
\caption{Cumulative energy flux distributions (uncorrected for non-uniform sensitivity and detection/association efficiency). Top: FSRQs (red), BL Lacs (blue), blazars of unknown type (green). The solid histograms are for 2LAC, the dashed ones for 1LAC. Bottom: 2LAC FSRQs (red), unassociated 2FGL sources with $|b|>$ 10$\arcdeg$ and $\Gamma>2.2$ (black), sum of two histograms (magenta).
}
\label{fig:logn_logse}
\end{figure}

Threshold sensitivity in terms of photon flux is strongly dependent on source spectral index (Figure \ref{fig:index_flux}), whereas energy flux is not (Figure \ref{fig:index_S}). BL Lacs and FSRQs are both complete to an energy flux of $\sim  5\times 10^{-12}$ erg cm$^{-2}$ s$^{-1}$.  The $\log N - \log S$  energy-flux distribution of unassociated 2FGL sources with $|b| > 10^{\circ}$ and $\Gamma>2.2$ that are potential FSRQ candidates is displayed in Figure \ref{fig:logn_logse} (bottom; black histogram). Adding the $\log N - \log S$  distribution for these sources to that for FSRQs results in the magenta histogram, which exhibits a steeper slope at low fluxes than the case with FSRQs alone.  Thus we conclude that the unassociated sources are likely to be a mixture of FSRQs and BL Lacs, including possibly other source types.

\subsection{Aligned and Misaligned Sources}

The {\it Fermi}-LAT has increased the number of known, high-confidence  $\gamma$-ray emitting BL~Lacs by a factor of $\sim 20$ over the number detected with EGRET \citep{3EGcatalog,mhr01,Din01,Sow03,Sow04}. The number of BL~Lacs has increased by  43\% (395 versus 275) from the 1LAC to 2LAC Clean Samples, while the number of FSRQs has increased by only $\sim  25$\% (310 versus 248). This discrepancy might be even larger due to the evident lack of cataloged southern hemisphere BL~Lac objects, as noted above. Yet the number of misaligned AGNs observed at large, $\gtrsim 10^\circ$ angles to the jet axis, remains small---only 11 were reported in the dedicated Fermi paper on these sources \citep{magn}. Three of these, 3C~78, 3C~111, and 3C~120,  are not now in the 2LAC, evidently due to variability (\S \ref{sec:RG}), illustrating that the jetted component can make a dominant contribution to the $\gamma$-ray emission in radio galaxies.  Two other radio galaxies---Centaurus B and Fornax A---are, however, now included. 

The LAT-detected Fanaroff-Riley II (FR II) radio galaxies and steep spectrum radio quasars (SSRQs) have $\gamma$-ray luminosities $\sim  10^{45}$ -- $10^{46}$ erg s$^{-1}$,  and are found at the faint end of the luminosity distribution of FSRQs, which extends upwards to $\gtrsim 10^{49}$ erg s$^{-1}$.  In comparison, the LAT-detected Fanaroff-Riley I (FR I) radio galaxies have $\gamma$-ray luminosities 2 -- 4 orders of magnitude lower than the lowest typical $\gamma$-ray luminosities, $\sim  10^{44}$ erg s$^{-1}$, of BL~Lac objects \citep[see Figure \ref{fig:index_L} and][]{magn}.  Besides the slow increase in numbers, this raises the interesting and possibly related question why the ratio of measured $\gamma$-ray luminosities of FR I galaxies and BL Lac objects span a much larger range than that for FR II galaxies and FSRQs.  If SSRQs are FSRQs seen at slightly larger angle to the jet axis, then the low-luminosity range of FSRQs could be a mixture of sources with lower-power jets and those with powerful jets, but with slight misalignment. One possibility is that this could be due to different $\gamma$-ray emission beaming factors, with the emission being more beamed in the latter case due to external Compton scattering \citep{dermer1995,Geo2001}. The more rapid fall-off in off-axis flux, combined with the relative paucity of nearby FR II galaxies, could therefore make detection of FR IIs less likely than for the FR Is. Another possibility is that the preferential detection of FR Is over FR IIs reflects the difference in jet structure in FSRQs and BL Lac objects \citep[e.g.,][]{Chiaberge2000,Mey11}, with broader emission cones in BL Lacs that consequently favor the detection of FR Is. Furthermore, extended jet or lobe emission  could be present in the FR Is that is missing in FR II galaxies.  The situation is further complicated in that some LSP BL Lac objects have properties associated with FR II rather than FR I radio galaxies \citep{Kollgaard1992}.

\subsection{Variability}

Monthly light curves established for the whole 2LAC have enabled the confirmation of trends
obtained over a more limited source sample and shorter time span, namely that:
\begin{enumerate}
\item The mean fractional variability on time scales sampled by our data, as given by the normalized excess variance, is higher for FSRQs than for BL Lacs. The normalized excess variance for BL Lacs decreases from LSP to
ISP and HSP BL Lac objects.
\item With the definition of duty cycle used in Section \ref{sec:var} based on monthly-averaged time bin light curves, bright FSRQs and BL Lac objects both have duty cycles of about 0.05 - 0.10.
\item The Power Density Spectra in the frequency range $\sim   (0.033$ -- 0.5) month$^{-1}$ for bright FSRQs and bright BL Lacs of all types are each described by a power law with mean index of $\sim  1.2$ (Figure \ref{fig:PDS}). The discrete auto-correlation and structure function analyses shows that FSRQs display slightly longer correlation timescales and steeper and more broadly distributed structure function indices than HSP BL Lac sources (Figure \ref{fig:ACF} and \ref{fig:SF}). Thus the FSRQs tend to be slightly more ``Brownian-variable," i.e., driven by longer-memory processes, than HSP objects.
\end{enumerate}

Differences between variability properties of BL Lac objects and FSRQs at GeV energies are important for understanding the jet location and jet radiation mechanisms, considering that rapid variability is more likely to be related to emission sites near the central nucleus, whereas extended ($\gtrsim$ kpc) jets can only make weakly variable or quiescent emission. Earlier analysis of GeV light curves indicate that FSRQs have larger variability amplitudes than BL Lacs \citep{Abdo_var}, and this result is confirmed here by considering the normalized excess variance (Figure \ref{fig:index_var}), which also follows from a comparison of BL Lac and FSRQ light curves with similar photon statistics (Figure \ref{fig:varind_relunc}).  The larger variability amplitudes in FSRQs than BL Lacs can be interpreted as a result of shorter cooling timescales of electrons making GeV emission through external Compton processes in FSRQs above the $\nu F_\nu$ peak compared with the longer cooling timescales of the  lower-energy electrons making GeV emission through synchrotron self-Compton processes in HSP BL Lac objects at frequencies below the $\nu F_\nu$ peak \citep{1997ARA&A..35..445U}. This assumes, however, that the jet is long-lived and not subject to adiabatic expansion that would make  achromatic variability at all frequencies.  Radiation from extended jets in BL Lac objects \citep{2008ApJ...679L...9B}, which might be less important in the relatively younger but more powerful FSRQs, could also make a weakly varying high-energy radiation component, as could cascade emission induced by ultra-high energy cosmic-ray protons \citep{Ess09,Ess10,2011ApJ...731...51E}, or the cascade emission from TeV $\gamma$ rays interacting with photons of the extragalactic background light \citep[e.g.,][]{2007A&A...469..857D}.

\subsection{EBL and High Redshift AGNs}

The number of high-energy ($> 10$ GeV) photons from $z> 0.5$ sources that can constrain EBL models  has increased by a factor $\sim 2$ in 2LAC compared with the 11 month data \citep{Abdo_EBL}, due to increased exposure and better background rejection. This should increase further with improvements in our capability to reconstruct event tracks and reject background at high energy \citep{Baldini}. The detection of 30  photons with $E>10$~GeV and $z>0.5$ in the 2LAC that are also above the $\tau_{\gamma\gamma} = 3$ opacity curve predicted by the \cite{Ste06} ``baseline model''  will further constrain this high EBL model. EBL models that produce lower opacity \cite[e.g.][]{Fran08,Fin10b,Gil09} in the $(E,z)$ phase space cannot, however, be constrained in this manner.  The detection of 5 photons in 2LAC with $E>100$~GeV and from $z > 0.5$ sources can probe the EBL at much longer wavelengths than was previously possible with {\it Fermi}-LAT data. 

Remarkably, no source detected in the 2LAC  
is at higher redshift than in the 1LAC, even though the exposure has more than 
doubled. The most distant blazar detected is still at $z = 3.10$. Thus the lower flux limits in the 
2LAC have helped detect fainter objects at lower redshifts, rather than finding objects with comparable 
luminosities as those found in the 1LAC but farther away. 
With the detection limits of the 2LAC,  FSRQs with a $\gamma$-ray luminosity 
of $\sim  10^{48}$ erg s$^{-1}$ (many of which are present 
in the 2LAC at $z \geq 1$, see Figure \ref{fig:L_redshift}) would have been detected up to $z\sim 6$,
and up to $z\sim 4$ for luminosities as low as $10^{47.5}$ erg s$^{-1}$. 
Thus the lack of high-redshift objects is not due to luminosity selection. A change of SED 
properties for blazars at high redshift is suggested by comparing the overlapping sources from the 
BAT survey in the hard X-ray band \citep{Aje09,2010A&A...524A..64C} with LAT samples (7 above $z=2$ and $|b|>10^\circ$, among 14 and 30 sources in the BAT 58-month and 2LAC samples, respectively)
and the fact that none of the more than 50 known luminous FSRQs above $z = 3.10$ in the BZCAT \citep{bzcat}
is detected in the 2LAC. These are likely characterized by a much lower 
$\nu_{peak}$ frequency of the SED \citep[see, e.g.,][]{2010MNRAS.402..497G}, with the $\gamma$-ray 
peak near 1 MeV rather than at $\sim  10$ -- 100 MeV. A source with this type of SED would be very difficult 
to detect with {\it Fermi}, since the LAT band would be sampling the $\gamma$-ray cutoff of the SED, 
but should be easily detectable in the hard X-ray band with upcoming missions like NuSTAR \citep{2010SPIE.7732E..21H}
and Astro-H \citep{2010SPIE.7732E..27T}.

\subsection{Summary of Results}

The 2LAC represents a significant advance with respect to the 1LAC, including many more sources and reduced uncertainties thanks to the doubling of exposure and refinement of the analysis. This has resulted in an $\sim  52$\% (1017 versus 671) increase in the number of associated sources, better localization, more accurate time-averaged spectra, and more detailed light curves and characterization of variability patterns. Despite the problems outlined above concerning the incomplete classification of the 2LAC Clean Sample, the following results---most of which were already found in 1LAC---can be stated with confidence: 

\begin{enumerate}
\item $\gamma$-ray AGNs are almost exclusively blazars, with $\gtrsim 95$\% of the 2LAC sources associated with members of this class. The number of non-blazar sources in the Clean Sample has dropped from 26 to 24 between 1LAC and 2LAC, though part of this reduction is due to variability of sources previously classified as radio galaxies.  There is no compelling evidence for $\gamma$-ray emission from radio-quiet AGNs.

\item BL~Lacs outnumber FSRQs. BL~Lacs, with generally harder spectra, can be detected more easily with the {\it Fermi}-LAT than FSRQs at a given significance limit with increased exposure (as was also the case in the LBAS and 1LAC samples). 

\item A strong correlation is found between spectral index and blazar class for sources with measured redshift. This effect is most clearly visible in the flux-limited sample shown in Figure \ref{fig:index_c}. For that sample, the average photon spectral index
$\langle \Gamma \rangle$ continuously shifts to lower values (i.e., harder
spectra) as the class varies from FSRQs ($\langle \Gamma \rangle = 2.42$) to
LSP-BL~Lacs ($\langle \Gamma \rangle = 2.17$), ISP-BL~Lacs
($\langle \Gamma \rangle = 2.14$), and HSP-BL~Lacs
($\langle \Gamma \rangle = 1.90$). These values are systematically slightly lower, by $\sim$ 0.06 units, than those found in 1LAC.

\item  Among BL~Lacs, HSP sources dominate over ISPs and LSPs. The percentages, $\sim  20$\% 27\%, 53\% for LSPs, ISPs, HSPs, respectively, are essentially the same as for the 1LAC.

\item Due to the flattening of the $\log N - \log S$  distribution for FSRQs (Figure \ref{fig:logn_logse}), increased exposure should yield only a modest addition to the number of such sources.

\item  BL Lac objects and FSRQs display significantly different variability properties. The differences are weaker than those found in the bright LBAS sample \citep{Abdo_var}, probably due to the use of coarser time binning (one month instead of one week) and the inclusion in the larger 2LAC sample of fainter or less variable sources.

\item Most of the 45 TeV AGN have now been detected with {\it Fermi}. Of these, 39 are
in the 2FGL and 34 of these are in the 2LAC Clean Sample. The six that
have not been detected with {\it Fermi} are HSPs. The increase in the break
between the spectral index measured by {\it{Fermi}} and that reported
in the TeV regime as a function of the redshift of the AGNs \citep{FermiTeV} has been confirmed with
this larger sample of GeV-TeV AGNs.

\end{enumerate}

The fact that many sources lack proper classification or a measured redshift calls for a large multiwavelength effort by the blazar community, emphasizing optical spectroscopy when the jet activity is low and the emission line flux is not hidden by nonthermal jet radiation. The general trends identified in the 1LAC, many of them  already apparent in the LBAS, are confirmed. Overall, the 2LAC should allow for a deeper understanding of the blazar phenomenon and the relations between blazar classes.

\section{Acknowledgments}
\acknowledgments The \textit{Fermi} LAT Collaboration acknowledges generous
ongoing support from a number of agencies and institutes that have supported
both the development and the operation of the LAT as well as scientific data
analysis.  These include the National Aeronautics and Space Administration and the Department of Energy in the United States, the Commissariat \`a l'Energie
Atomique and the Centre National de la Recherche Scientifique / Institut
National de Physique Nucl\'eaire et de Physique des Particules in France, the
Agenzia Spaziale Italiana and the Istituto Nazionale di Fisica Nucleare in
Italy, the Ministry of Education, Culture, Sports, Science and Technology
(MEXT), High Energy Accelerator Research Organization (KEK) and Japan
Aerospace Exploration Agency (JAXA) in Japan, and the K.~A.~Wallenberg
Foundation, the Swedish Research Council and the Swedish National Space Board
in Sweden. Additional support for science analysis during the operations phase is gratefully acknowledged from the Istituto Nazionale di Astrofisica in Italy and the Centre National d'\'Etudes Spatiales in France.

This work is partly based on optical spectroscopy observations performed at Telescopio Nazionale Galileo, La Palma, Canary Islands (proposal AOT20/09B and AOT21/10A). Part of this work is based on archival data, software or on-line
services provided by the ASI Science Data Center (ASDC).
This research has made use of the NASA/IPAC Extragalactic Database
(NED) which is operated by the Jet Propulsion Laboratory, California
Institute of Technology, under contract with the National Aeronautics
and Space Administration.

{\it Facilities:} \facility{{\it Fermi} LAT}.

\clearpage

\clearpage

\clearpage

\clearpage

\clearpage

\clearpage

\clearpage
\begin{deluxetable}{lccc}
\tablecolumns{4}
\tablewidth{0pt}
\tabletypesize{\normalsize}
\tablecaption{\label{tbl-ab}Likelihood-ratio parametrization}
\tablehead{
\colhead{Survey} & 
\colhead{$a$} & 
\colhead{$b$} &
\colhead{log(LR$_c$)}
}
\startdata
NVSS & 0.162 $\pm$ 0.001 & 0.744 $\pm$0.004 & $-$0.28\\
SUMSS & 0.50 $\pm$ 0.03 & 0.88 $\pm$ 0.02 & 0.79\\
RASS & 0.70 $\pm$ 0.03 & 0.79 $\pm$0.02 &  1.71\\
PMN & 0.59 $\pm$ 0.03 & 0.88 $\pm$0.02 &  1.36\\
AT20G & 0.59 $\pm$ 0.07 & 0.25 $\pm$0.02 & 2.91\\
\enddata
\end{deluxetable}

\begin{deluxetable}{lccccccccccc}
\tablecolumns{12}
\tabletypesize{\normalsize}
\tablecaption{\label{tbl-assoc}Comparison of association methods}
\rotate
\tablewidth{0pt}
\tablehead{
\colhead{Sample}&
\colhead{Total}&
\colhead{$N_{false}$}&
\colhead{Bayesian}&
\colhead{$N_{false}$}&
\colhead{N$_S$}&
\colhead{LR}&
\colhead{$N_{false}$}&
\colhead{N$_S$}&
\colhead{$\log N - \log S$}&
\colhead{$N_{false}$}&
\colhead{N$_S$}
}
\startdata
 All& 1017& 16.3& 846& 12.5& 2 & 1007& 27.4&113 & 763& 22.7& 6\\
 Clean Sample& 886& 11.7&754& 9.1& 2 & 877& 21.0 &  82 & 691& 19.1 & 5\\
\enddata
\end{deluxetable}

\begin{deluxetable}{llrrcccccccccc}
\setlength{\tabcolsep}{0.025in}
\tablecolumns{14}
\tabletypesize{\scriptsize}
\rotate
\tablewidth{650pt}
\tablecaption{2LAC sample (high latitude)}
\tablehead{
\colhead{2FGL source name} &
\colhead{Counterpart name} &
\colhead{RA} &
\colhead{DEC} &
\colhead{AngSep}&
\colhead{$\theta_{95}$}&
\colhead{Optical Class} &
\colhead{SED Class} &
\colhead{Redshift} &
\colhead{Photon } &
\colhead{Probability} &
\colhead{Probability} &
\colhead{Reliability} &
\colhead{Reliability} \\
\colhead{} &
\colhead{} &
\colhead{($\arcdeg$)} &
\colhead{($\arcdeg$)} &
\colhead{($\arcdeg$)} &
\colhead{($\arcdeg$)} &
\colhead{} &
\colhead{} &
\colhead{}&
\colhead{Index}&
\colhead{Bayesian}&
\colhead{$\log N - \log S$}&
\colhead{LR\_RG}&
\colhead{LR\_XG}\\
}
\startdata
J0000.9$-$0748*&PMN J0001$-$0746&0.32502&$-$7.77411&0.099&0.181&BL Lac&ISP&0&2.39$\pm$0.14&0.98&0.83&0.97&0.81\\
J0001.7$-$4159*&1RXS J000135.5$-$41551&0.38794&$-$41.92392&0.082&0.118&AGU&HSP&0&2.14$\pm$0.19&\nodata&\nodata&0.81&0.89\\
J0004.7$-$4736*&PKS 0002$-$478&1.14842&$-$47.60567&0.022&0.104&FSRQ&LSP&0.88&2.45$\pm$0.09&1.00&1.00&0.99&0.95\\
J0006.1+3821*&S4 0003+38&1.48810&38.33754&0.032&0.133&FSRQ&LSP&0.229&2.60$\pm$0.08&1.00&1.00&0.99&\nodata\\
J0007.8+4713*&MG4 J000800+4712&1.99986&47.20213&0.033&0.058&BL Lac&LSP&0.28&2.10$\pm$0.06&1.00&0.98&0.98&0.96\\
J0008.7$-$2344&RBS 0016&2.14734&$-$23.65775&0.090&0.174&BL Lac&\nodata&0.147&1.62$\pm$0.25&0.99&\nodata&0.92&\nodata\\
J0008.7$-$2344$-$&PKS 0005$-$239&2.00159&$-$23.65512&0.196&0.174&FSRQ&\nodata&1.412&1.62$\pm$0.25&\nodata&\nodata&0.96&\nodata\\
J0009.0+0632$-$&GB6 J0009+0625&2.32097&6.43164&0.125&0.126&AGU&\nodata&\nodata&2.40$\pm$0.16&\nodata&\nodata&0.96&\nodata\\
J0009.0+0632&CRATES J0009+0628&2.26701&6.47266&0.070&0.126&BL Lac&LSP&0&2.40$\pm$0.16&0.99&0.97&0.98&0.91\\
J0009.1+5030*&NVSS J000922+503028&2.34475&50.50801&0.034&0.050&AGU&\nodata&\nodata&1.85$\pm$0.06&\nodata&0.88&\nodata&\nodata\\
J0009.9$-$3206&IC 1531&2.39901&$-$32.27696&0.180&0.147&AGU&LSP&0.025&2.17$\pm$0.16&\nodata&\nodata&0.97&\nodata\\
J0011.3+0054&PMN J0011+0058&2.87641&0.96429&0.078&0.199&FSRQ&LSP&1.4934&2.43$\pm$0.13&0.99&0.99&0.96&\nodata\\
J0012.9$-$3954*&PKS 0010$-$401&3.24980&$-$39.90718&0.007&0.107&BL Lac&\nodata&0&2.16$\pm$0.16&1.00&1.00&0.99&\nodata\\
J0013.8+1907*&GB6 J0013+1910&3.48510&19.17825&0.056&0.158&BL Lac&\nodata&0.473&2.06$\pm$0.19&0.99&1.00&0.97&\nodata\\
J0017.4$-$0018*&S3 0013$-$00&4.04574&$-$0.25404&0.322&0.280&FSRQ&LSP&1.574&2.60$\pm$0.14&\nodata&\nodata&0.97&\nodata\\
J0017.6$-$0510*&PMN J0017$-$0512&4.39900&$-$5.21179&0.030&0.071&FSRQ&LSP&0.226&2.44$\pm$0.07&1.00&1.00&0.99&0.97\\
J0018.5+2945*&RBS 0042&4.61563&29.79174&0.035&0.098&BL Lac&HSP&0&1.24$\pm$0.28&1.00&\nodata&0.95&0.99\\
J0018.8$-$8154*&PMN J0019$-$8152&4.84104&$-$81.88083&0.028&0.134&AGU&HSP&\nodata&2.14$\pm$0.12&\nodata&0.87&0.93&0.96\\
J0019.4$-$5645*&PMN J0019$-$5641&4.86058&$-$56.69525&0.061&0.174&AGU&\nodata&\nodata&2.66$\pm$0.28&0.98&0.88&0.89&\nodata\\
J0021.6$-$2551*&CRATES J0021$-$2550&5.38552&$-$25.84700&0.024&0.079&BL Lac&ISP&0&1.98$\pm$0.11&1.00&0.91&0.98&\nodata\\
J0022.2$-$1853*&1RXS 002209.2$-$185333&5.53816&$-$18.89249&0.020&0.063&AGU&HSP&\nodata&1.53$\pm$0.12&\nodata&0.95&0.97&0.96\\
J0022.3$-$5141*&1RXS 002159.2$-$514028&5.49937&$-$51.67408&0.062&0.150&AGU&HSP&\nodata&2.22$\pm$0.17&\nodata&\nodata&0.85&0.97\\
J0022.5+0607*&PKS 0019+058&5.63526&6.13457&0.013&0.059&BL Lac&LSP&0&2.09$\pm$0.06&1.00&1.00&0.99&\nodata\\
J0023.2+4454*&B3 0020+446&5.89755&44.94339&0.069&0.107&FSRQ&\nodata&1.062&2.36$\pm$0.12&1.00&1.00&0.97&\nodata\\
J0024.5+0346*&GB6 J0024+0349&6.18826&3.81761&0.055&0.166&FSRQ&\nodata&0.545&2.24$\pm$0.16&\nodata&0.97&0.91&\nodata\\
\enddata
\tablecomments{Column 1 and 2 are the 2FGL and counterpart names, columns 3 -- 4 are the coordinates, column 5  gives the angular separation between the $\gamma$-ray position and that of the counterpart, column 6 is the 95\% error radius,  column 7 lists the optical class, column 8 is the  spectral energy distribution (SED) class (depending on the synchrotron peak frequency), column 9 gives the redshift and columns 10-12 report the three probabilities for Bayesian, Likelihood Ratio and $\log N - \log S$ methods respectively. LR$_{RG}$ and LR$_{XG}$ are the reliability values (see Eq. \ref{rel}) for the radio--$\gamma$-ray match and the X-ray--$\gamma$-ray match respectively. * refers to sources in the Clean Sample, $^i$ refers to sources which have been firmly identified, $^-$ refers to counterparts not given  in the 2FGL catalog for sources with double associations. The full table is available in the on-line version and at: http://www.asdc.asi.it/fermi2lac/.}
\label{tab:clean}
\end{deluxetable}

\begin{deluxetable}{llrrcccccccccc}
\tablecolumns{14}
\setlength{\tabcolsep}{0.025in}
\tabletypesize{\scriptsize}
\rotate
\tablewidth{650pt}
\tablecaption{Low Latitude ($|b|<10$) sources}
\tablehead{
\colhead{2FGL source name} &
\colhead{Counterpart name} &
\colhead{RA} &
\colhead{DEC} &
\colhead{AngSep}&
\colhead{$\theta_{95}$}&
\colhead{Optical Class} &
\colhead{SED Class} &
\colhead{Redshift}&
\colhead{Photon }&
\colhead{Probability}&
\colhead{Probability}&
\colhead{Reliability}&
\colhead{Reliability} \\
\colhead{} &
\colhead{} &
\colhead{($\arcdeg$)} &
\colhead{($\arcdeg$)} &
\colhead{($\arcdeg$)} &
\colhead{($\arcdeg$)} &
\colhead{} &
\colhead{} &
\colhead{}&
\colhead{Index}&
\colhead{Bayesian}&
\colhead{$\log N - \log S$}&
\colhead{LR\_RG}&
\colhead{LR\_XG} \\
}
\startdata
J0010.5+6556&GB6 J0011+6603&2.91238&66.06075&0.168&0.190&AGU&\nodata&\nodata&2.41$\pm$0.23&0.87&\nodata&0.91&\nodata\\
J0035.8+5951&1ES 0033+595&8.96930&59.83486&0.019&0.040&BL Lac&HSP&0&1.87$\pm$0.07&1.00&\nodata&0.99&1.00\\
J0047.2+5657&GB6 J0047+5657&11.75224&56.96170&0.031&0.064&BL Lac&\nodata&0&2.06$\pm$0.07&1.00&1.00&0.99&\nodata\\
J0102.7+5827&TXS 0059+581&15.69076&58.40321&0.059&0.059&FSRQ&LSP&0.644&2.28$\pm$0.05&0.99&1.00&0.99&\nodata\\
J0103.5+5336&1RXS 010325.9+533721&15.85868&53.62000&0.026&0.067&AGU&HSP&\nodata&1.75$\pm$0.16&\nodata&\nodata&0.97&0.99\\
J0109.9+6132&TXS 0106+612&17.44394&61.55816&0.026&0.044&FSRQ&LSP&0.785&2.19$\pm$0.06&1.00&1.00&0.99&\nodata\\
J0110.3+6805&4C +67.04&17.55254&68.09483&0.011&0.052&AGU&ISP&\nodata&2.13$\pm$0.08&1.00&1.00&1.00&0.98\\
J0131.1+6121&1RXS 013106.4+612035&22.77986&61.34246&0.014&0.041&AGU&HSP&\nodata&1.91$\pm$0.08&\nodata&\nodata&0.98&1.00\\
J0137.7+5811&1RXS 013748.0+581422&24.45948&58.23698&0.039&0.094&AGU&HSP&\nodata&2.33$\pm$0.12&\nodata&\nodata&0.98&0.99\\
J0241.3+6548&NVSS J024121+654311&40.34080&65.71981&0.089&0.071&AGU&HSP&\nodata&1.97$\pm$0.16&\nodata&\nodata&0.97&0.96\\
J0250.7+5631&NVSS J025047+562935&42.69858&56.49304&0.033&0.108&AGU&\nodata&\nodata&2.25$\pm$0.13&\nodata&\nodata&0.95&0.97\\
J0253.5+5107&NVSS J025357+510256&43.48992&51.04909&0.096&0.087&FSRQ&\nodata&1.732&2.44$\pm$0.07&0.93&0.86&0.98&\nodata\\
J0303.5+4713&4C +47.08&45.89702&47.27117&0.054&0.061&BL Lac&LSP&0&2.24$\pm$0.07&1.00&0.99&1.00&0.95\\
J0303.5+6822&TXS 0259+681&46.09134&68.36020&0.076&0.138&AGU&\nodata&\nodata&2.77$\pm$0.11&0.98&0.99&0.99&0.91\\
J0334.3+6538&TXS 0329+654&53.48632&65.61562&0.046&0.074&AGU&ISP&\nodata&1.82$\pm$0.14&0.99&0.98&0.99&0.96\\
J0359.1+6003&TXS 0354+599&59.76081&60.08954&0.035&0.103&FSRQ&ISP&0.455&2.30$\pm$0.08&0.99&1.00&0.99&0.97\\
J0423.8+4149&4C +41.11&65.98325&41.83412&0.023&0.036&BL Lac&\nodata&0&1.80$\pm$0.06&1.00&1.00&1.00&\nodata\\
J0503.3+4517&1RXS 050339.8+451715&75.91498&45.28299&0.048&0.089&AGU&\nodata&\nodata&1.85$\pm$0.14&\nodata&\nodata&0.95&0.98\\
J0512.9+4040&B3 0509+406&78.21907&40.69547&0.031&0.102&AGU&\nodata&\nodata&1.89$\pm$0.12&0.99&1.00&0.99&0.96\\
J0517.0+4532&4C +45.08&79.36892&45.61742&0.111&0.127&FSRQ&LSP&0.839&2.13$\pm$0.11&0.93&0.93&0.99&\nodata\\
J0521.7+2113&VER J0521+211&80.44167&21.21429&0.009&0.023&BL Lac&ISP&0&1.93$\pm$0.03&1.00&1.00&1.00&1.00\\
J0533.0+4823&TXS 0529+483&83.31617&48.38132&0.039&0.058&FSRQ&LSP&1.16&2.31$\pm$0.05&1.00&1.00&0.99&0.95\\
J0622.9+3326&B2 0619+33&95.71749&33.43628&0.026&0.043&AGU&\nodata&\nodata&2.13$\pm$0.04&1.00&0.99&0.99&\nodata\\
J0643.2+0858&PMN J0643+0857&100.86013&8.96074&0.049&0.069&FSRQ&\nodata&0.882&2.49$\pm$0.09&0.98&0.99&0.99&\nodata\\
\enddata
\tablecomments{Columns 1 and 2 are the 2FGL and counterpart names, columns 3 -- 4 are the coordinates, column 5  gives the angular separation between the $\gamma$-ray position and that of the counterpart, column 6 is the 95\% error radius,  column 7 lists the optical class, column 8 is the  spectral energy distribution (SED) class (depending on the synchrotron peak frequency), column 9 gives the redshift and columns 10-12 report the three probabilities for Bayesian, Likelihood Ratio and $\log N - \log S$  methods respectively. LR$_{RG}$ and LR$_{XG}$ are the reliability values (see Eq. \ref{rel}) for the radio--$\gamma$-ray match and the X-ray--$\gamma$-ray match respectively. $^i$ refers to sources which have been firmly identified, $^-$ refers to counterparts not given  in the 2FGL catalog for sources with double associations. The full table is available in the on-line version and at: http://www.asdc.asi.it/fermi2lac/.}
\label{tab:lowlat}
\end{deluxetable}

\clearpage
\begin{deluxetable}{lccc}
\tablecolumns{4}
\tabletypesize{\footnotesize}
\tablecaption{\label{tab:census}Census of sources}
\tablewidth{0pt}
\tablehead{
\colhead{AGN type}&
\colhead{Entire 2LAC}&
\colhead{2LAC Clean Sample\tablenotemark{a}}&
\colhead{Low-lat sample}
}
\startdata
{\bf All}&{\bf 1017}&{\bf 886}&{\bf 104}\\
\\
{\bf FSRQ}&{\bf 360}&{\bf 310}&{\bf 19}\\
{\ldots}LSP&\phn246&\phn221&\phn7\\
{\ldots}ISP&\phn\phn4&\phn\phn3&\phn2\\
{\ldots}HSP&\phn\phn2&\phn\phn0&\phn0\\
{\ldots}no classification &108&\phn86&10\\

\\
{\bf BL Lac}&{\bf 423}&{\bf 395}&{\bf 16}\\
{\ldots}LSP&\phn65&\phn61&\phn3\\
{\ldots}ISP&\phn82&\phn81&\phn3\\
{\ldots}HSP&174&160&\phn5\\
{\ldots}no classification &102&\phn93&\phn5\\
\\
{\bf Blazar of Unknown type}&{\bf 204}&{\bf 157}&{\bf 67}\\
{\ldots}LSP&\phn24&\phn19&10\\
{\ldots}ISP&\phn13&\phn11&\phn3\\
{\ldots}HSP&\phn65&\phn 53&13\\
{\ldots}no classification &102&74&41\\
\\
{\bf Other AGN}&{\bf \phn30}&{\bf \phn24}&{\bf \phn2}\\
\enddata

\tablenotetext{a}{Sources with single counterparts and without analysis flags. See \S~\ref{sec:cat} for the definitions of this sample.}
\end{deluxetable}

\begin{deluxetable}{llrrccccccccccccc}
\setlength{\tabcolsep}{0.025in}
\tabletypesize{\scriptsize}
\rotate
\tablewidth{500pt}
\tablecaption{1LAC sources missing in 2LAC}
\tablehead{
\colhead{1FGL source name} &
\colhead{1LAC Counterpart name} &
\colhead{RA} &
\colhead{DEC} &
\colhead{Optical Class} &
\colhead{SED Class} &
\colhead{Redshift}&
\colhead{1LAC }&
\colhead{1LAC }&
\colhead{1LAC }&
\colhead{Flags} \\
\colhead{} &
\colhead{} &
\colhead{($\arcdeg$)} &
\colhead{($\arcdeg$)} &
\colhead{} &
\colhead{} &
\colhead{}&
\colhead{Note}&
\colhead{Clean}&
\colhead{Prob}&
\colhead{} \\
}
\startdata
   J0013.7$-$5022 & BZB~J0014$-$5022 & 3.54675 & $-$50.37575 & BLL & HSP & \nodata & S & Y & 1.00& C\\
  J0019.3+2017 & PKS~0017+200 & 4.90771 & 20.36267 & BLL & LSP & \nodata & S & Y & 0.99 & C\\
  J0041.9+2318 & PKS~0039+230 & 10.51896 & 23.33367 & FSRQ & \nodata & 1.426 & S & Y & 0.98 & C\\
  J0202.1+0849 & RX~J0202.4+0849 & 30.61000 & 8.82028 & BLL & LSP & \nodata & S & Y & 0.99 & C\\
  J0208.6+3522 & BZB~J0208+3523 & 32.15913 & 35.38686 & BLL & HSP & 0.318 & S & Y & 1.00& C\\
  J0305.0$-$0601 & CRATES~J0305$-$0607 & 46.25238 & $-$6.12819 & BLL & \nodata & \nodata & S & Y & 0.95 & NC, V\\
  J0308.3+0403 & NGC~1218 & 47.10927 & 4.11092 & AGN & \nodata & 0.029 & S & Y & 0.98 & C\\
  J0343.4$-$2536 & PKS~0341$-$256 & 55.83138 &$-$25.50480 & FSRQ & LSP & 1.419 & S & Y & 0.97 & C\\
  J0422.1+0211 & PKS~0420+022 & 65.71754 & 2.32414 & FSRQ & LSP & 2.277 & S & Y & 0.86 & NC, V\\
  J0457.9+0649 & 4C~+06.21 & 74.28212 & 6.75203 & FSRQ & LSP & 0.405 & S & Y & 0.84 & UnA \\
  J0622.3$-$2604 & CRATES~J0622-2606 & 95.59888 & $-$26.10767 & \nodata & \nodata & \nodata & S & Y & 0.99 & S\\
  J0625.9$-$5430 & CGRaBS~J0625$-$5438 & 96.46771 & $-$54.64739 & FSRQ & LSP & 2.051 & S & Y & 0.99 & BC\\
  J0626.6$-$4254 & CRATES~J0626$-$4253 & 96.53292 & $-$42.89219 & \nodata & \nodata & \nodata & S & Y & 0.89 & CC \\
  J0645.5+6033 & BZU~J0645+6024 & 101.25571 & 60.41175 & AGN & \nodata & 0.832 & S & Y & 0.87 & UnA \\
  J0722.3+5837 & BZB~J0723+5841 & 110.80817 & 58.68844 & BLL & HSP & \nodata & S & Y & 0.95 & NC, V\\
  J0809.4+3455 & B2~0806+35 & 122.41204 & 34.92700 & BLL & HSP & 0.082 & S & Y & 0.99 & C\\
  J0835.4+0936 & CRATES~J0835+0937 & 128.93008 & 9.62167 & BLL & \nodata & \nodata & S & Y & 0.96 & NC, V\\
  J0842.2+0251 & BZB~J0842+0252 & 130.6063 & 2.88131 & BLL & HSP & 0.425 & S & Y & 0.99 & BC\\
  J0850.2+3457 & RX~J0850.6+3455 & 132.65083 & 34.92305 & BLL & ISP & 0.149 & S & Y & 0.99 & C\\
  J0952.2+3926 & BZB~J0952+3936 & 148.06129 & 39.60442 & BLL & HSP & \nodata & S & Y & 0.82 & NC, V\\
  J1007.0+3454 & BZB~J1006+3454 & 151.73527 & 34.91255 & BLL & HSP & \nodata & S & Y & 1.00& NC, V\\
  J1119.5$-$3044 & BZB~J1119$-$3047 & 169.91458 & $-$30.78894 & BLL & HSP & 0.412 & S & Y & 1.00& C\\
  J1220.2+3432 & CGRaBS~J1220+3431 & 185.03454 & 34.52269 & BLL & ISP & \nodata & S & Y & 1.00& C\\
  J1226.8+0638 & BZB~J1226+0638 & 186.68428 & 6.64811 & BLL & HSP & \nodata & S & Y & 0.99 & C\\
  J1253.7+0326 & CRATES~J1253+0326 & 193.44588 & 3.44178 & BLL & HSP & 0.065 & S & Y & 0.99 & C\\
  J1331.0+5202 & CGRaBS~J1330+5202 & 202.67750 & 52.03761 & AGN & \nodata & 0.688 & S & Y & 0.99 & C\\
  J1341.3+3951 & BZB~J1341+3959 & 205.27127 & 39.99595 & BLL & HSP & 0.172 & S & Y & 0.93 & C\\
  J1422.2+5757 & 1ES~1421+582 & 215.66206 & 58.03208 & BLL & HSP & \nodata & S & Y & 0.95 & C\\
  J1422.7+3743 & CLASS~J1423+3737 & 215.76921 & 37.62516 & BLL & \nodata & \nodata & S & Y & 0.90 & S\\
  J1442.1+4348 & CLASS~J1442+4348 & 220.52979 & 43.81020 & BLL & \nodata & \nodata & S & Y & 0.99 & CC \\
  J1503.3+4759 & CLASS~J1503+4759 & 225.94999 & 47.99195 & BLL & LSP & \nodata & S & Y & 0.96 & UnA \\
  J1531.8+3018 & BZU~J1532+3016 & 233.00929 & 30.27468 & BLL & HSP & 0.065 & S & Y & 0.99 & C\\
  J1536.6+8200 & CLASS~J1537+8154 & 234.25036 & 81.90862 & \nodata & \nodata & \nodata & S & Y & 0.82 & CC \\
  J1616.1+4637 & CRATES~J1616+4632 & 244.01571 & 46.54033 & FSRQ & \nodata & 0.95 & S & Y & 0.96 & C\\
  J1624.7$-$0642 & 4C~$-$06.46 & 246.13717 & $-$6.83047 & \nodata & \nodata & \nodata & S & Y & 0.94 & NC\\
  J1635.4+8228 & NGC~6251 & 248.13325 & 82.53789 & AGN & \nodata & 0.025 & S & Y & 0.88 & O\\
  J1735.4$-$1118 & CRATES~J1735$-$1117 & 263.86325 & $-$11.29292 & \nodata & \nodata & \nodata & S & Y & 1.00& C\\
  J1804.1+0336 & CRATES~J1803+0341 & 270.9845 & 3.68544 & FSRQ & \nodata & 1.42 & S & Y & 0.95 & BC\\
  J1925.1$-$1018 & CRATES~J1925$-$1018 & 291.26333 & $-$10.30344 & BLL & \nodata & \nodata & S & Y & 1.00& S\\
  J2006.6$-$2302 & CRATES~J2005$-$2310 & 301.48579 & $-$23.17417 & FSRQ & LSP & 0.833 & S & Y & 0.91 & UnA \\
  J2008.6$-$0419 & 3C~407 & 302.10161 & $-$4.30814 & AGN & \nodata & 0.589 & S & Y & 0.99 & NC, V\\
  J2025.9$-$2852 & CGRaBS~J2025$-$2845 & 306.47337 & $-$28.76353 & \nodata & LSP & \nodata & S & Y & 0.97 & C\\
  J2117.8+0016 & CRATES~J2118+0013 & 319.57250 & 0.22133 & FSRQ & \nodata & 0.463 & S & Y & 0.91 & C\\
  J2126.1$-$4603 & PKS~2123$-$463 & 321.62846 & $-$46.09633 & FSRQ & \nodata & 1.67 & S & Y & 0.98 & S\\
  J2322.3$-$0153 & PKS~2320$-$021 & 350.76929 & $-$1.84669 & FSRQ & \nodata & 1.774 & S & Y & 0.84 & C\\
\enddata
\tablecomments{C= Confirmed 1FGL sources; NC = not  confirmed 1FGL sources \citep[see ][]{2FGL}; BC =  1FGL sources confirmed by the 11-m binned likelihood analysis; S = the 1FGL source was split/resolved in one or more seeds;  O = overlapping $\theta_{99.9}$ error regions with one or more seeds; V = variable source  visible only in the first 11 months; UnA=  while the $\gamma$-ray source is in 2FGL, it is now unassociated due to the displacement of the $\gamma$-ray centroid, CC=  while the $\gamma$-ray source is in 2FGL, its counterpart has changed due to the displacement of the $\gamma$-ray centroid.}
\label{tab:1LAC}
\end{deluxetable}

\begin{center}
\begin{deluxetable}{llrrccrrr}
\setlength{\tabcolsep}{0.025in}
\tabletypesize{\scriptsize}
\rotate
\tablewidth{410pt}
\tablecaption{2LAC sources: Flux table (high-latitude sources)}
\tablehead{
\colhead{2FGL source name} &
\colhead{Counterpart name} &
\colhead{TS}&
\colhead{Radio flux} &
\colhead{X-ray flux} &
\colhead{USNO B1} &
\colhead{SDSS} &
\colhead{$\alpha_{ox}$} &
\colhead{$\alpha_{ro}$}\\
\colhead{} &
\colhead{} &
\colhead{} &
\colhead{mJy} &
\colhead{$10^{-13}$ erg cm$^{-2}$ s$^{-1}$} &
\colhead{Vmag} &
\colhead{Vmag} &
\colhead{} &
\colhead{}\\}
\startdata
J0000.9$-$0748*&PMN J0001$-$0746&46&209&8.10&17.61&\nodata&1.33&0.53\\
J0001.7$-$4159*&1RXS J000135.5$-$41551&45&12&11.01&18.97&\nodata&1.12&0.17\\
J0004.7$-$4736*&PKS 0002$-$478&173&995&14.70&17.30&\nodata&1.12&0.67\\
J0006.1+3821*&S4 0003+38&164&573&14.10&17.72&\nodata&1.20&0.65\\
J0007.8+4713*&MG4 J000800+4712&326&61&14.80&18.28&\nodata&1.08&0.51\\
J0008.7$-$2344&RBS 0016&25&36&\nodata&16.64&\nodata&\nodata&0.42\\
J0008.7$-$2344$-$&PKS 0005$-$239&25&375&\nodata&16.51&\nodata&\nodata&0.55\\
J0009.0+0632$-$&GB6 J0009+0625&43&180&\nodata&19.49&19.19&\nodata&\nodata\\
J0009.0+0632&CRATES J0009+0628&43&247&13.00&18.70&18.10&1.17&0.63\\
J0009.1+5030*&NVSS J000922+503028&310&12&\nodata&19.52&\nodata&\nodata&\nodata\\
J0009.9$-$3206&IC 1531&35&389&5.01&8.91&\nodata&2.78&$-$0.09\\
J0011.3+0054&PMN J0011+0058&49&167&5.41&20.17&20.40&0.86&0.78\\
J0012.9$-$3954*&PKS 0010$-$401&50&495&\nodata&18.09&\nodata&\nodata&0.74\\
J0013.8+1907*&GB6 J0013+1910&25&161&\nodata&18.41&\nodata&\nodata&0.61\\
J0017.4$-$0018*&S3 0013$-$00&38&1086&3.19&19.99&19.17&1.11&0.84\\
J0017.6$-$0510*&PMN J0017$-$0512&185&178&17.40&18.09&\nodata&1.10&0.63\\
J0018.5+2945*&RBS 0042&31&34&143.00&17.47&\nodata&0.90&0.36\\
J0018.8$-$8154*&PMN J0019$-$8152&69&83&29.70&16.35&\nodata&1.32&0.30\\
J0019.4$-$5645*&PMN J0019$-$5641&37&61&\nodata&20.36&\nodata&\nodata&\nodata\\
J0021.6$-$2551*&CRATES J0021$-$2550&116&69&1.72&17.22&\nodata&1.63&0.49\\
J0022.2$-$1853*&1RXS 002209.2$-$185333&141&23&10.90&17.45&\nodata&1.34&0.32\\
J0022.3$-$5141*&1RXS 002159.2$-$514028&36&20&50.30&16.58&\nodata&1.15&0.23\\
J0022.5+0607*&PKS 0019+058&391&340&2.45&19.51&\nodata&1.04&0.82\\
J0023.2+4454*&B3 0020+446&76&141&\nodata&\nodata&\nodata&\nodata&\nodata\\
J0024.5+0346*&GB6 J0024+0349&32&22&\nodata&19.81&\nodata&\nodata&0.61\\
\enddata
\tablecomments{* refers to sources in the Clean Sample, $^-$ refers to counterparts not given  in the 2FGL catalog for source with double associations. The full table is available in the on-line version and at: http://www.asdc.asi.it/fermi2lac/.}
\label{tab:prob1}
\end{deluxetable}

\begin{deluxetable}{llrrccrrr}
\setlength{\tabcolsep}{0.025in}
\tabletypesize{\scriptsize}
\rotate
\tablewidth{410pt}
\tablecaption{Flux table, low-latitude sources}
\tablehead{
\colhead{2FGL source name} &
\colhead{Counterpart name} &
\colhead{TS}&
\colhead{Radio flux} &
\colhead{X-ray flux} &
\colhead{USNO B1} &
\colhead{SDSS} &
\colhead{$\alpha_{ox}$} &
\colhead{$\alpha_{ro}$}\\
\colhead{} &
\colhead{} &
\colhead{} &
\colhead{mJy} &
\colhead{$10^{-13}$ erg cm$^{-2}$ s$^{-1}$} &
\colhead{Vmag} &
\colhead{Vmag} &
\colhead{} &
\colhead{}\\}
\startdata
J0010.5+6556&GB6 J0011+6603&71&64&\nodata&19.70&\nodata&\nodata&\nodata\\
J0035.8+5951&1ES 0033+595&243&148&318.00&18.21&\nodata&1.01&0.37\\
J0047.2+5657&GB6 J0047+5657&201&190&\nodata&19.58&\nodata&\nodata&0.62\\
J0102.7+5827&TXS 0059+581&298&849&3.83&18.06&\nodata&1.61&0.64\\
J0103.5+5336&1RXS 010325.9+533721&44&31&63.70&16.09&\nodata&1.43&0.15\\
J0109.9+6132&TXS 0106+612&1102&305&2.60&19.10&\nodata&1.70&0.48\\
J0110.3+6805&4C +67.04&145&1707&23.20&17.13&\nodata&1.69&0.42\\
J0131.1+6121&1RXS 013106.4+612035&276&20&471.00&19.29&\nodata&1.03&0.15\\
J0137.7+5811&1RXS 013748.0+581422&65&171&252.00&18.40&17.04&1.23&0.29\\
J0241.3+6548&NVSS J024121+654311&70&191&41.60&19.43&\nodata&1.22&0.44\\
J0250.7+5631&NVSS J025047+562935&41&36&34.30&\nodata&\nodata&\nodata&\nodata\\
J0253.5+5107&NVSS J025357+510256&141&430&\nodata&20.24&\nodata&\nodata&0.71\\
J0303.5+4713&4C +47.08&218&964&3.59&17.45&\nodata&1.63&0.68\\
J0303.5+6822&TXS 0259+681&81&1208&\nodata&\nodata&\nodata&\nodata&\nodata\\
J0334.3+6538&TXS 0329+654&51&288&16.60&18.57&\nodata&1.41&0.45\\
J0359.1+6003&TXS 0354+599&90&953&38.80&17.25&\nodata&1.46&0.48\\
J0423.8+4149&4C +41.11&335&1756&\nodata&19.78&\nodata&\nodata&0.72\\
J0503.3+4517&1RXS 050339.8+451715&45&35&75.20&\nodata&\nodata&\nodata&\nodata\\
J0512.9+4040&B3 0509+406&35&877&\nodata&15.81&\nodata&\nodata&\nodata\\
J0517.0+4532&4C +45.08&42&1336&1.55&20.04&\nodata&1.54&0.70\\
J0521.7+2113&VER J0521+211&1542&530&60.20&16.29&\nodata&1.52&0.37\\
J0533.0+4823&TXS 0529+483&400&435&10.80&19.18&\nodata&1.16&0.66\\
J0622.9+3326&B2 0619+33&566&240&\nodata&\nodata&\nodata&\nodata&\nodata\\
J0643.2+0858&PMN J0643+0857&267&543&\nodata&\nodata&17.85&\nodata&0.46\\
\enddata
\tablecomments{$^i$ refers to sources which have been firmly identified, $^-$ refers to counterparts not given  in the 2FGL catalog for source with double associations. The full table is available at: http://www.asdc.asi.it/fermi2lac/}
\label{tab:prob2}
\end{deluxetable}

\begin{deluxetable}{lcccccccccccccccc}
\setlength{\tabcolsep}{0.025in}
\tabletypesize{\scriptsize}
\tablewidth{390pt}
\tablecaption{Properties of the TeV AGN detected by the {\it{Fermi}}
  LAT. The top section of the table shows the 34 AGN that are in the
  clean sample of 2LAC. The bottom section shows the 5 TeV AGN and 1
  TeV unidentified source that are in 2FGL but not in the 2LAC Clean
  Sample.}
\tablehead{
\colhead{} &
\colhead{} &
\colhead{Source} &
\colhead{SED} &
\colhead{} &
\colhead{Spectrum} &
\colhead{}\\
\colhead{TeV Name} &
\colhead{2FGL Name} &
\colhead{Class} &
\colhead{Type} &
\colhead{Redshift} &
\colhead{Type$^a$} &
\colhead{1LAC$^b$}\\}
\startdata
RGB\,J0152$+$017         &   J0152.6$+$0148 & BL Lac       & HSP       & 0.08$\dagger$       & PL       &  ---     \\
3C 66A                   &   J0222.6$+$4302 & BL Lac       & ISP       & ---                 & LP       &  Y       \\
RBS\,0413$^*$            &   J0319.6$+$1849 & BL Lac       & HSP       & 0.19                & PL       &  Y       \\
NGC\,1275$^*$            &   J0319.8$+$4130 & Radio Gal    & ISP       & 0.018               & LP       &  Y       \\
1ES\,0414$+$009          &   J0416.8$+$0105 & BL Lac       & ---       & 0.287               & PL       &  Y       \\
PKS\,0447$-$439          &   J0449.4$-$4350 & BL Lac       & ---       & 0.205               & PL       &  Y       \\
1ES\,0502$+$675$^*$      &   J0508.0$+$6737 & BL Lac       & HSP       & 0.416               & PL       &  Y       \\
RGB\,J0710$+$591         &   J0710.5$+$5908 & BL Lac       & HSP       & 0.125$\dagger$      & PL       &  Y       \\
S5\,0716$+$714           &   J0721.9$+$7120 & BL Lac       & ISP       & 0.31$^c$$\dagger$   & LP       &  Y       \\
1ES\,0806$+$524          &   J0809.8$+$5218 & BL Lac       & HSP       & 0.137097$\dagger$   & PL       &  Y       \\
1ES\,1011$+$496          &   J1015.1$+$4925 & BL Lac       & HSP       & 0.212$\dagger$      & LP       &  Y       \\
1ES\,1101$-$232          &   J1103.4$-$2330 & BL Lac       & ---       & 0.186$\dagger$      & PL       &  Y       \\
Markarian\,421           &   J1104.4$+$3812 & BL Lac       & HSP       & 0.031$\dagger$      & PL       &  Y       \\
Markarian\,180           &   J1136.7$+$7009 & BL Lac       & HSP       & 0.046$\dagger$      & PL       &  Y       \\
1ES\,1215$+$303          &   J1217.8$+$3006 & BL Lac       & HSP       & 0.13                & PL       &  Y       \\
1ES\,1218$+$304          &   J1221.3$+$3010 & BL Lac       & HSP       & 0.18365$\dagger$    & PL       &  Y       \\
W\,Comae                 &   J1221.4$+$2814 & BL Lac       & ISP       & 0.102891$\dagger$   & PL       &  Y       \\
4C\,$+$21.35$^*$         &   J1224.9$+$2122 & FSRQ         & LSP       & 0.433507            & LP       &  Y       \\
M\,87                    &   J1230.8$+$1224 & Radio Gal    & LSP       & 0.0036$\dagger$     & PL       &  Y       \\
3C\,279                  &   J1256.1$-$0547 & FSRQ         & LSP       & 0.536               & LP       &  Y       \\
Centaurus\,A             &   J1325.6$-$4300 & Radio Gal    & ---       & 0.0008$^d$$\dagger$ & PL       &  Y       \\
PKS\,1424$+$240$^*$      &   J1427.0$+$2347 & BL Lac       & ISP       & ---                 & PL       &  Y       \\
H\,1426$+$428            &   J1428.6$+$4240 & BL Lac       & HSP       & 0.129172$\dagger$   & PL       &  Y       \\
1ES\,1440$+$122          &   J1442.7$+$1159 & BL Lac       & HSP       & 0.16309             & PL       &  Y       \\
PKS\,1510$-$089          &   J1512.8$-$0906 & FSRQ         & LSP       & 0.36                & LP       &  Y       \\
AP\,Lib$^*$              &   J1517.7$-$2421 & BL Lac       & LSP       & 0.048               & PL       &  Y       \\
PG\,1553$+$113           &   J1555.7$+$1111 & BL Lac       & HSP       & ---                 & PL       &  Y       \\
Markarian\,501           &   J1653.9$+$3945 & BL Lac       & HSP       & 0.0337$\dagger$     & PL       &  Y       \\
1ES\,1959$+$650          &   J2000.0$+$6509 & BL Lac       & HSP       & 0.047$\dagger$      & PL       &  Y       \\
PKS\,2005$-$489          &   J2009.5$-$4850 & BL Lac       & ---       & 0.071$\dagger$      & PL       &  Y       \\
PKS\,2155$-$304          &   J2158.8$-$3013 & BL Lac       & HSP       & 0.116$\dagger$      & PL       &  Y       \\
BL\,Lacertae             &   J2202.8$+$4216 & BL Lac       & ISP       & 0.0686$\dagger$     & LP       &  Y       \\
B3\,2247$+$381           &   J2250.0$+$3825 & BL Lac       & HSP       & 0.119               & PL       &  Y       \\
H\,2356$-$309            &   J2359.0$-$3037 & BL Lac       & HSP       & 0.165$\dagger$      & PL       &  Y       \\
\hline
IC\,310                  &   J0316.6$+$4119 & Radio Gal    & HSP       & 0.018849            & PL       &  ---     \\
VER\,J0521$+$211$^*$     &   J0521.7$+$2113 & BL Lac       & ISP       & ---                 & PL       &  L       \\
VER\,J0648$+$152$^{*,\star}$     &   J0648.9$+$1516 & AGU          & HSP       & ---                 & PL       &  ---     \\
1RXS\,J101015.9$-$311909 &   J1009.7$-$3123 & BL Lac       & HSP       & 0.143               & PL       &  ---     \\
MAGIC\,J2001$+$435$^*$   &   J2001.1$+$4352 & BL Lac       & ISP       & ---                 & PL       &  L       \\
1ES\,2344$+$514          &   J2347.0$+$5142 & BL Lac       & HSP       & 0.044$\dagger$      & PL       &  L       \\
\enddata

$^*$ Sources for which {\it{Fermi}} LAT data motivated the observations leading to their discovery at TeV energies.
$^\dagger$ The sources used to make Figure~\ref{fig:GeVTeVEBL}.
$^\star$ VER\,J0648$+$152 is listed as an unidentified source in TeVCat. It is spatially consistent with the 2LAC AGN, 2FGL\,J0648$+$1516.
$^a$ The shape of the best fit spectrum: power-law (PL); LogParabola(LP).
$^b$ Sources that are flagged with a ``Y'' were in the 1LAC Clean Sample; those 
with an ``L'' were in 1FGL but not in 1LAC due to their low Galactic latitude . All others were not in 1LAC. 
$^c$ The redshift assumed for this source is uncertain at $z = 0.31 \pm 0.08$ 
and is therefore not listed in 2LAC \citep{2009ApJ...704L.129A}. 
$^d$ The redshift is not in the 2LAC table because, as a member of the local group, 
the redshift does not provide a reliable estimate of its distance. \citet{2007ApJ...654..186F} 
used Cepheid variables to calculate its distance and derived a value of 
$3.42 \pm 0.18$ (random) $\pm 0.25$ (systematic) Mpc, which we converted to redshift 
of $z = 0.0008$, with the tool at this URL: http://www.astro.ucla.edu/$\sim$wright/CosmoCalc.html 
assuming the cosmological values quoted in \S~\ref{sec:intro}.
\label{TAB:GeVTeV}
\end{deluxetable}
\end{center}

\end{document}